\documentclass[journal]{IEEEtran}
\usepackage{cite}
\usepackage{verbatim}
\usepackage{algpseudocode}
\usepackage[ruled]{algorithm2e}

  \usepackage[pdftex]{graphicx}
  \usepackage{caption}
  \usepackage{subcaption}
  \usepackage[font=footnotesize,labelfont=footnotesize]{subcaption}
  \usepackage[font=footnotesize,labelfont=footnotesize]{caption}
  \usepackage{epstopdf}

\usepackage[cmex10]{amsmath}
\usepackage{mathrsfs}
\usepackage{amssymb}
\usepackage{amsfonts}
\DeclareMathOperator\sign{sgn}
\usepackage{relsize}
\usepackage[capitalise]{cleveref}
\usepackage{cleveref}
\makeatletter
\newcommand{\leqnomode}{\tagsleft@true}
\newcommand{\reqnomode}{\tagsleft@false}
\makeatother

\usepackage{amsthm}

\newtheorem{lemma}{Lemma}
\newtheorem{prop}{Proposition}
\newtheorem{remark}{Remark}
\Crefname{figure}{Fig.}{Figs.}
{
      \theoremstyle{plain}
      
}

\usepackage{array}
\usepackage{mdwmath}
\usepackage{mdwtab}

\usepackage{stfloats}

\usepackage{amsfonts}
\usepackage[fleqn,tbtags]{mathtools}
\usepackage{psfrag}

\usepackage{multirow}

\def\dashfill{\cleaders\hbox{-~-}\hfill}

\usepackage[]{algorithm2e}

\newcommand{\blue}{\color{blue}}
\usepackage{color,setspace}
\definecolor{blue}{rgb}{0.2,0.3,0.8}
\newcommand{\red}{\color{red}}

\def\dashfill{\cleaders\hbox{-~-}\hfill}
\makeatletter
\newcommand*{\rom}[1]{\expandafter\@slowromancap\romannumeral #1@}
\makeatother
\begin{document}
\title{On Distributed Estimation in Hierarchical Power Constrained Wireless Sensor Networks}

\author{Mojtaba~Shirazi, 
	Azadeh~Vosoughi,~\IEEEmembership{Senior~Member,~IEEE}
	%\thanks{This research is supported by the NSF under grants CCF-1341966 and CCF-1319770.}
	}

\maketitle

\begin{abstract}
We consider distributed estimation of a random source in a hierarchical power constrained wireless sensor network. Sensors within each cluster send their  measurements  to a cluster head (CH). 
%which collects data from sensors within its cluster and fuses the received signals. CHs send pilot symbols to a remote fusion center (FC) over orthogonal channels, to enable the FC estimate the channels between the CHs and the FC, and then transmit their signals over the same orthogonal channels to the FC that forms the linear minimum mean square error (LMMSE) estimate of the unknown.
CHs optimally fuse the received signals and transmit to the fusion center (FC) over orthogonal fading channels. {\blue To enable channel estimation at the FC, CHs send pilots,} prior to data transmission.  
We derive the mean square error (MSE) corresponding to the linear minimum mean square error (LMMSE) estimator of the source at the FC, {\blue and obtain }
the Bayesian Cram\'{e}r-Rao bound (CRB). 
%establish lower bounds on the MSE, including the Bayesian Cram\'{e}r-Rao bound (CRB). 
%To minimize the MSE subject to network transmit power constraint, we propose a computationally favorable power scheduling scheme which allocates power among i) the CHs for CH-FC pilot transmissions, ii) the sensors within each cluster for sensor-CH data transmissions, iii) the CHs for CH-FC data transmissions. 
Our goal is to find (i) the optimal training power, (ii) the optimal power that sensors in a cluster spend to transmit their amplified measurements to their CH, and (iii) the optimal weight vector employed by each CH for its linear signal fusion, such that the MSE is minimized, subject to a network power constraint.
%We show the superior performance of our proposed power allocation scheme with respect to each optimization variable via comparing it with the scheme that allocates power equally with respect to that optimization variable and optimally with respect to the other two optimization variables, namely: 1) total training power of CHs is a fixed percentage of network transmit power and equal power allocation among CHs for training, 2) equal power allocation among clusters for sensor-CH data communication, 3) equal power allocation among CHs for CH-FC data transmission. 
To untangle the performance gain that optimizing each set of these variables provide, we also analyze three special cases of the original problem, where in each special case, only two sets of variables are optimized across clusters. We define three factors that allow us to quantify the effectiveness of each power allocation scheme in achieving an MSE-power tradeoff that is close to that of the Bayesian CRB. 
%%%%%%%%%%%%%%%%%%%%%%%
%Therefore, the factors $g_t, g_c, g_d$ allow us to quantify how much a particular power allocation scheme is effective in reducing the MSE (compared with another scheme). 
%We can quantify the MSE reduction that each optimization problem offer, %
{\blue Combining the information gained from the factors and Bayesian CRB with our computational complexity analysis  
%we have conduced in Section VI (in Section VI we compare the computational complexity of the algorithm corresponding to ``our proposed power allocation scheme'' and the computational complexity of the algorithms corresponding to the special cases (P1-SC1),(P1-SC2),(P1-SC3))
%
provides the system designer with quantitative complexity-versus-MSE improvement tradeoffs offered by different power allocation schemes.}
%
%
%%%%%%%%%%%%%%%%%%%%%%%%%%%%%%%%%
%
%Our numerical results demonstrate that power allocations among CHs for training as well as CH-FC data transmission are always beneficial for low-region of $P_{tot}$, and power allocation among clusters for sensor-CH data transmission is beneficial for low-region to moderate-region of $P_{tot}$. 
%
%We also numerically investigate how the power allocation obtained from solving the original problem depends on the sensors’ observation qualities, physical layer parameters and $P_{tot}$. 
%
\end{abstract}
%\begin{IEEEkeywords}
%Distributed estimation, random source, hierarchical power constrained WSN, fading channels, channel estimation, LMMSE estimator, transmit power optimization.
%\end{IEEEkeywords}
%
\IEEEpeerreviewmaketitle
\vspace{-.2cm}
\section{Introduction} \label{Introduction}
\vspace{-.1cm}
The plethora of wireless sensor network (WSN) applications, with stringent power constraints, raises challenging technical problems for system-level engineers, one of which is distributed estimation (DES) in a power constrained WSN \cite{mojtaba_azadeh_TSIPN,Shirazi_PIMRC2014,Shirazi_asilomar2014,Vosoughi_Sani_2016,Vosoughi_Sani_2018}. In this work, 
we address DES of a random signal $\theta$ in a WSN, where sensors {\blue are deployed in a large field} and
make noisy measurements of $\theta$. {\blue Due to limited communication range}, however, the battery-powered sensors cannot directly communicate with the fusion center (FC). {\blue Hence, the field is divided into $L$ geographically disjoint zones (clusters) and hierarchically into three tiers: sensors, cluster-heads (CHs) one per cluster, and
the FC \cite{Fang-TWC-2009,Lin-IET-2012,Chaudhary-TSP-2013,dist_det_cluster_2019}.  The implicit assumption is that the communication ranges of CHs are larger, {\red and their energy and computational resources are higher (compared with sensors).}}
%
%than those of sensors. Also, the CHs have higher energy and computational  resources (compared with sensors).
%
%and  act  as  high-power  relays  in  amplify-and-forward (AF) or decode-and-forward (DF) fashion \cite{dist_det_cluster_2019}.
%
%We consider a hierarchical WSN in which sensors are grouped into $L$ clusters. 
%
%Each cluster has {\blue an assigned} CH, which acts as a local FC that collects data from sensors within its cluster, and fuses the collected signals. 
{\red After local signal processing}, CHs  transmit signals {\red received from sensors} over orthogonal fading channels to the FC, whose task is to find an estimate of $\theta$, based on the received signals from CHs \cite{Vetterli-TIT-2003,goldsmith-TSP-2007}. 

%One of the research thrusts in the field of DES is optimizing a network performance metric with respect to energy or power consumption during transmission. \cite{Fang-TWC-2009,Lin-IET-2012,Chaudhary-TSP-2013} studied this problem in a cluster-based sensor network. Assuming a linear minimum mean square error (LMMSE) fusion rule at the FC, \cite{Fang-TWC-2009,Lin-IET-2012} solved the problem of finding the optimal power allocation among different clusters as well as the optimal collaboration (amplification matrix) among the sensors within each cluster under a total power constraint, for noncoherent and coherent MAC, respectively. 
There is a rich body of literature on DES {\blue and distributed detection} in a power constrained WSN, where the researchers study and optimize {\blue an estimation-theoretic-based or a detection-theoretic-based performance metric},  subject to power constraints. {\blue Examples in the context of distributed detection are \cite{Ekman_TWC_2016,Salcedo_TSP_2015,Hamid-Azadeh-SP-letter,Hamid_Azadeh_TWC_2013,Hamid_Azadeh_TVT_2018}.} 
An alternative direction is to study the outlier contamination of the data in WSNs by outlier detection methods such as \cite{Mahleqa_M1,Mahleqa_M2,Matin_1} caused by imperfect sensors and power deteriorations \cite{Mahleqa_complementory,Matin_3}. We focus on power optimization and 
to conserve space, we  elaborate only the most related ones to our current work in the following. The authors in \cite{Fang-TWC-2009,Lin-IET-2012,Chaudhary-TSP-2013} studied DES in a {\blue three-layered} hierarchical power constrained WSN, assuming that the FC forms the linear minimum mean square error (LMMSE) estimate of random $\theta$, and the objective is to minimize the MSE of this estimator. 
%
%
%In particular, \cite{Fang-TWC-2009,Lin-IET-2012} found the optimal power allocation among the clusters and the optimal collaboration matrix among the sensors within each cluster, when CHs communicate with the FC over noncoherent and coherent multiple access channel (MAC). 
%In \cite{Chaudhary-TSP-2013}, the authors considered a different observation model where the sensors in cluster $l$ make noisy measurements of correlated $\theta_l, l=1,...,L$. Assuming CH$_l$ forms the local LMMSE estimate of $\theta_l$ and the FC forms the LMMSE estimate of all $\theta_l$'s, the authors in \cite{Chaudhary-TSP-2013} studied the optimal power allocation among sensors within each cluster and among CHs, subject to individual cluster and total network power constraints. 
%Banavar et al. focused on the effects of partial CSI at the sensors in a WSN deployed for the estimation of a global physical parameter, while they assumed perfect CSI at the FC, and calculated the variance of the estimate at the FC \cite{Banavar_TSP_2010}. They also studied the power allocation among sensors when perfect CSI is available at the sensors.
%
%
{\blue The authors in} \cite{Banavar_TSP_2010} considered DES in a WSN,
%with one FC only, 
where sensors transmit to the FC over orthogonal fading channels and the FC finds the LMMSE estimate of $\theta$. The authors studied how partial channel state information (CSI) at the sensors 
%(as opposed to perfect CSI) 
affects the MSE performance and the optimal power allocation among the sensors. 
%A similar scenario was considered in \cite{Dey_Eurasip_2011}, where the authors focused on a clustered WSN, 
{\blue The authors in} \cite{Dey_Eurasip_2011} considered DES in a hierarchical WSN, where the CHs amplify and forward their received signals over orthogonal Nakagami fading channels to the FC. %They investigated the optimum power allocation to minimize the distortion outage probability at the FC subject to an expected sum transmit power constraint, when different amounts of channel information are available at the CHs: 1) full CSI is available at the CHs, 2) only rate-limited channel feedback is available at the CHs. The works in \cite{Fang-TWC-2009,Lin-IET-2012,Chaudhary-TSP-2013,Banavar_TSP_2010,Dey_Eurasip_2011} do not 
Assuming the FC finds the LMMSE estimate of $\theta$, the authors studied how partial CSI at the CHs impacts the outage probability of the MSE.
%
%
%as well as the optimal power allocation among the CHs, subject to an average sum transmit power constraint. 
%
%
%We note that the underlying assumption in \cite{Banavar_TSP_2010,Dey_Eurasip_2011} to characterize partial CSI at the sensors (or the CHs) is that the FC feeds back a quantized version of the perfectly known fading coefficients to the sensors (or the CHs) for transmit power adaptation. 
%
None of the works in \cite{Banavar_TSP_2010,Dey_Eurasip_2011} consider the cost of channel estimation at the FC. 
To enable channel estimation at the FC, each CH needs to transmit a training (pilot) symbol, prior to data symbol. In a hierarchical WSN, where there is a cap on the network transmit power, the cost of channel estimation cannot be overlooked. 
%channel estimation introduces a new dimension to the overall network performance analysis and power allocation optimization. 
Note that training symbol transmission consumes the power that could have been used otherwise for data symbol transmission. Hence, {\blue {\it training and data transmit power should be optimized
%(in terms of power consumption for sending training symbols from the CHs to the FC, the power that could have been used otherwise for data transmission), and also the effects of channel estimation errors on the estimation performance as well as data power allocation. Since the network is power limited, both channel estimation and data transmission have to be performed 
judiciously, such that the estimation accuracy of $\theta$ at the FC is maximized}}. 
%To enable estimating the unknown fading coefficients at the FC, a common practice is {\it training-based channel estimation}, where each CH transmits a pilot symbol prior to its data symbol. The FC with the coherent receiver is tasked with estimating $\theta$ based on all the received signals from CHs. In a hierarchical power constrained sensor network, where there is a cap on the total transmit power for sensors-to-CH and CHs-to-FC communications, channel estimation introduces a new dimension to the power allocation problem: how should we allocate the power among different clusters for sensors-to-CH communications as well as different CHs for CHs-to-FC communications, and how should each CH allocate its transmit power to the training and data symbols? 

Assuming the FC employs the LMMSE estimator of $\theta$, we address this problem, by formulating and solving a new optimization problem that allows us to analyze the effect of channel estimation on the MSE performance and transmit power allocation. The optimization problem is novel, since considering training transmit power introduces a new dimension to the network performance analysis and power allocation optimization. 
In this regard, the most relevant works are \cite{Cihan_TSP_2008,Wu_Elsevier_2011}, where the authors considered channel estimation for DES in a WSN with one FC only.
%assuming LMMSE fusion rule at the FC with orthogonal and coherent MAC models, respectively, the authors investigated the effects of imperfect CSI at the FC for DES problem by using a two-phase approach, where in the first phase, sensors transmit pilots and the FC estimates the fading channels, and in the second phase, sensors transmit amplified noisy observations of the source and the FC estimates the source using these observations and the channel estimates. These works obtained analytical closed-form solutions for the problems of optimal power allocation scheme in which training power and data power for each sensor are optimized, and equal power allocation scheme in which training power is optimized while data power for each sensor is set equal, under a network power constraint.
%The aforementioned works \cite{Cihan_TSP_2008,Wu_Elsevier_2011} follow a two-phase procedure where, in the first phase, sensors transmit pilots and the FC estimates the fading channels. In the second phase, sensors transmit amplified noisy observations of the source, and the FC estimates the source using these observations and the channel estimates.
%Despite the works in \cite{Cihan_TSP_2008,Wu_Elsevier_2011}, in our work we consider hierarchical WSN and thus the power consumed for transmission of observations from sensors to CH is also considered in the power allocation problem. 
Our work is different from \cite{Cihan_TSP_2008,Wu_Elsevier_2011}, since in the hierarchical WSN, our problem formulation considers power distribution among different clusters for sensor-CH data transmission as well as power allocation among different CHs for CH-FC data and training transmissions. 
Moreover, we obtain the optimal linear fusion rules at CHs as the by-product of solving the network power allocation problem{\blue \footnote{{\blue We note that there is a rich body of literature on clustering algorithms and 
energy efficient routing protocols \cite{Vergados_CST_2013}.  Similar to \cite{Fang-TWC-2009,Lin-IET-2012,Chaudhary-TSP-2013,dist_det_cluster_2019}, we assume that clusters and their CHs are given.  Given this network structure, our goal is designing
(sub-)optimal distributed signal processing such that the MSE distortion at the FC is minimized, under a network power constraint.}}}.  
{\blue {\bf Contribution}}: {\blue We derive the MSE corresponding to the LMMSE estimator of $\theta$ at the FC, denoted as $D$,} and establish lower bounds on $D$, including the Bayesian Cram\'{e}r-Rao bound (CRB). We then formulate a new constrained optimization problem that minimizes $D$, subject to network transmit power constraint $P_{tot}$, where the optimization variables are: i) training power for CH$_l$, ii) total power that sensors in cluster $l$ spend to transmit their amplified measurements to CH$_l$ (which we refer to as intra-cluster power), iii) power that CH$_l$ spends to send its fused signal to the FC. 
We demonstrate the superior performance of our proposed power allocation scheme with respect to the following spacial case schemes: 
scheme (i) allots a fixed percentage of $P_{tot}$ for training power and distributes this power equally among CHs, however, it optimally allocates intra-cluster power among clusters, and optimally allocates power among CHs for data transmission, scheme (ii) optimally allocates power among CHs for training, equally allocates intra-cluster power among clusters, and optimally allocates power among CHs for data transmission, scheme (iii) optimally allocates power among CHs for training, optimally allocates intra-cluster power among clusters, and equally allocates power among CHs for data transmission. We 
%also investigate how the proposed power allocation varies as the parameters of the observation and communication models, and the network transmit power constraint change.
analytically and numerically compare the power allocation scheme obtained from solving the original problem with the special case schemes, and show their effectiveness in providing an MSE-power tradeoff that is close to that of the Bayesian CRB. 
Our numerical results demonstrate that power allocations among CHs for training and CH-FC data transmission are always beneficial for low-region of $P_{tot}$, and power allocation among clusters for sensor-CH data transmission is beneficial for low-region to moderate-region of $P_{tot}$. 

{\blue {\bf Organization}: The rest of the paper is organized as follows. Section \ref{System Model and Problem Formulation} describes our system model and power constraints and states the problem we aim to solve (i.e., the constrained minimization of MSE $D$ at the FC, with respect to three sets of optimization variables). Section \ref{LMMSE estimation} characterizes $D$ and its lower bounds. We also derive the Bayesian CRB. In Section \ref{solving max problem} we solve our proposed constrained MSE  minimization problem. We also briefly discuss the constrained minimization of MSE lower bounds. In Section \ref{Special Cases of P1} we solve three special cases of the original problem, where in each special case, only two (of three) sets of variables are optimized across clusters. This analysis allows us to entangle the performance gain that optimizing each set of these variables provide. Section \ref{comp complexity} compares the computational complexity of the proposed algorithms for solving the original problem as well as its three special cases. 
In Section \ref{conv of alg to solve P_A} we discuss the convergence analysis of our proposed algorithms. Section \ref{simulation} presents our numerical and simulation results. Section \ref{conclusions} concludes the work and outlines our future research directions. }

{\bf Notations}: Matrices are denoted by bold uppercase letters, vectors by bold lowercase letters, and scalars by normal letters. $\mathbb{E}$ denotes the mathematical expectation operator, $[.]^T$ represents the matrix-vector transpose operation, and $|\cal A|$ is the cardinality of set $\cal A$. 
The real and imaginary parts of a complex random variable $x$ are represented by $x_r\!=\!{\cal{R}}e\{x\}$ and $x_i\!=\!{\cal{I}}m\{x\}$. The probability distribution function (pdf) of $x$, denoted as $f(x)$, is defined as the joint pdf of $x_r$ and $x_i$, i.e., we have $f(x)\!=\!f(x_r,x_i)$ \cite{proakis_book}.
%
%
%%%%%%%%%%%%%%%%%%%%%%%%%%%%%
\vspace{-.2cm}
\begin{table}[h!]
  \begin{center}
    {\blue \caption{ Notations and their corresponding definitions.}
    \vspace{-1mm}
    \label{table-symbol}
    \begin{footnotesize}
    \begin{tabular}{l|l} % <-- Alignments: 1st column left, 2nd middle and 3rd right, with vertical lines in between
      \textbf{Notation} & \textbf{Vector and Matrix Definitions} \\
      \hline \hline
      $\boldsymbol{x}_l, \boldsymbol{t}_l$ & $\boldsymbol{x}_l\!=\![x_{l,1},...,x_{l,K_l}]^T$, $\boldsymbol{t}_l\!=\![t_{l,1},...,t_{l,K_l}]^T$\\
       $\boldsymbol{n}_l, \boldsymbol{q}_l$ & $\boldsymbol{n}_l\!=\![n_{l,1},...,n_{l,K_l}]^T$, $\boldsymbol{q}_l\!=\![q_{l,1},...,q_{l,K_l}]^T$ \\
     $\sqrt{\boldsymbol{A}_l}$ & $\sqrt{\boldsymbol{A}_l} \!=\!\text{diag}(\sqrt{{\alpha}_{l,1}},...,\sqrt{{\alpha}_{l,K_l}})$ \\                
       $\boldsymbol{x}, \boldsymbol{t}$ &  $\boldsymbol{x}\!=\![{\boldsymbol{x}_1}^T,...,{\boldsymbol{x}_L}^T]^T$, $\boldsymbol{t}\!=\![{\boldsymbol{t}_1}^T,...,{\boldsymbol{t}_L}^T]^T$  \\       
      $\boldsymbol{y}, \boldsymbol{z}$ &  $\boldsymbol{y}\!=\![y_1,...,y_L]^T$, $\boldsymbol{z}\!=\![z_1,...,z_L]^T$ \\ $\boldsymbol{n}, \boldsymbol{q}$ & $\boldsymbol{n}\!=\![{\boldsymbol{n}_1}^T,...,{\boldsymbol{n}_L}^T]^T$, $\boldsymbol{q}\!=\![{\boldsymbol{q}_1}^T,...,{\boldsymbol{q}_L}^T]^T$  \\                        
      $\boldsymbol{v}, \boldsymbol{H}$ & $\boldsymbol{v}\!=\![v_1,...,v_L]^T$, $\boldsymbol{H}\!=\!\text{diag}([h_1, ..., h_L])$  \\  
      $\boldsymbol{M}, \boldsymbol{W}$ & $\boldsymbol{M}\!=\!\text{diag}(\sqrt{\boldsymbol{A}_1}, ..., \sqrt{\boldsymbol{A}_L})\!$, $\boldsymbol{W}\!=\!\text{diag}({\boldsymbol{w}_1}^T, ..., {\boldsymbol{w}_L}^T)$  \\ $\boldsymbol{\Sigma}_{n}, \boldsymbol{\Sigma}_{n_l}$ & $\boldsymbol{\Sigma}_{n}\!=\!\text{diag}(\boldsymbol{\Sigma}_{n_1}, ..., \boldsymbol{\Sigma}_{n_L})$, $\boldsymbol{\Sigma}_{n_l}$ is arbitrary \\   
      $\boldsymbol{\Sigma}_{q}, \boldsymbol{\Sigma}_{q_l}$ &  $\boldsymbol{\Sigma}_{q}\!=\!\text{diag}(\boldsymbol{\Sigma}_{q_1}, ..., \boldsymbol{\Sigma}_{q_L})$, $\boldsymbol{\Sigma}_{q_l} \!=\!\text{diag}(\sigma_{q_{l,1}}^2,...,\sigma_{q_{l,K_l}}^2)$ \\$\boldsymbol{\Sigma}_{v}$ &  $\boldsymbol{\Sigma}_{v}\!=\!\text{diag}([2\sigma_{v_{1}}^2, ..., 2\sigma_{v_{L}}^2])$ \\          
      $\hat{\boldsymbol{H}}, \tilde{\boldsymbol{H}}$ & $\hat{\boldsymbol{H}}\!=\!\text{diag}([\hat{h}_1, ..., \hat{h}_L]), \tilde{\boldsymbol{H}}\!=\!\text{diag}([\tilde{h}_1, ..., \tilde{h}_L])$ \\ $\boldsymbol{\Gamma}, \boldsymbol{\Sigma}$ & $\boldsymbol{\Gamma}\!=\!\text{diag}([\zeta_1, ..., \zeta_L])$, $\boldsymbol{\Sigma}\!=\!\text{diag}(\boldsymbol{\Sigma}_1, ..., \boldsymbol{\Sigma}_L)$ \\ $\boldsymbol{\mu}, \boldsymbol{\Lambda}_1$ & $\boldsymbol{\mu}\!=\![{\boldsymbol{\mu}_1}^H,...,{\boldsymbol{\mu}_L}^H]^H$, $\boldsymbol{\Lambda}_1\!=\!\text{diag}(\boldsymbol{\Lambda}_{1_1}, ..., \boldsymbol{\Lambda}_{1_L})$ \\
      $\boldsymbol{D}_l$ & $\boldsymbol{D}_l\!=\!\text{diag}([\sqrt{d_{l,1}}, ..., \sqrt{d_{l,K_l}}])$
    \end{tabular}
    \end{footnotesize}   
    }
   \end{center}     
   \vspace{-6.5mm}  
\end{table}
\begin{figure*}[t]
	\centering
	\includegraphics[width=5.6in,height=1.65in]{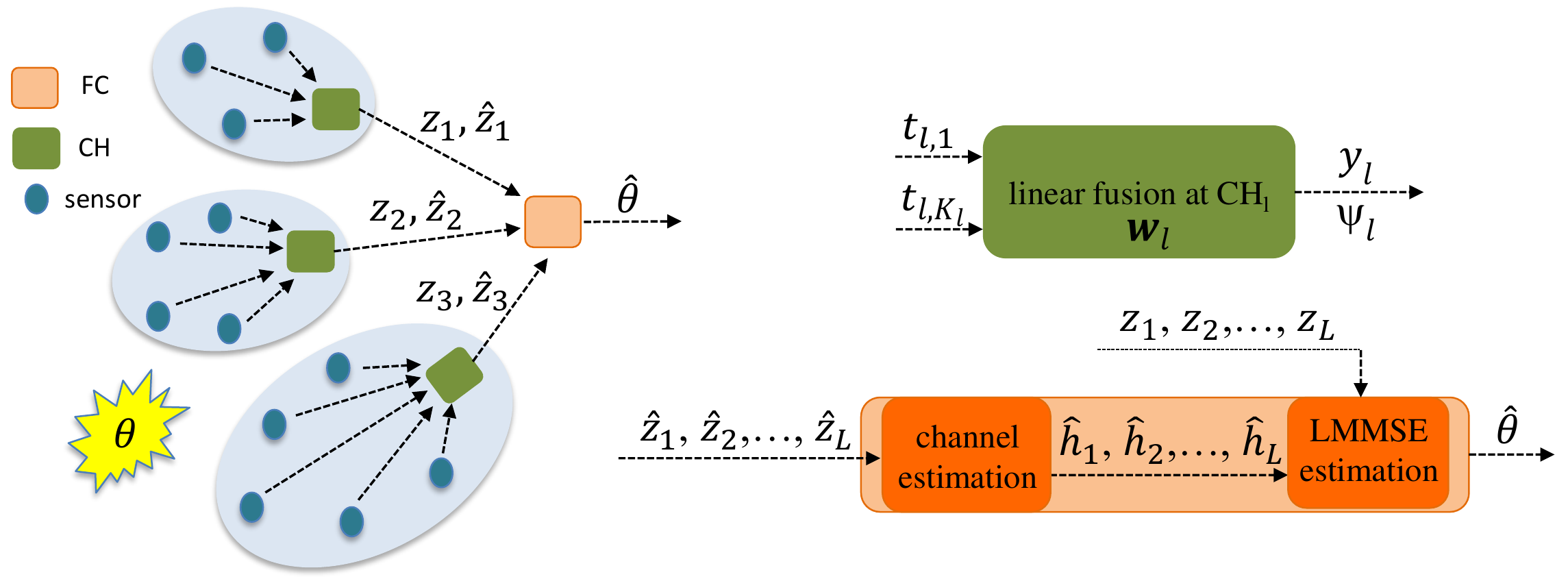}
	\vspace{-0.1cm}
	\caption{Our system model consists of $L$ clusters, each with a CH, and a FC that is tasked with estimating a random scalar {\boldmath$\theta$}.}
	\label{system_model}
	\vspace{-.5cm}
\end{figure*}
%=======================================
%=======================================
%=======================================

%\vspace{-.1cm}
%=======================================
%=======================================
%=======================================
\vspace{-.15cm}
\section{System Model and Problem Formulation} \label{System Model and Problem Formulation}
\vspace{-.05cm}
\subsection{System Model Description} \label{System Model}
\vspace{-.1cm}
We consider a DES problem in a hierarchical power constrained WSN (see Fig.~\ref{system_model}), consisting of $K$ spatially-distributed sensors deployed in $L$ disjoint clusters, $L$ cluster heads (CHs), and a FC. Each sensor makes a noisy measurement of an unknown random variable $\theta$, that we wish to estimate at the FC. Cluster $l$ includes $K_l$ sensors and its associated CH, denoted as CH$_l$, and we have $\sum_{l=1}^{L}K_l=K$. 
We assume $\theta$ is zero-mean with variance $\sigma^2_{\theta}$. Let $x_{l,k}$ denote the measurement of sensor $k$ in cluster $l$. We have:
\reqnomode
\vspace{-.2cm}
\begin{equation} \label{obs of sensors}
x_{l,k}=\theta+n_{l,k},\ \ l=1,..., L,\ \ k=1,..., K_l,
\vspace{-.1cm}
\end{equation}
where $n_{l,k}\!$ denotes zero-mean additive measurement noise with variance $\sigma_{n_{l,k}}^2\!$. {\blue We assume that $n_{l,k}$'s are correlated across sensors, due to their proximity within cluster $l$.}
Sensors within a cluster amplify and forward their measurements to their respective CH over orthogonal AWGN channels{\blue \footnote{{\blue The AWGN channel model is equivalent to  the channel model with a static and known channel gain. Given the channel gain, CH$_l$ can equalize it, which is equivalent to scaling the communication noise variance $\sigma_{q_{l,k}}^2$. The AWGN channel model for communication channels within a cluster is reasonable, since sensors 
are  closely  located  and  typically there  are  direct  line  of  sight  transmissions between sensors and their CH \cite{Sayeed_JSAC_2005,dist_det_cluster_2019}. On the other hand, we  model  the  communication channels  between
CHs  and  the  FC  as  randomly-varying  fading  channels  that  require  channel  estimation.  The reason is that  the  transmission  distances  between  CHs  and  the  FC  are large   and  hence  communication  becomes  subject  to
multipath  fading effect.}}}, such that the received signal at CH$_l$ from sensor $k$ within cluster $l$ is:
\vspace{-.25cm}
\begin{equation} \label{transmitted obs from sensor to CH}
t_{l,k}=\sqrt{{\alpha}_{l,k}}x_{l,k}+q_{l,k},\ \ l=1,..., L,\ \ k=1,..., K_l,
\vspace{-.1cm}
\end{equation}
where ${\alpha}_{l,k}\geq 0$ is an amplifying factor (to be determined) used by sensor $k$, and $q_{l,k}\sim {\cal N}(0,\sigma_{q_{l,k}}^2)$ is the additive communication channel noise. 
%is zero-mean noise\footnote{} with variance $\sigma_{q_{l,k}}^2$. 
We assume that $q_{l,k}$'s are uncorrelated across the sensors. 
%and also are uncorrelated with $\theta$ and $n_{l,k}$'s, $\forall k, l$. 
For a compact representation, we define the column vectors $\boldsymbol{x}_l$ and $\boldsymbol{t}_l$ {\blue in Table \ref{table-symbol}} corresponding to cluster $l$ and rewrite \eqref{obs of sensors} and \eqref{transmitted obs from sensor to CH} as:
\vspace{-.2cm}
\begin{equation} \label{x_l and t_l in cluster l}
\boldsymbol{x}_l=\theta\boldsymbol{1}_l+\boldsymbol{n}_l,\ \ \boldsymbol{t}_l=\sqrt{\boldsymbol{A}_l}\boldsymbol{x}_l+\boldsymbol{q}_l,\ \ l=1,..., L.
\vspace{-.1cm}
\end{equation}
%
%in which $\boldsymbol{1}_l\!=\![1,...,1]^T$ is an $K_l\times 1$ vector of all ones, and $\boldsymbol{n}_l\!=\![n_{l,1},...,n_{l,K_l}]^T$ is zero-mean measurement noise vector with covariance matrix $\boldsymbol{\Sigma}_{n_l}$ that captures the correlation among measurement noises in cluster $l$, $\sqrt{\boldsymbol{A}_l}\!=\!\text{diag}([\sqrt{{\alpha}_{l,1}}, ..., \sqrt{{\alpha}_{l,K_l}}])$ is the amplification matrix, and $\boldsymbol{q}_l\!=\![q_{l,1},...,q_{l,K_l}]^T$ is zero-mean communication channel noise vector with covariance matrix $\boldsymbol{\Sigma}_{q_l}\!=\!\text{diag}([\sigma_{q_{l,1}}^2, ..., \sigma_{q_{l,K_l}}^2])$. 
%We adopt the following assumptions for this setting:
%The auto-covariance matrices $\boldsymbol{\Sigma}_{n_l}$ and $\boldsymbol{\Sigma}_{q_l}$ are available at CH$_l$ and the FC. 
%
%
%
{\blue where $\boldsymbol{1}_l$ is a column vector of $K_l$  ones, 
column vectors $\boldsymbol{n}_l$, $\boldsymbol{q}_l$, with covarinace matrices $\boldsymbol{\Sigma}_{n_l}, \boldsymbol{\Sigma}_{q_l}$,  
matrix $\sqrt{\boldsymbol{A}_l}$ defined in Table \ref{table-symbol}.
}
We assume $\boldsymbol{n}_l$, $\boldsymbol{q}_l$,  $\theta$ are uncorrelated, i.e.,   $\mathbb{E}\{\boldsymbol{n}_l{\boldsymbol{q}_l}^T\}\!=\!\boldsymbol{0},~ {\blue \mathbb{E}\{\boldsymbol{n}_l \theta \}\!=\!\boldsymbol{0},\ \mathbb{E}\{\boldsymbol{q}_l \theta \}\!=\!\boldsymbol{0}},~  \forall l$, and the noise vectors across different clusters are mutually uncorrelated, i.e., $\mathbb{E}\{\boldsymbol{n}_i{\boldsymbol{n}_j}^T\}\!=\!\boldsymbol{0}$ and $\mathbb{E}\{\boldsymbol{q}_i{\boldsymbol{q}_j}^T\}\!=\!\boldsymbol{0},\ \forall i\!\ne\!j$. 
%Also, $\boldsymbol{n}_l$ and $\boldsymbol{q}_l$ are uncorrelated with $\theta$, $\forall l$.

\begin{comment}
In the second phase, the CH forms a linear combination of all local observations to which it has access, and transmits the resultant local message to the FC. In particular, the transmitted signal from CH $l$ to the FC is:
%
\begin{equation} \label{transmitted signal from CH l}
y_l={\boldsymbol{w}_l}^T\boldsymbol{t}_l,\ \ l=1,..., L,
\end{equation}
%
where $\boldsymbol{w}_l$ is the fusion vector used by CH $l$. The CH then amplifies and transmits $y_l$ to the FC over power constrained orthogonal wireless channels subject to fading and noise, such that the signal at the FC received from $l$-th cluster is given by:

CH$_l$ forwards an optimized linear combination of the signals received from all sensors within its cluster to the FC. In particular, CH$_l$ transmits scalar $y_l\!=\!{\boldsymbol{w}_l}^T\boldsymbol{t}_l$ to the FC, where vector $\boldsymbol{t}_l$ is defined in \eqref{x_l and t_l in cluster l} and $\boldsymbol{w}_l\!\in\!\mathbb{R}^{K_l}$ is the optimal linear weight vector. CHs transmit over power constrained 
\end{comment}
%
%
Each CH linearly fuses the signals received from the sensors within its cluster. Let $y_l\!=\!{\boldsymbol{w}_l}^T\boldsymbol{t}_l$, where $y_l$ is the scalar fused signal at CH$_l$ and $\boldsymbol{w}_l$ is the linear weight vector employed by CH$_l$ for linear fusion{\blue \footnote{{\blue When the pdf of $\theta$ is unknown, it is reasonable to assume that CH$_l$ applies a  linear  fusion  rule  $\boldsymbol{w}_l$ and  we  seek  the
best  $\boldsymbol{w}_l$. In Section \ref{Solving P2-1} we show that $\boldsymbol{w}_l^{opt}$ is equal to the linear operator corresponding to the LMMSE estimation of
$\theta$
based on $\boldsymbol{t}_l$, multiplied by an optimized scalar $\chi_l$. When $\theta \sim N(0,\sigma_{\theta}^2)$ the MMSE and LMMSE estimates of $\theta$
based on $\boldsymbol{t}_l$ coincide.}}} (to be optimized). CHs transmit these fused signals to the FC over orthogonal Rayleigh fading channels, such that the received signal at the FC from CH$_l$ is:
\vspace{-.35cm}
\begin{equation} \label{z_l at the FC}
z_l=h_ly_l+v_l,\ \ l=1,..., L,
\vspace{-.1cm}
\end{equation}
where $h_l\sim\mathcal{CN}\left(0,2\sigma_{h_l}^2\right)$ is fading channel coefficient corresponding to the link between CH$_l$ and the FC and $v_l\sim\mathcal{CN}\left(0,2\sigma_{v_l}^2\right)$ is the additive communication channel noise. We assume $v_l$ is uncorrelated with $\theta, \boldsymbol{n}_l, \boldsymbol{q}_l$, $\forall l$. 
To enable estimating $h_l$ at the FC, CH$_l$ transmits a pilot symbol \cite{Hamid-Azadeh-SP-letter} with power $\psi_l$ to the FC, prior to sending its signal $y_l$. Without loss of generality, we assume training symbols are all ones. Assuming $h_l$ does not change during transmission of $y_l$ and the training symbol{\red\footnote{
{\red We assume time-division-duplex transmission and channel reciprocity. We also  assume that the channel coherence time is larger than the overall duration  of pilot transmission, channel estimation, power optimization, information feedback, and data transmission.}}}, the received signal at the FC from CH$_l$ corresponding to the training symbol is:
\vspace{-.2cm}
\begin{equation}\label{pilot symbol at the FC}
\hat{z}_l=h_l\sqrt{\psi_l}+\nu_l,\ \ l=1,..., L,
\vspace{-.1cm}
\end{equation}
where $\nu_l$ in \eqref{pilot symbol at the FC} is independent of $v_l$ in \eqref{z_l at the FC} and is identically distributed. The FC adopts the following two-stage strategy to process the received signals from the CHs and reconstruct $\theta$: stage 1) the FC uses the received signals $\{\hat{z}_l\}_{l=1}^{L}$ corresponding to training symbol transmissions to estimate $\{h_l\}_{l=1}^{L}$ and obtain the channel estimates $\{\hat{h}_l\}_{l=1}^{L}$, 
%we assume the FC provides CH_l with hat_h_l through a feedback channel,
stage 2) the FC uses these channel estimates and the received signals $\{z_l\}_{l=1}^{L}$ corresponding to $\{y_l\}_{l=1}^{L}$ transmissions and find the LMMSE estimate of $\theta$, denoted as $\hat{\theta}$. 
%The FC processes the received signals from the CHs to first estimate the unknown channels and then uses these channel estimates to estimate $\theta$, using a linear estimator. 
Finding the LMMSE estimator has a lower computational complexity, compared with the optimal MMSE estimator, and it requires only the knowledge of first and second order statistics. Let $D\!=\!\mathbb{E}\{(\theta-\hat{\theta})^2\}$ denote the MSE corresponding to the LMMSE estimator $\hat{\theta}$. 
Our main objective is to study power allocation among different clusters, subject to a network transmit power constraint (including power for training and data transmissions), such that $D$ is minimized. Section \ref{Power Constraints and Problem Statement} provides a formal description of our constrained optimization problem, including the power constraints and the set of our optimization variables{\red \footnote{{\red Comparing orthogonal channel model and multiple-access channel (MAC) model adopted in \cite{Wu_Elsevier_2011,Lin-IET-2012,Salcedo_TSP_2015}, the former consumes more time or bandwidth for transmission, however,  it does not require symbol-level synchronization for compensating complex channel phase at transmitter. We note that the complexity of the sequence of operation in our work (pilot transmission, channel estimation, power optimization, information feedback, and data transmission) is comparable with that of those works that rely upon perfect CSI, since implementing power allocation solutions obtained based on perfect CSI \cite{goldsmith-TSP-2007,Fang-TWC-2009,Chaudhary-TSP-2013,Lin-IET-2012,Salcedo_TSP_2015} requires pilot transmission and channel estimation, prior to data transmission.}}}.

%%%%%%%%%%%%%%%%%%%%%%%%%%%%
\vspace{-.25cm} 
\subsection{{\blue Power Constraints}} \label{Power Constraints and Problem Statement}
\vspace{-.1cm}
We describe our power constraints. Let $P_{l,k}$ denote the average power that sensor $k$ consumes to send its amplified measurement to CH$_l$ and $P_l\!=\!\sum_{k=1}^{K_l}P_{l,k}$ be the total power that sensors in cluster $l$ spend to send their amplified measurements to CH$_l$. From \eqref{transmitted obs from sensor to CH} we have:
\vspace{-.1cm}
\begin{equation} \label{power of transmission form sensor to CH}
P_{l,k}\!=\!{\alpha}_{l,k}\mathbb{E}\{x_{l,k}^2\}\!=\!{\alpha}_{l,k}(\sigma^2_{\theta}+\sigma_{n_{l,k}}^2),\ k=1,..., K_l.
\vspace{-.1cm}
\end{equation}
% 
%via an orthogonal MAC subject to fading and noise. 
{\blue For tractability, similar to \cite{Wang_2011,dist_det_cluster_2019} we assume $P_l$ is equally divided between sensors within cluster $l$, i.e., $P_{l,k}\!=\!P_l/K_l$. Under this assumption from \eqref{power of transmission form sensor to CH} we obtain ${\alpha}_{l,k}=P_l d_{l,k}$} where $d_{l,k}\!=\!\frac{1}{K_l(\sigma^2_{\theta}+\sigma_{n_{l,k}}^2)}$, {\blue or equivalently in matrix form, }
we find %$\sqrt{{\alpha}_{l,k}}\!=\!\frac{\sqrt{P_l}}{K_l(\sigma^2_{\theta}+\sigma_{n_{l,k}}^2)}$.\\ 
$\sqrt{\boldsymbol{A}_l}\!=\!\sqrt{P_l}\boldsymbol{D}_l$, where $\boldsymbol{D}_l$ {\blue is given in Table \ref{table-symbol}}. 
Let ${\mathcal P}_l$ represent the average power that CH$_l$ spends to send its fused signal $y_l$ to the FC. We have: 
{\blue 
\vspace{-.1cm}
\begin{equation} \label{new-before-7}
{\mathcal P}_l\!=\!\mathbb{E}\{y_l^2\}\!=\!{\boldsymbol{w}_l}^T \underbrace{\mathbb{E}\{\boldsymbol{t}_l{\boldsymbol{t}_l}^T \}}_{=\boldsymbol{R}_{t_l}}\boldsymbol{w}_l. 
\vspace{-.1cm}
\end{equation}
Applying \eqref{x_l and t_l in cluster l} and noting that $\boldsymbol{x}_l, \boldsymbol{q}_l$ in \eqref{x_l and t_l in cluster l} are zero mean and uncorrelated, it is easy to verify that:
\vspace{-.1cm}
\begin{equation}\label{defs needed for power constraints}
\boldsymbol{R}_{t_l}= P_l\boldsymbol{\Omega}_l \!+\! \boldsymbol{\Sigma}_{q_l},
\vspace{-.1cm}
\end{equation}
where
\vspace{-.25cm}
\begin{align*} 
\boldsymbol{\Omega}_l\!=\!\boldsymbol{\Delta}_l+\sigma^2_{\theta}\boldsymbol{\Pi}_l,\ \boldsymbol{\Delta}_l\!=\!\boldsymbol{D}_l\boldsymbol{\Sigma}_{n_l}\boldsymbol{D}_l,\ \boldsymbol{\Pi}_l\!=\!\boldsymbol{\rho}_l{\boldsymbol{\rho}_l}^T,\ \boldsymbol{\rho}_l\!=\!\boldsymbol{D}_l\boldsymbol{1}_l. 
\vspace{-.1cm}
\end{align*}
Combining  (\ref{new-before-7}) and (\ref{defs needed for power constraints}) we obtain:
\vspace{-.1cm}
\begin{equation} \label{transmission power of CH l}
{\mathcal P}_l={\boldsymbol{w}_l}^T
(P_l {\boldsymbol{\Omega}_l}+ \boldsymbol{\Sigma}_{q_l})\boldsymbol{w}_l.
\vspace{-.1cm}
\end{equation}
} 
Let $P_{trn}\!=\!\sum_{l=1}^{L}\psi_l$ be the total power that CHs spend to transmit their pilot symbols to the FC for channel estimation. We assume there is a constraint on the network transmit power, such that:
\vspace{-.1cm}
\begin{equation}\label{cnstrt in terms of P and cal_P}
P_{trn}+\sum_{l=1}^{L}P_l+{\mathcal P}_l\leq P_{tot}.
\vspace{-.1cm}
\end{equation}
Substituting ${\mathcal P}_l$ in \eqref{transmission power of CH l} into the constraint in \eqref{cnstrt in terms of P and cal_P} we reach:
\vspace{-.1cm}
\begin{equation} \label{cnstrt in terms of P and w}
P_{trn}+\sum_{l=1}^{L}{\boldsymbol{w}_l}^T\boldsymbol{\Sigma}_{q_l}\boldsymbol{w}_l+P_l(1\!+\!{\boldsymbol{w}_l}^T\boldsymbol{\Omega}_l\boldsymbol{w}_l)\leq P_{tot}.
\vspace{-.1cm}
\end{equation}
%%%%%%%%%%%%%%%%%%%%%%%%
%\vspace{-.55cm} 
\subsection{{\blue Problem Statement}} \label{Problem Statement-sub}
\vspace{-.15cm}
Under the network power constraint {\red in (\ref{cnstrt in terms of P and cal_P})}, our goal is to find the optimal $P_{trn},\{P_l, {\cal P}_l\}_{l=1}^L$ such that $D$ is minimized. 
The constraint in \eqref{cnstrt in terms of P and w} shows that finding the optimal $\{P_l, {\cal P}_l\}_{l=1}^L$ in our problem is equivalent to finding the optimal $\{P_l, \boldsymbol{w}_l\}_{l=1}^L$, since given $\{P_l, \boldsymbol{w}_l\}_{l=1}^L$ one can find $\{{\cal P}_l\}_{l=1}^L$ using \eqref{transmission power of CH l}.
Therefore, our goal is to find the optimal total training power $P_{trn}$, 
%and for each cluster, sensors' total transmission power consumption $P_l$ and fusion vector $\boldsymbol{w}_l$ used by CH$_l$, that minimizes $D$. 
the optimal total power that sensors in cluster $l$ spend to transmit their measurements to their CH $P_l$, and the optimal $\boldsymbol{w}_l$ employed by CH$_l$ for its linear fusion, such that $D$ is minimized. In other words, we are interested in solving the following constrained optimization problem:
%\leqnomode
%
\vspace{-.15cm}
\begin{align} \label{original min problem}
\mathop{\text{min}}_{P_{trn},\{P_l, \boldsymbol{w}_l\}_{l=1}^L}& \ D(P_{trn},\{P_l, \boldsymbol{w}_l\}_{l=1}^L)\\
\text{s.t.}\ \ \ \ \ \ &P_{trn}\!+\!\sum_{l=1}^{L}\!{\boldsymbol{w}_l}^T\boldsymbol{\Sigma}_{q_l}\boldsymbol{w}_l\!+\!P_l(1\!+\!{\boldsymbol{w}_l}^T\boldsymbol{\Omega}_l\boldsymbol{w}_l)\!\leq\!P_{tot},\nonumber\\ 
&P_{trn}\in\mathbb{R}^{+},P_l\in \mathbb{R}^{+},\boldsymbol{w}_l\in \mathbb{R}^{K_l},\ \forall l.\nonumber%\tag{P0}
\vspace{-.1cm}
\end{align}
%
%\reqnomode
We note that $\boldsymbol{\Sigma}_{q_l}$ and $\boldsymbol{\Omega}_l$ in the network transmit power constraint do not depend on our optimization variables. 
%Once optimal $\{P_l,\boldsymbol{w}_l\}_{l=1}^L$ are found $\{{\cal P}_l\}_{l=1}^L$ can be found using \eqref{transmission power of CH l}.
% % % % % % % % % % % % % % % % % % % % %
% % % % % % % % % % % % % % % % % % % % % % %
\vspace{-.15cm}
\section{Characterizing $D$ and its Lower Bounds} \label{LMMSE estimation}
\vspace{-.1cm}
\subsection{Characterization of $D$ in terms of Channel Estimates}
\vspace{-.1cm}
We characterize the objective function $D$ in \eqref{original min problem}, in terms of our optimization variables. Before delving into the derivations of $D$, we introduce the following notations. Considering our signal model in Section \ref{System Model and Problem Formulation}, we define column vectors $\boldsymbol{x}$, $\boldsymbol{t}$, $\boldsymbol{y}$, $\boldsymbol{z}$ {\blue in Table \ref{table-symbol}}, which are obtained from stacking the signals corresponding to all clusters. We have:
\vspace{-.2cm}
\begin{subequations} \label{all clusters in a compact form}
\begin{align} 
\boldsymbol{x}=\theta\boldsymbol{1}+\boldsymbol{n},\ \ \boldsymbol{t}&=\boldsymbol{M}\boldsymbol{x}+\boldsymbol{q},\ \ \boldsymbol{y}=\boldsymbol{W}\boldsymbol{t},\label{x_t_y_vec}\\
\boldsymbol{z}&=\boldsymbol{H}\boldsymbol{y}+\boldsymbol{v},\label{z_vec}
\vspace{-.1cm}
\end{align}
\end{subequations}
where {\blue $\boldsymbol{1}$ is a column vector of $K$ ones}, column vectors $\boldsymbol{n}$, $\boldsymbol{q}$, $\boldsymbol{v}$, and matrices $\boldsymbol{M}$, $\boldsymbol{W}$, $\boldsymbol{H}$ are defined {\blue in Table \ref{table-symbol}}. The noise vectors $\boldsymbol{n}$, $\boldsymbol{q}$, $\boldsymbol{v}$ are zero-mean with covariance matrices $\boldsymbol{\Sigma}_{n}$, $\boldsymbol{\Sigma}_{q}$, $\boldsymbol{\Sigma}_{v}$, respectively, {\blue given in Table \ref{table-symbol}}. 
We model the fading coefficient as $h_l\!=\!\hat{h}_l+\tilde{h}_l$, where $\hat{h}_l$ is the MMSE channel estimate %in terms of training power $\psi_l$ \cite{Kay_book_est_theory}:
and $\tilde{h}_l$ is the corresponding zero-mean estimation error with the variance $\zeta_l^2$. The expressions for $\hat{h}_l$ and $\zeta_l^2$ in terms of training power $\psi_l$ are \cite{Kay_book_est_theory}:
\vspace{-.15cm}
\begin{equation} \label{channel estimate and channel estimation error variance}
\hat{h}_l=\frac{\sigma_{h_{l}}^2\sqrt{\psi_l}\hat{z}_l}{\sigma_{v_{l}}^2+\psi_l\sigma_{h_{l}}^2},\ \ \zeta_l^2=\frac{2\sigma_{h_{l}}^2\sigma_{v_{l}}^2}{\sigma_{v_{l}}^2+\psi_l\sigma_{h_{l}}^2}.
\vspace{-.1cm}
\end{equation}
We define matrices $\hat{\boldsymbol{H}}, \tilde{\boldsymbol{H}}$ {\blue in Table \ref{table-symbol}} and thus we have $\boldsymbol{H}\!=\!\hat{\boldsymbol{H}}\!+\!\tilde{\boldsymbol{H}}$. Substituting this channel model into \eqref{z_vec}, we can rewrite the received signal $\boldsymbol{z}$ as the following:
\vspace{-.1cm}
\begin{equation} \label{z_vec given h_hat}
\boldsymbol{z}\!=\!\underbrace{[\hat{\boldsymbol{H}}\boldsymbol{W}\boldsymbol{M}\boldsymbol{1}]\theta}_{=\boldsymbol{z}_1}\!+\!\underbrace{(\tilde{\boldsymbol{H}}\boldsymbol{W}\boldsymbol{M}\boldsymbol{1})\theta}_{=\boldsymbol{z}_2}\!+\!\underbrace{(\hat{\boldsymbol{H}}\!+\!\tilde{\boldsymbol{H}})\boldsymbol{W}(\boldsymbol{q}\!+\!\boldsymbol{M}\boldsymbol{n})\!+\!\boldsymbol{v}}_{=\boldsymbol{z}_3}.
\end{equation}
We proceed with characterizing $D$ in terms of the channel estimates. From optimal linear estimation theory, we have:
\vspace{-.15cm}
\begin{eqnarray}\label{theory of optimal linear estimation}
\hat{\theta}&=&{\boldsymbol{g}}^H\boldsymbol{z},\ \text{where}\ \boldsymbol{g}\!=\!(\mathbb{E}\{\boldsymbol{z}{\boldsymbol{z}}^H\})^{-1}\mathbb{E}\{\theta\boldsymbol{z}\},\nonumber\\
D&=&\sigma^2_{\theta}-{\mathbb{E}\{\theta\boldsymbol{z}\}}^H(\mathbb{E}\{\boldsymbol{z}{\boldsymbol{z}}^H\})^{-1}\mathbb{E}\{\theta\boldsymbol{z}\}.
\vspace{-.15cm}
\end{eqnarray}
where $\hat{\theta}$ and $D$ depend on the channel estimates $\{\hat{h}_l\}_{l=1}^L$. In the following, we find $\mathbb{E}\{\boldsymbol{z}{\boldsymbol{z}}^H\}$ and $\mathbb{E}\{\theta\boldsymbol{z}\}$ in \eqref{theory of optimal linear estimation} by examining the statistics of channel estimation error.
By the orthogonality principle of  LMMSE estimation \cite{Yupeng-Azadeh-TWC-2011}, $\tilde{h}_l$ is orthogonal to $\hat{h}_l$, that is $\mathbb{E}\{\tilde{h}_l\hat{h}_l\}\!=\!0,\ \forall l$, and therefore, $\mathbb{E}\{\boldsymbol{z}_1{\boldsymbol{z}_2}^H\}\!=\!\boldsymbol{0}$. Using the fact that $\theta$, $\boldsymbol{n}$, $\boldsymbol{q}$, $\boldsymbol{v}$ are mutually uncorrelated, we have $\mathbb{E}\{\boldsymbol{z}_1{\boldsymbol{z}_3}^H\}\!=\!\boldsymbol{0},\ \mathbb{E}\{\boldsymbol{z}_2{\boldsymbol{z}_3}^H\}\!=\!\boldsymbol{0}$. Combined these with the fact that $\mathbb{E}\{\boldsymbol{z}\}\!=\!\boldsymbol{0}$, the covariance matrix $\boldsymbol{C}_{\boldsymbol{z}}\!=\!\mathbb{E}\{\boldsymbol{z}{\boldsymbol{z}}^H\}$ given $\hat{\boldsymbol{H}}$ can be expressed as:
\vspace{-.1cm}
\begin{align} \label{C_z in extended form}
\boldsymbol{C}_{\boldsymbol{z}}&=\sigma^2_{\theta}\hat{\boldsymbol{H}}\boldsymbol{W}\boldsymbol{M}\boldsymbol{1}{\boldsymbol{1}}^T\boldsymbol{M}{\boldsymbol{W}}^T\hat{\boldsymbol{H}}^{H}+\sigma^2_{\theta}(\boldsymbol{\Gamma}\boldsymbol{W}\boldsymbol{M}\boldsymbol{\Sigma}\boldsymbol{M}{\boldsymbol{W}}^T\boldsymbol{\Gamma})\nonumber\\
&+\hat{\boldsymbol{H}}\boldsymbol{W}(\boldsymbol{\Sigma}_{q}+\boldsymbol{M}\boldsymbol{\Sigma}_{n}\boldsymbol{M}){\boldsymbol{W}}^T\hat{\boldsymbol{H}}^{H}\nonumber\\
&+\boldsymbol{\Gamma}\boldsymbol{W}(\boldsymbol{\Sigma}_{q}+\boldsymbol{M}\boldsymbol{\Sigma}_{n}\boldsymbol{M}){\boldsymbol{W}}^T\boldsymbol{\Gamma}+\boldsymbol{\Sigma}_{v},
\vspace{-.1cm}
\end{align}
where $\boldsymbol{\Gamma}$ and $\boldsymbol{\Sigma}$ {\blue are defined in Table \ref{table-symbol} and $\boldsymbol{\Sigma}_l$ is a $K_l \times K_l$ matrix of all ones. }
%
%Let $\boldsymbol{\Lambda}_1\!=\!\text{diag}(\boldsymbol{\Lambda}_{1_1}, ..., \boldsymbol{\Lambda}_{1_L})$ where $\boldsymbol{\Lambda}_{1_l}$ is:
We define $\boldsymbol{\Lambda}_{1_l}$ and $\boldsymbol{\Lambda}_2$ as bellow:
\vspace{-.2cm}
\begin{align} \label{l-th block of Lambda_1, Lambda_2, mu_l}
&\boldsymbol{\Lambda}_{1_l}=\sigma^2_{\theta}\zeta_l^2P_l\boldsymbol{\Pi}_l+({|\hat{h}_l|}^2+\zeta_l^2)(\boldsymbol{\Sigma}_{q_l}+P_l\boldsymbol{\Delta}_l),\\
&\boldsymbol{\Lambda}_2=|\boldsymbol{\mu}|{|\boldsymbol{\mu}|}^T,\ \boldsymbol{\mu}_l\!=\!\sqrt{P_l}\hat{h}_l\boldsymbol{\rho}_l,\ \forall l.\nonumber
\vspace{-.1cm}
\end{align}
% 
%and $\boldsymbol{\mu}_l\!=\!\sqrt{P_l}\hat{h}_l\boldsymbol{D}_l\boldsymbol{G}_l\boldsymbol{1}_l$. 
{\blue where $\boldsymbol{\mu}$ is defined in Table \ref{table-symbol}.}
It is straightforward to simplify \eqref{C_z in extended form} and write it as the following:
\vspace{-.2cm}
\begin{equation} \label{final formula of C_z}
\boldsymbol{C}_{\boldsymbol{z}}=\boldsymbol{W}(\boldsymbol{\Lambda}_1+\sigma^2_{\theta}\boldsymbol{\Lambda}_2){\boldsymbol{W}}^T+\boldsymbol{\Sigma}_{v}.
\vspace{-.1cm}
\end{equation}
%
%We will now argue that $\hat{h}_l\!=\!\mathbb{E}_{h_l|\hat{z}_l}\{h_l\}\!=\!\mathbb{E}_{h_l|\hat{h}_l}\{h_l\}$, which is because $\hat{h}_l$ is a linear function of $\hat{z}_l$ in LMMSE estimation, and thus we have: 
{\blue where $\boldsymbol{\Lambda}_1$ is defined in Table \ref{table-symbol}}. 
To find $\mathbb{E}\{\theta\boldsymbol{z}\}$ we consider \eqref{z_vec given h_hat} and we realize that $\mathbb{E}\{\theta\boldsymbol{z}_3\}\!=\!\boldsymbol{0}$. Therefore:
\vspace{-.2cm}
\begin{equation} \label{E(theta Z)}
\mathbb{E}\{\theta\boldsymbol{z}\}\!=\!\mathbb{E}\{\theta\boldsymbol{z}_1\}\!+\!\mathbb{E}\{\theta\boldsymbol{z}_2\}\!\overset{(a)}{=}\!\sigma^2_{\theta}\hat{\boldsymbol{H}}\boldsymbol{W}\boldsymbol{M}\boldsymbol{1}\!=\!\sigma^2_{\theta}\boldsymbol{W}\boldsymbol{\mu},
\end{equation}
where ($a$) in \eqref{E(theta Z)} is obtained from the fact that $\mathbb{E}\{\tilde{\boldsymbol{H}}\}\!=\!\boldsymbol{0}$. Based on \eqref{final formula of C_z}, \eqref{E(theta Z)}, the LMMSE estimator $\hat{\theta}$ and its corresponding MSE in \eqref{theory of optimal linear estimation} can be written as:
\vspace{-.15cm}
\begin{eqnarray} \label{LMMSE of theta and its MSE}
\hat{\theta}&=&\sigma^2_{\theta}{\boldsymbol{\mu}}^H{\boldsymbol{W}}^T\boldsymbol{C}^{-1}_{\boldsymbol{z}}\boldsymbol{z},\notag\\
D&=&\sigma^2_{\theta}-\sigma^4_{\theta}{\boldsymbol{\mu}}^H{\boldsymbol{W}}^T\boldsymbol{C}^{-1}_{\boldsymbol{z}}\boldsymbol{W}\boldsymbol{\mu}.
\vspace{-.1cm}
\end{eqnarray}
in which $\boldsymbol{\mu}$ and $\boldsymbol{C}_{\boldsymbol{z}}$ depend on the channel estimates. Substituting \eqref{final formula of C_z} in \eqref{LMMSE of theta and its MSE} and applying the Woodbury identity\footnote{For matrices $\boldsymbol{A}, \boldsymbol{B}, \boldsymbol{C}$, $\boldsymbol{D}$ the Woodbury identity states that ${(\boldsymbol{A}+\boldsymbol{B}\boldsymbol{C}\boldsymbol{D})}^{-1}\!=\!{\boldsymbol{A}}^{-1}-{\boldsymbol{A}}^{-1}\boldsymbol{B}{({\boldsymbol{C}}^{-1}+\boldsymbol{D}{\boldsymbol{A}}^{-1}\boldsymbol{B})}^{-1}\boldsymbol{D}{\boldsymbol{A}}^{-1}$ \cite{Matrix_Analysis}.} yields: 
\vspace{-.5cm}
\begin{eqnarray} \label{MSE in simple form}
D&=&(\sigma^{-2}_{\theta}+{\boldsymbol{\mu}}^H{\boldsymbol{W}}^T(\boldsymbol{W}\boldsymbol{\Lambda}_1{\boldsymbol{W}}^T+\boldsymbol{\Sigma}_{v})^{-1}\boldsymbol{W}\boldsymbol{\mu})^{-1}\nonumber\\
&=&(\sigma^{-2}_{\theta}+\sum_{l=1}^{L}\frac{P_l{|\hat{h}_l|}^2{\boldsymbol{w}_l}^T\boldsymbol{\Pi}_l\boldsymbol{w}_l}{\sigma_{v_{l}}^2+{\boldsymbol{w}_l}^T\boldsymbol{\Lambda}_{1_l}\boldsymbol{w}_l})^{-1},
\vspace{-.1cm}
\end{eqnarray}
%
%in which ($a$) in \eqref{MSE in simple form} is obtained from exploiting the block-diagonal structures of matrices $\boldsymbol{W}$, $\boldsymbol{\Lambda}_1$, and $\boldsymbol{\Sigma}_{v}$. 
Examining $D$ in \eqref{MSE in simple form} we notice that $\boldsymbol{\Pi}_l$ does not depend on our optimization variables. However, $\boldsymbol{\Lambda}_{1_l}$ depends on $P_l$ and $\psi_l$ (through the channel estimate ${|\hat{h}_l|}^2$ and the channel estimation error variance $\zeta_l^2$). Clearly, $D$ depends on $\boldsymbol{w}_l$. 
%%%%%%%%%%%%%%%%%%%%%%%%%%%%%%%%%%%%
\vspace{-.2cm}
\subsection{Three Lower Bounds on $D$} \label{3 lower bounds}
\vspace{-.05cm}
We provide three lower bounds on $D$, denoted as $D_1, D_2, D_3$. To obtain $D_1$ we consider the scenario when $\{h_l\}_{l=1}^{L}$ are available at the FC (perfect CSI). This implies $\hat{h}_l\!=\!h_l$ and $\zeta_l^2\!=\!0,\forall l$, in \eqref{MSE in simple form}, and the MSE becomes:
\vspace{-.2cm}
\begin{equation} \label{MSE for full CSI}
D_1\!=\!(\sigma^{-2}_{\theta}\!+\!\sum_{l=1}^{L}\frac{P_l{|h_l|}^2{\boldsymbol{w}_l}^T\boldsymbol{\Pi}_l\boldsymbol{w}_l}{\sigma_{v_{l}}^2+{|h_l|}^2{\boldsymbol{w}_l}^T(\boldsymbol{\Sigma}_{q_l}\!+\!P_l\boldsymbol{\Delta}_l)\boldsymbol{w}_l})^{-1}.
\vspace{-.1cm}
\end{equation}
To obtain $D_2$ we consider the scenario when in addition to perfect CSI, sensors' noisy measurement vector $\boldsymbol{x}_l$ is available at CH$_l$ (i.e., error-free channels between sensors and their CHs). Therefore, $\boldsymbol{A}_l\!=\!\boldsymbol{I}_l$, where $\boldsymbol{I}_l$ denotes the identity matrix, and $\boldsymbol{\Sigma}_{q_l}\!=\!\boldsymbol{0},\forall l$. In this scenario \eqref{MSE for full CSI} simplifies to:
\vspace{-.2cm}
\begin{equation} \label{MSE for full CSI and error free sensor-CH}
D_2=(\sigma^{-2}_{\theta}+\sum_{l=1}^{L}\frac{{|h_l|}^2{\boldsymbol{w}_l}^T\boldsymbol{\Sigma}_l\boldsymbol{w}_l}{\sigma_{v_{l}}^2+{|h_l|}^2{\boldsymbol{w}_l}^T\boldsymbol{\Sigma}_{n_l}\boldsymbol{w}_l})^{-1}.
\vspace{-.1cm}
\end{equation}
%
%As a baseline, we can obtain the MSE of clairvoyant estimator in which, for all $l$,
To obtain $D_3$ we consider the scenario when $\boldsymbol{x}_l$ is available at CH$_l$ and $y_l$ is available at the FC. This is equivalent to having all measurements $\{\boldsymbol{x}_l\}_{l=1}^{L}$ available at the FC (i.e., error-free channels between sensors and their CHs, and between CHs and the FC). Therefore, the MSE becomes:
\vspace{-.1cm}
\begin{equation} \label{MSE for clairvoyany estimator}
D_3=(\sigma^{-2}_{\theta}+\sum_{l=1}^{L}{\boldsymbol{1}_l}^T\boldsymbol{\Sigma}^{-1}_{n_l}\boldsymbol{1}_l)^{-1}.
\vspace{-.1cm}
\end{equation}
Clearly, we have $D_3 < D_2 < D_1 < D$.
\begin{comment}
In the following, we consider the optimal allocation of power for training and for data transmission, to minimize $D$ in \eqref{MSE in simple form} under the following network power constraint:
%
\begin{equation*} 
\sum_{l=1}^{L}P_l+{\mathcal P}_l+\psi_l\leq P_{tot},
\end{equation*}
%
where the optimization parameters $P_l$, ${\mathcal P}_l$, and $\psi_l$, respectively, are total power consumed by sensors whitin cluster $l$ to send their measurements to CH $l$, power consumed by CH $l$ for transmission of $y_l$ to the FC, and amount of power consumed by CH $l$ for its channel estimation phase. 
\end{comment}
%By substituting $P_l$, ${\mathcal P}_l$, and $\psi_l$ from Assumption \ref{assumption for orthogonal MAC in a cluster}, \eqref{transmission power of CH l}, and \eqref{final formula psi_l}, respectively, and since $\sum_{l=1}^{L}\psi_l\!=\!P_{trn}$, 
%Recall the constraint $P_{trn}+\sum_{l=1}^{L}P_l+{\mathcal P}_l\leq P_{tot}$ in \eqref{original min problem}. Substituting ${\mathcal P}_l$ with its expression given in \eqref{transmission power of CH l}, the power constraint further simplifies to:
%%%%%%%%%%%%%%%%%%%%%%%%%%%%%%%%%%%%%%%%%%%%%%%
\vspace{-.2cm}
\subsection{Bayesian CRB} \label{BCRB}
%\vspace{-.1cm}
{\blue Let $G$ denote the Bayesian Fisher information corresponding to estimating $\theta$, given $\boldsymbol{z}$ and the vector of channel estimates $\hat{\boldsymbol{h}}\!=\![\hat{h}_1, ..., \hat{h}_L]$ at the FC. The inverse of $G$ is the Bayesian CRB and it sets an estimation-theoretic lower bound on the MSE of any Bayesian estimation of $\theta$, given $\boldsymbol{z}, \hat{\boldsymbol{h}}$ \cite{Van_Trees_Bayesian_bnd_book,Vosoughi2006sp1,Vosoughi2006sp2}. Using the definition in \cite{Van_Trees_Bayesian_bnd_book,Vosoughi2006sp1,Vosoughi2006sp2} in our problem $G\!=\!\mathbb{E}\{{\!(\frac{\partial \ln f(\boldsymbol{z},\hat{\boldsymbol{h}},\theta)}{\partial \theta})\!}^2\}$, where}
$f(\boldsymbol{z},\hat{\boldsymbol{h}},\theta)$ denotes the joint pdf of $\boldsymbol{z},\hat{\boldsymbol{h}},\theta$ and 
%
%\vspace{-0.15cm}
%\begin{equation*} %\label{FI_Van_Trees}
%G=\mathbb{E}\{{(\frac{\partial \ln f(\boldsymbol{z},\hat{\boldsymbol{h}},\theta)}{\partial \theta})}^2\},
%\end{equation*}
%
and the expectation is taken over $f(\boldsymbol{z},\hat{\boldsymbol{h}},\theta)$. 
{\blue 
\begin{lemma}\label{lemma for crb}
\textnormal{
The Bayesian Fisher information corresponding to estimating $\theta$, given $\boldsymbol{z}, \hat{\boldsymbol{h}}$ is:
\vspace{-0.1cm}
\begin{align} \label{FI in decomposed form-2}
G\!=\!\mathbb{E}\{G_1(\theta)\}\!+\!\mathbb{E}\{G_2(\theta)\},
\vspace{-.1cm}
\end{align}
where $G_1(\theta)=-\frac{\partial^2 \ln f(\theta)}{\partial \theta^2}$ and $G_2(\theta)$ is given below. For $\theta \sim N(0,\sigma_{\theta}^2)$ we have $\mathbb{E}\{G_1(\theta)\}=\sigma_{\theta}^{-2}$. Both expectations in (\ref{FI in decomposed form-2}) are taken over $f(\theta)$, which represents the pdf of $\theta$. 
\vspace{-0.15cm}
\begin{equation} \label{G_2(theta) distributed}
G_2(\theta)=\sum_{l=1}^{L}\int_{\hat{h}_l}\int_{z_l}\frac{f(\hat{h}_l)}{f(z_l|\hat{h}_l,\theta)}{(\frac{\partial f(z_l|\hat{h}_l,\theta)}{\partial \theta})}^2dz_ld\hat{h}_l,
\vspace{-.1cm}
\end{equation}
where $f(z_l|\hat{h}_l,\theta)$ and its derivative with respect to $\theta$ are:
\vspace{-.25cm}
\begin{align} \label{f(z_l|h_hat_l,theta)}
f(z_l|\hat{h}_l,\theta)\!&=\! a_1e^{-a_2\theta^2}\!\sum_{m=0}^{\infty}\sum_{n=0}^{m}\sum_{p=0}^{m-n}\!c_{m,n,p}(\theta)\nonumber\\
&\times\!\!\int_{-\infty}^{\infty}\!\int_{-\infty}^{\infty}\!\!s_{m,n,p,b}(\theta)\exp(-\frac{{|z_l-b|}^2}{2\sigma_{v_l}^2})db,
\vspace{-.1cm}
\end{align}
%
%The derivative of $f(z_l|\hat{h}_l,\theta)$ in \eqref{f(z_l|h_hat_l,theta)} w.r.t $\theta$ becomes:
%\begin{figure*}[b]
\vspace{-.2cm}
\begin{align} \label{der wrt theta f(z_l|h_hat_l,theta)}
&\frac{\partial f(\!z_l|\hat{h}_l,\theta)}{\partial \theta}\!=\!a_1e^{-a_2\theta^2}\!\sum_{m=0}^{\infty}\!\sum_{n=0}^{m}\!\sum_{p=0}^{m\!-\!n}[(\!\frac{m\!-\!n\!+\!p}{\theta}-2a_2\theta)\nonumber\\
&\times c_{m,n,p}(\theta)\!\int_{-\infty}^{\infty}\!\int_{-\infty}^{\infty}\!\!s_{m,n,p,b}(\theta)\exp(-\frac{{|z_l-b|}^2}{2\sigma_{v_l}^2})db],
\vspace{-.1cm}
\end{align}
and the parameters $a_1, a_2, a_3, c_{m,n,p}(\theta), s_{m,n,p,b}(\theta), \bar{\phi}$ are:
\vspace{-.2cm}
\begin{align}\label{parameters for f and der f}
\!\!\!\!\!\!\!\!\!\!\!\!\!\!a_1\!&=\!\frac{\exp(-{|\hat{h}_l|}^2/{\zeta_l^2})}{\pi^2\zeta_l^2\bar{\sigma}_l^2\sigma_{v_l}^2}, a_2\!=\!\frac{{a_3}^2}{\bar{\sigma}_l^2}, a_3\!=\!{\boldsymbol{w}_l}^T\sqrt{\boldsymbol{A}_l}\boldsymbol{1}_l,\\
\!\!\!\!\!\!\!\!\!\!\!\!\!\!c_{m,n,p}(\theta)\!&=\!\frac{{|\hat{h}_l|}^{m+n-p}{|a_3\theta|}^{m-n+p}}{m!n!p!(m-n-p)!{\zeta_l}^{2m+n-p}{\bar{\sigma}_l}^{2m-n+p}},\nonumber\\
\!\!\!\!\!\!\!\!\!\!\!\!\!\!s_{m,n,p,b}(\theta)\!&=\!{|b|}^mK_{n-p}(\frac{2|b|}{\bar{\sigma}_l\zeta_l}){(2\cos(\bar{\phi}\!-\!\frac{\pi}{2}(1\!-\!\sign(a_3\theta))))}^{m-n-p},\nonumber\\
\!\!\!\!\!\!\!\!\!\!\!\!\!\!\bar{\phi}\!&=\!\angle b-\angle\hat{h}_l,\ \bar{\sigma}_l^2\!=\!{\boldsymbol{w}_l}^T(\sqrt{\boldsymbol{A}_l}\boldsymbol{\Sigma}_{n_l}\sqrt{\boldsymbol{A}_l}+\boldsymbol{\Sigma}_{q_l})\boldsymbol{w}_l.\nonumber
\vspace{-.1cm}
\end{align}
%
%\eqref{der wrt theta f(z_l|h_hat_l,theta)} shows the derivative of $f(z_l|\hat{h}_l,\theta)$ in \eqref{f(z_l|h_hat_l,theta)} w.r.t. $\theta$. %
%\end{figure*}
%Substituting \eqref{f(z_l|h_hat_l,theta)} and \eqref{der wrt theta f(z_l|h_hat_l,theta)} in \eqref{G_2(theta) distributed}, we compute $G_2(\theta)$.
}
\end{lemma}
\vspace{-.3cm}
\begin{proof}
	See Appendix \ref{deriving crb}.
\end{proof}
}
%=======================================
%=======================================
\vspace{-.3cm}
\section{{\blue Solving the Constrained Minimization of $D$} } \label{solving max problem} 
%\vspace{-.1cm}
We consider the constrained optimization problem in \eqref{original min problem}, where $D$ is provided in \eqref{MSE in simple form}. We define:
\vspace{-.2cm}
\begin{eqnarray}\label{def of J_l and C_l}
{\cal J}_l(P_{trn},P_l,\boldsymbol{w}_l)&=&\frac{P_l{|\hat{h}_l|}^2{\boldsymbol{w}_l}^T\boldsymbol{\Pi}_l\boldsymbol{w}_l}{\sigma_{v_{l}}^2+{\boldsymbol{w}_l}^T\boldsymbol{\Lambda}_{1_l}\boldsymbol{w}_l},\\
C_l(P_l,\boldsymbol{w}_l)&=&{\boldsymbol{w}_l}^T\boldsymbol{\Sigma}_{q_l}\boldsymbol{w}_l+P_l(1+{\boldsymbol{w}_l}^T\boldsymbol{\Omega}_l\boldsymbol{w}_l).\nonumber
\vspace{-.1cm}
\end{eqnarray}
%
%According to \eqref{MSE in simple form}, the constrained minimization problem in \eqref{original min problem}
Using the two definitions in \eqref{def of J_l and C_l} we can replace the problem in \eqref{original min problem} with its equivalent, problem \eqref{max problem to minimize D}, that has a simpler presentation. In particular, we can write $D^{-1}\!=\!\sigma^{-2}_{\theta} + \sum_{l=1}^{L}{\cal J}_l(P_{trn},P_l,\boldsymbol{w}_l)$. Hence, problem \eqref{max problem to minimize D} becomes:
\leqnomode
\vspace{-.2cm}
\begin{align} \label{max problem to minimize D}
~~~~~~&\mathop{\text{max}}_{P_{trn},\{P_l, \boldsymbol{w}_l\}_{l=1}^L}\ \ \sum_{l=1}^{L}{\cal J}_l(P_{trn},P_l,\boldsymbol{w}_l)\tag{P1}\\
\text{s.t.}\ \ &P_{trn}\!+\!\sum_{l=1}^{L}\!C_l(\!P_l,\boldsymbol{w}_l\!)\!\leq \!P_{tot}, P_{trn},P_l\!\in\! \mathbb{R}^{+}\!,\boldsymbol{w}_l\!\in\! \mathbb{R}^{K_l}\!, \forall l.\nonumber 
\vspace{-.2cm}
\end{align}
\reqnomode
It is easy to show that the solution of \eqref{max problem to minimize D} holds with active constraint $P_{trn}+\sum_{l=1}^{L}C_l(P_l,\boldsymbol{w}_l)=P_{tot}$. 
%The reasoning follows. If the solution $\{P_l,\boldsymbol{w}_l\}_{l=1}^{L}$ is such that strict inequality holds, we can equally scale up $P_l,\boldsymbol{w}_l$ so that the equality holds. In addition, if we equally scale up $P_l,\boldsymbol{w}_l$, we obtain a larger value for the cost function in \eqref{max problem to minimize D}. Consequently, with the optimal solution $\{P_l,\boldsymbol{w}_l\}_{l=1}^L$, the inequality constraint must be active. 
We further note that due to the cap on the network transmit power, only a subset of the clusters may become active at each observation period. We refer to this active subset as ${\cal A}\!=\!\{l:\!P_l\!>\!0,\ l\!=\!1, \dots, L\}$, where $|{\cal A}| \leq L$. 
%Our strategy to solve \eqref{max problem to minimize D} follows: we decompose \eqref{max problem to minimize D} into two sub-problems (P$_A$) and (P$_B$). In the first sub-problem (P$_A$) we find the optimal $\{P_l, \boldsymbol{w}_l\}_{l=1}^L$ given $P_{trn}$. In the second sub-problem (P$_B$) we find the optimal 
Regarding the objective function ${\cal J}_l$ in \eqref{max problem to minimize D} we note that it depends on $\hat{h}_l$ (through ${|\hat{h}_l|}^2$ in the numerator and $\boldsymbol{\Lambda}_{1_l}$ in the denominator of \eqref{def of J_l and C_l}). Regarding the optimization variables in \eqref{max problem to minimize D} we notice that, since pilot transmission proceeds data transmission, $P_{trn}$ cannot depend on the channel estimates $\{\hat{h}_l\}_{l=1}^L$ and can only depend on the statistical information of communication channels and the observation model. Examining \eqref{max problem to minimize D}, we note however, that solving it for $P_{trn}$ provides an answer that depends on $\hat{h}_l$ (which is unrealizable). On the other hand, the variables $P_l, \boldsymbol{w}_l$ should be chosen according to the available CSI $\hat{h}_l$. Based on these observations, we propose to {\blue consider} two problems \eqref{max problem to find P_l,w_l all l} and (P$_B$) stemming from \eqref{max problem to minimize D}. problem \eqref{max problem to find P_l,w_l all l} finds the optimal $\{P_l, \boldsymbol{w}_l\}_{l=1}^L$ that minimizes $D$, given $P_{trn}$. Let $\sigma\in(0,1)$ such that $P_{trn}\!=\!(1-\sigma)P_{tot}$. 
%power is expended in all clusters on forwarding observations from the sensors to their respective CH and from the CHs to the FC. 
Given $P_{trn}$ (and thus $\sigma$), we define ${\cal F}_l(P_l,\boldsymbol{w}_l)\!=\!{\cal J}_l(P_{trn},P_l,\boldsymbol{w}_l)$. Problem \eqref{max problem to find P_l,w_l all l} becomes:
\leqnomode
\vspace{-.2cm}
\begin{align} \label{max problem to find P_l,w_l all l}
\ \ \ \ \ \ \ \ \ &\text{given}\ P_{trn},\ \ \mathop{\text{max}}_{\{P_l, \boldsymbol{w}_l\}_{l=1}^L}\ \ \sum_{l=1}^{L}{\cal F}_l(P_l,\boldsymbol{w}_l)\tag{P$_A$}\\
&\text{s.t.}\ \ \sum_{l=1}^{L}\!C_l(P_l,\boldsymbol{w}_l)\!\leq\!\sigma P_{tot},\ P_l\!\in \!\mathbb{R}^{+}, \boldsymbol{w}_l\!\in \!\mathbb{R}^{K_l}, \forall l.\nonumber
\vspace{-.2cm}
\end{align}
\reqnomode
Section \ref{find opt P_l and w_l} is devoted to solving \eqref{max problem to find P_l,w_l all l}. 
%We note that solving \eqref{max problem to minimize D} for $P_{trn}$ would lead into an answer that would depend on $\hat{h}_l$ (which is impractical). Hence, to find $P_{trn}$ we consider a modified objective function
Problem (P$_B$) finds the optimal $P_{trn}$ that, instead of minimizing $D$, it minimizes a modified objective function $\mathbb{E}\{D\}$, where an average is taken over the channel estimates. In Section \ref{find opt P_trn and psi_l} we address (P$_B$) and find $P_{trn}$ as well as training power distribution $\{\psi_l\}_{l=1}^L$ among the CHs such that $\sum_{l=1}^L \psi_l\!=\!P_{trn}$. 
%we first assume that $P_{trn}$ is given\footnote{Given $P_{trn}$, the FC computes $\{\hat{h}_l,\zeta_l^2\}_{l=1}^{L}$ using the procedure described in section \ref{finding psi_l}.} and we find optimal $\{P_l, \boldsymbol{w}_l\}_{l=1}^L$. Later in Section \ref{find opt P_trn and psi_l} we propose our approach to find optimal value of $P_{trn}$ as well as training powers $\{\psi_l\}_{l=1}^L$. Therefore, we need to deal with two sub-problems: sub-problem (P$_A$): Finding optimal values of $\{P_l, \boldsymbol{w}_l\}_{l=1}^L$ given $P_{trn}$, sub-problem (P$_B$): Finding optimal value of $P_{trn}$. Sections \ref{find opt P_l and w_l} and \ref{finding P_{trn}^{*}} describe our approaches to tackle (P$_A$) and (P$_B$), respectively. Having the solution to (P$_A$), we summarize our approach to address \eqref{max problem to minimize D} in Section \ref{proposed algorithm} in Algorithm \ref{solution to problem P1}.
% -------------------------------------------------------------
\vspace{-.2cm}
\subsection{Finding Optimal $\{P_l, \boldsymbol{w}_l\}_{l=1}^L$ {\blue Given Total Training Power}} \label{find opt P_l and w_l}
%\vspace{-.1cm}
We start with \eqref{max problem to find P_l,w_l all l}. By taking the second derivative of $\sum_{l=1}^{L}{\cal F}_l(P_l,\boldsymbol{w}_l)$ w.r.t $\{P_l,\boldsymbol{w}_l\}$, it is straightforward to show that \eqref{max problem to find P_l,w_l all l} is not jointly concave over the optimization variables. Alternatively, we propose  {\red a solution} approach {\blue that converges to a stationary point of \eqref{max problem to find P_l,w_l all l}}. Problem \eqref{max problem to find P_l,w_l all l} contains the constraint $\sum_{l=1}^{L}C_l(P_l,\boldsymbol{w}_l)\leq\sigma P_{tot}$, which is referred to as coupling or complicating constraint in the literature \cite{primal_dual_decomp}. By introducing additional auxiliary variables $\{{\cal V}_l\}_{l=1}^L$, problem \eqref{max problem to find P_l,w_l all l} becomes:
\leqnomode
\vspace{+.1cm}
\begin{align} \label{max problem to find P_l,w_l,V_l}
\ \ \ \ \ \ \ \ \ &\text{given}\ P_{trn},\ \mathop{\text{max}}_{\{{\cal V}_l
, P_l, \boldsymbol{w}_l\}_{l=1}^L}\ \ \sum_{l=1}^{L}{\cal F}_l(P_l,\boldsymbol{w}_l)\tag{P2}\\
\!\!\text{s.t.}\ C_l(\!P_l,\boldsymbol{w}_l\!)&\!\leq\!{\cal V}_l, \sum_{l=1}^{L}\!{\cal V}_l\!\leq\!\sigma P_{tot}, {\cal V}_l, P_l\!\in\!\mathbb{R}^{+}, \boldsymbol{w}_l\!\in\!\mathbb{R}^{K_l}, \forall l.\nonumber
\vspace{-.1cm}
\end{align}
\reqnomode
Note that the auxiliary variable ${\cal V}_l$ represents the total amount of power allocated to cluster $l$ (for sensors within cluster $l$ to transmit their observations to CH$_l$ and for CH$_l$ to transmit $y_l$ to the FC). According to the {\it primal decomposition} \cite{primal_dual_decomp}, problem \eqref{max problem to find P_l,w_l,V_l} can be decomposed as the following:   
\leqnomode
\vspace{-.1cm}
\begin{align} \label{max problem to find P_l,w_l}\tag{SP2-1}
~~~~~~&\text{given}\ P_{trn},{\cal V}_l,\ \mathop{\text{max}}_{P_l,\boldsymbol{w}_l}\ {\cal F}_l(P_l,\boldsymbol{w}_l)\\
&\text{s.t.}\ \ C_l(P_l,\boldsymbol{w}_l)\!\leq\!{\cal V}_l,\ P_l\in \mathbb{R}^{+},\ \boldsymbol{w}_l\in\mathbb{R}^{K_l},\nonumber
\end{align}
\vspace{-.8cm}
\begin{align} \label{max problem to find V_l}\tag{SP2-2}
~~~~~~~~~~&\text{given}\ P_{trn},\{\!P_l,\boldsymbol{w}_l\!\}_{l=1}^L, \mathop{\text{max}}_{\{{\cal V}_l\}_{l=1}^{L}} \sum_{l=1}^{L}\!{\cal F}^{opt}_l\\
&\text{s.t.}\ \ \sum_{l=1}^{L}{\cal V}_l\!\leq\!\sigma P_{tot},{\cal V}_l\in\mathbb{R}^{+}, \forall l,\nonumber
\vspace{-.1cm}
\end{align}
\reqnomode
where ${\cal F}^{opt}_l$ denotes the maximum of ${\cal F}_l(P_l,\boldsymbol{w}_l)$, which depends on ${\cal V}_l$. The  solution can be reached by iteratively solving sub-problems \eqref{max problem to find P_l,w_l} and \eqref{max problem to find V_l}. 
%At each iteration, first \eqref{max problem to find P_l,w_l} is solved for each cluster to find $P_l,\boldsymbol{w}_l$ given $P_{trn},{\cal V}_l$. Second, sub-problem \eqref{max problem to find V_l} is solved to find $\{{\cal V}_l\}_{l=1}^{L}$ given $P_{trn},\{P_l,\boldsymbol{w}_l\}_{l=1}^L$. 
In the following, we provide the detailed solutions for \eqref{max problem to find P_l,w_l} and \eqref{max problem to find V_l}. 
%\vspace{-2cm}
\subsubsection{Solving Optimization Problem \eqref{max problem to find P_l,w_l}} \label{Solving P2-1}
We start with a brief overview of this section. Let ${\boldsymbol{w}}^{opt}_l, P^{opt}_l$ denote the solution of \eqref{max problem to find P_l,w_l}. We will show how to compute ${\boldsymbol{w}}^{opt}_l$ in terms of $P_l$ using \eqref{opt w_l and f_l as a FN of P_l} and how to compute $P^{opt}_l$ in terms of $\boldsymbol{w}_l$ using \eqref{P_opt in terms of w_opt and v_l}. Having two equations \eqref{opt w_l and f_l as a FN of P_l}, \eqref{P_opt in terms of w_opt and v_l}, we substitute $\boldsymbol{w}_l$ from \eqref{opt w_l and f_l as a FN of P_l} into \eqref{P_opt in terms of w_opt and v_l} to reach \eqref{eqn to solve to obtain P_l}, which is a function of $P^{opt}_l$ only. Employing a numerical line search method we obtain $P^{opt}_l$ from \eqref{eqn to solve to obtain P_l}. Having $P^{opt}_l$, we find ${\boldsymbol{w}}^{opt}_l$ using \eqref{opt w_l and f_l as a FN of P_l}. The detailed explanations follow.

Examining ${\cal F}_l(P_l,\boldsymbol{w}_l)$ and $C_l(P_l,\boldsymbol{w}_l)$ expressions given in \eqref{def of J_l and C_l}, it is evident that scaling up equally $P_l, \boldsymbol{w}_l$ increases both ${\cal F}_l(P_l,\boldsymbol{w}_l)$ and $C_l(P_l,\boldsymbol{w}_l)$. Therefore, \eqref{max problem to find P_l,w_l} is equivalent to its converse formulation, where $C_l(P_l,\boldsymbol{w}_l)$ is minimized subject to a constraint on ${\cal F}_l(P_l,\boldsymbol{w}_l)$:
\leqnomode
\vspace{-.1cm}
\begin{align} \label{converse min prob to find P_l,w_l} 
~~~~~~~~~&\text{given}\ P_{trn},{\cal U}_l,\ \mathop{\text{min}}_{P_l,\boldsymbol{w}_l}\ \ C_l(P_l,\boldsymbol{w}_l)\tag{CSP2-1}\\
&\text{s.t.}\ \ {\cal F}_l(\!P_l,\boldsymbol{w}_l\!)\!\geq\!{\cal U}_l,P_l\!\in\!\mathbb{R}^{+}\!,\boldsymbol{w}_l\!\in\!\mathbb{R}^{K_l}\!.\nonumber
\vspace{-.1cm}
\end{align}
\reqnomode
%in the sense that the optimal solutions ${\cal F}^{opt}_l({\cal V}_l)$ (of \eqref{max problem to find P_l,w_l}) and $C^{opt}_l({\cal U}_l)$ (of \eqref{converse min prob to find P_l,w_l}) are inverses of one another. Substituting $\boldsymbol{\Lambda}_{1_l}$ from \eqref{l-th block of Lambda_1, Lambda_2, mu_l}, the constraint in \eqref{converse min prob to find P_l,w_l} can further be simplified to: 
Let $C^{opt}_l$ be the minimum of $C_l(P_l,\boldsymbol{w}_l)$, which depends on ${\cal U}_l$. 
%Since \eqref{converse min prob to find P_l,w_l} is converse of \eqref{max problem to find P_l,w_l} the optimal solutions of \eqref{converse min prob to find P_l,w_l} and \eqref{max problem to find P_l,w_l} are inversely related. 
To solve \eqref{converse min prob to find P_l,w_l} we simplify its constraint by substituting $\boldsymbol{\Lambda}_{1_l}$ from \eqref{l-th block of Lambda_1, Lambda_2, mu_l} into ${\cal F}_l(P_l,\boldsymbol{w}_l)$ in \eqref{def of J_l and C_l}. Let $\boldsymbol{B}_l\!=\!\sigma^2_{\theta}\zeta_l^2\boldsymbol{\Pi}_l+({|\hat{h}_l|}^2+\zeta_l^2)\boldsymbol{\Delta}_l$. The constraint in \eqref{converse min prob to find P_l,w_l} becomes:
%
%\vspace{-.15cm}
\begin{equation} \label{expanded cnstrt in CSP2-1}
P_l{\boldsymbol{w}_l}^T\!({|\hat{h}_l|}^2\boldsymbol{\Pi}_l-{\cal U}_l\boldsymbol{B}_l)\boldsymbol{w}_l-({|\hat{h}_l|}^2\!+\zeta_l^2){\cal U}_l\boldsymbol{w}_l^T\!\boldsymbol{\Sigma}_{q_l}\boldsymbol{w}_l-\sigma_{v_{l}}^2{\cal U}_l\!\geq\! 0.
\end{equation}
%
%========single column equation===============================
\begin{figure*}[b]
\begin{align} \label{lagrangian of P6}
\mathcal L(\!\gamma,\eta,P_l,\boldsymbol{w}_l\!)\!=\!{\boldsymbol{w}_l}^T\!\boldsymbol{\Sigma}_{q_l}\boldsymbol{w}_l\!+\!P_l(1\!+\!{\boldsymbol{w}_l}^T\!\boldsymbol{\Omega}_l\boldsymbol{w}_l)\!+\!\gamma(({|\hat{h}_l|}^2\!+\!\zeta_l^2){\cal U}_l\boldsymbol{w}_l^T\!\boldsymbol{\Sigma}_{q_l}\boldsymbol{w}_l\!+\!\sigma_{v_{l}}^2{\cal U}_l\!-\!P_l{\boldsymbol{w}_l}^T\!({|\hat{h}_l|}^2\boldsymbol{\Pi}_l\!-\!{\cal U}_l\boldsymbol{B}_l)\boldsymbol{w}_l)\!-\!\eta P_l,
\end{align}
\end{figure*}
%=============================================================
Consider \eqref{converse min prob to find P_l,w_l} where its constraint is now replaced with the inequality in \eqref{expanded cnstrt in CSP2-1}. To solve \eqref{converse min prob to find P_l,w_l} we use the Lagrange multiplier method. Let ${\mathcal L}(\gamma,\eta,P_l,\boldsymbol{w}_l)$ be the Lagrangian for this problem and $\gamma$ and $\eta$ be the lagrange multipliers for the constraint in \eqref{expanded cnstrt in CSP2-1} and the constraint $P_l\!\geq \!0$, respectively. 
Equation \eqref{lagrangian of P6} shows ${\mathcal L}(\gamma,\eta,P_l,\boldsymbol{w}_l)$. The corresponding Karush-Kuhn-Tucker (KKT) optimality conditions are \cite[pp. 243-244]{boyd_convex_opt_book}:
\vspace{-.15cm}
\begin{subequations}\label{KKT of P6}
\begin{align}
\!\!\!\!\!\!\!\!\!\frac{\partial {\mathcal L}}{\partial \boldsymbol{w}_l}\!=&[\boldsymbol{R}_{t_l}\!+\!\gamma(({|\hat{h}_l|}^2\!+\!\zeta_l^2){\cal U}_l\boldsymbol{\Sigma}_{q_l}\!-\!P_l({|\hat{h}_l|}^2\boldsymbol{\Pi}_l\!-\!{\cal U}_l\boldsymbol{B}_l))]\boldsymbol{w}_l\!=\!\boldsymbol{0};\label{eq:KKT_P6_der_w}\\
\!\!\!\!\!\!\!\!\!\gamma(\!P_l\boldsymbol{w}_l^T&\!(\!{|\hat{h}_l|}^2\boldsymbol{\Pi}_l\!-\!{\cal U}_l\!\boldsymbol{B}_l)\boldsymbol{w}_l\!-\!(\!{|\hat{h}_l|}^2\!+\!\zeta_l^2){\cal U}_l\boldsymbol{w}_l^T\boldsymbol{\Sigma}_{q_l}\boldsymbol{w}_l\!-\!\sigma_{v_{l}}^2{\cal U}_l\!)\!=\!0;\label{eq:KKT_P6_der_gamma}\\
\!\!\!\!\!\!\!\!\!\frac{\partial {\mathcal L}}{\partial P_l}\!=&1\!+\!\boldsymbol{w}_l^T\boldsymbol{\Omega}_l\boldsymbol{w}_l\!-\!\gamma\boldsymbol{w}_l^T({|\hat{h}_l|}^2\boldsymbol{\Pi}_l\!-\!{\cal U}_l\boldsymbol{B}_l)\boldsymbol{w}_l\!-\!\eta\!=\!0;\label{eq:KKT_P6_der_P}\\
\!\!\!\!\!\!\!\!\!\eta P_l\!=&0,\label{eq:KKT_P6_der_eta}
%\vspace{-.1cm}
\end{align}
\end{subequations}
where $\boldsymbol{R}_{t_l}$, defined in {\red \eqref{defs needed for power constraints}}, depends on $P_l$. Similar to the solution of \eqref{max problem to minimize D}, one can show that the solutions of \eqref{max problem to find P_l,w_l} and \eqref{converse min prob to find P_l,w_l} must satisfy the equality constraints $C_l(P_l,\boldsymbol{w}_l)\!=\!{\cal V}_l$ and ${\cal F}_l(\!P_l,\boldsymbol{w}_l\!)\!=\!{\cal U}_l$ (or equivalently \eqref{eq:KKT_P6_der_gamma}), respectively. Thus we find:
\vspace{-.2cm}
\begin{subequations}\label{active constraints for P6}
\begin{align} 
&\!\!\!\!{\boldsymbol{w}_l}^T\boldsymbol{R}_{t_l}\boldsymbol{w}_l\!=\!{\cal V}_l-P_l,\label{active constraint P6 1}\\
&\!\!\!\!{\boldsymbol{w}_l}^T[P_l({|\hat{h}_l|}^2\boldsymbol{\Pi}_l\!-\!{\cal U}_l\!\boldsymbol{B}_l)\!-\!({|\hat{h}_l|}^2\!\!+\!\zeta_l^2){\cal U}_l\boldsymbol{\Sigma}_{q_l}]\boldsymbol{w}_l\!=\!\sigma_{v_{l}}^2{\cal U}_l.\label{active constraint P6 2} 
\vspace{-.1cm}
\end{align}
\end{subequations}
Combining \eqref{active constraint P6 1} and \eqref{active constraint P6 2} we reach:
\vspace{-.1cm}
\begin{equation} \label{pre-req eq to find gamma}
{\boldsymbol{w}_l}^T\![\boldsymbol{R}_{t_l}\!+\frac{{\cal V}_l\!-\!P_l}{\sigma_{v_{l}}^2{\cal U}_l}(({|\hat{h}_l|}^2\!+\zeta_l^2){\cal U}_l\boldsymbol{\Sigma}_{q_l}\!-\!P_l({|\hat{h}_l|}^2\boldsymbol{\Pi}_l-{\cal U}_l\boldsymbol{B}_l))]\boldsymbol{w}_l\!=\!\boldsymbol{0}.
\vspace{-.1cm}
\end{equation}
From \eqref{eq:KKT_P6_der_w} and \eqref{pre-req eq to find gamma} we find the lagrange multiplier $\gamma$:
\vspace{-.1cm}
\begin{equation} \label{gamma}
\gamma=\frac{{\cal V}_l\!-\!P_l}{\sigma_{v_{l}}^2{\cal U}_l}.
\vspace{-.1cm}
\end{equation}
%
%Let $\beta_l=\frac{{\cal V}_l\!-\!P_l}{\sigma_{v_{l}}^2}$. 
%We start with a brief overview of this section. Let ${\boldsymbol{w}}^{opt}_l, P^{opt}_l$ denote the solution of \eqref{max problem to find P_l,w_l}. We will show how to compute ${\boldsymbol{w}}^{opt}_l$ in terms of $P_l$ using \eqref{opt w_l and f_l as a FN of P_l} and how to compute $P^{opt}_l$ in terms of $\boldsymbol{w}_l$ using \eqref{P_opt in terms of w_opt and v_l}. Having two equations \eqref{max problem to find P_l,w_l}, \eqref{P_opt in terms of w_opt and v_l}, we substitute $\boldsymbol{w}_l$ from \eqref{opt w_l and f_l as a FN of P_l} into \eqref{P_opt in terms of w_opt and v_l} to reach \eqref{eqn to solve to obtain P_l}, which is a function of $P^{opt}_l$ only. Employing a numerical line search method we obtain $P^{opt}_l$ from \eqref{eqn to solve to obtain P_l}. Having $P^{opt}_l$, we find ${\boldsymbol{w}}^{opt}_l$ using \eqref{opt w_l and f_l as a FN of P_l}. The detailed explanations follow.
%%%%%%%%%%%%%%%%%%%%% 
$\bullet$ {\red Computing} ${\boldsymbol{w}}^{opt}_l$ given $P_l$: 
Substituting \eqref{gamma} into \eqref{eq:KKT_P6_der_w} and conducting some mathematical manipulations result in:
\vspace{-.15cm}
\begin{align} \label{eig problem}
{\cal U}_l[\underbrace{\frac{\sigma_{v_{l}}^2\boldsymbol{R}_{t_l}}{{\cal V}_l\!-\!P_l}\!+\!({|\hat{h}_l|}^2\!+\!\zeta_l^2)\boldsymbol{\Sigma}_{q_l}\!+\!P_l\boldsymbol{B}_l}_{=\boldsymbol{\cal B}_1}]\boldsymbol{w}_l\!=\!|\boldsymbol{\mu}_l|{|\boldsymbol{\mu}_l|}^T\boldsymbol{w}_l,
\vspace{-.1cm}
\end{align}
where $\boldsymbol{\mu}_l$ is defined in \eqref{l-th block of Lambda_1, Lambda_2, mu_l}. 
%Therefore the optimal tradeoff between distortion ${\cal F}_l$ and transmit power $C_l$ and also the optimal weights $\boldsymbol{w}_l$ that achieve that tradeoff are obtained through the solution of the generalized eigenvalue problem stated in \eqref{eig problem}. In particular, the function ${\cal F}^{opt}_l({\cal V}_l)$ and its inverse $C^{opt}_l({\cal U}_l)$ are \cite{varshney_TIT_2016}:
Since $\boldsymbol{R}_{t_l}\!\succ\!\boldsymbol{0}, \boldsymbol{\Sigma}_{q_l}\!\succ\!\boldsymbol{0}, \boldsymbol{B}_l\!\succ\!\boldsymbol{0}$, the matrix $\boldsymbol{\cal B}_1$ is positive definite and full rank and hence invertible. Multiplying both sides of \eqref{eig problem} with ${\boldsymbol{\cal B}_1}^{-1}$, we find:
\vspace{-.2cm}
\begin{equation} \label{simplified eig problem 1}
\!{\cal U}_l\boldsymbol{w}_l\!=\!{\boldsymbol{\cal B}_1}^{-1}|\boldsymbol{\mu}_l|{|\boldsymbol{\mu}_l|}^T\!\boldsymbol{w}_l.
\vspace{-.1cm}
\end{equation}
Also, multiplying both sides of \eqref{eig problem} with $\frac{1}{{\cal U}_l}{\boldsymbol{R}_{t_l}}^{-1}$ we reach: 
\vspace{-.25cm}
\begin{equation} \label{simplified eig problem 2}
\frac{\sigma_{v_{l}}^2}{{\cal V}_l\!-\!P_l}\boldsymbol{w}_l\!=\!\underbrace{{\boldsymbol{R}_{t_l}}\!^{-1}[\frac{|\boldsymbol{\mu}_l|{|\boldsymbol{\mu}_l|}^T}{{\cal U}_l}\!-\!({|\hat{h}_l|}^2\!\!+\!\zeta_l^2)\boldsymbol{\Sigma}_{q_l}\!\!-\!P_l\boldsymbol{B}_l]}_{=\boldsymbol{\cal B}_2}\!\boldsymbol{w}_l.
\vspace{-.1cm}
\end{equation}
Inspecting \eqref{simplified eig problem 1} and \eqref{simplified eig problem 2}, and aiming at finding vector $\boldsymbol{w}_l$, we realize that \eqref{simplified eig problem 1} and \eqref{simplified eig problem 2} are ordinary eigenvalue problems. 
%Let ${\cal F}^{opt}_l$ and $C^{opt}_l$ denote the maximum value that ${\cal F}_l(\!P_l,\boldsymbol{w}_l\!)$ and the minimum value that $C_l(P_l,\boldsymbol{w}_l)$ can attain, respectively. 
Since the solutions to \eqref{max problem to find P_l,w_l} and \eqref{converse min prob to find P_l,w_l} satisfy the equality constraints $C_l(P_l,\boldsymbol{w}_l)\!=\!{\cal V}_l$ and ${\cal F}_l(\!P_l,\boldsymbol{w}_l\!)\!=\!{\cal U}_l$, respectively, 
%the optimal tradeoff between distortion ${\cal F}_l(\!P_l,\boldsymbol{w}_l\!)$ and transmit power $C_l(P_l,\boldsymbol{w}_l)$, i.e., the maximum value that ${\cal F}_l(\!P_l,\boldsymbol{w}_l\!)$ attains, that is ${\cal F}^{opt}_l({\cal V}_l)$, and the minimum value that $C_l(P_l,\boldsymbol{w}_l)$ attains, that is $C^{opt}_l\!({\cal U}_l)$, respectively, are equal to:
from \eqref{simplified eig problem 1} and \eqref{simplified eig problem 2} we find:
\vspace{-.1cm}
\begin{eqnarray} \label{solution of eig-val problem}
{\cal F}^{opt}_l\!=\!\lambda_{max}({\boldsymbol{\cal B}_1}^{-1}|\boldsymbol{\mu}_l|{|\boldsymbol{\mu}_l|}^T),\ C^{opt}_l\!=\!\frac{\sigma_{v_{l}}^2}{\lambda_{max}(\boldsymbol{\cal B}_2)}\!+\!P_l.
%\vspace{-.1cm}
\end{eqnarray}
%
%respectively\footnote{The operators $\lambda(\boldsymbol{A},\boldsymbol{B})$ and $\boldsymbol{s}(\boldsymbol{A},\boldsymbol{B})$, respectively, denote the eigenvalue and eigenvector to the generalized eigenvalue problem $\boldsymbol{A}\boldsymbol{s}=\lambda\boldsymbol{B}\boldsymbol{s}$. Operator $\lambda_{max}(.,.)$ denotes the maximum among all real eigenvalues and $\lambda_{min}^{pos}(.,.)$ denotes the minimum among all positive eigenvalues (different from zero). Note that when $\boldsymbol{B}$ is full rank, then $\lambda(\boldsymbol{A},\boldsymbol{B})$ are the eigenvalues of ${\boldsymbol{B}}^{-1}\boldsymbol{A}$.}. 
Let ${\boldsymbol{s}}^{opt}_l$ be the eigenvector corresponding to $\lambda_{max}({\boldsymbol{\cal B}_1}^{-1}|\boldsymbol{\mu}_l|{|\boldsymbol{\mu}_l|}^T)$. We note that ${\cal F}^{opt}_l$ is achieved when $\boldsymbol{w}_l$ is an appropriately scaled version of ${\boldsymbol{s}}^{opt}_l$, i.e., ${\boldsymbol{w}}^{opt}_l\!=\!r_l{\boldsymbol{s}}^{opt}_l$, where scalar $r_l$ is such that \eqref{active constraint P6 1} is satisfied. Also recall $\boldsymbol{\Pi}_l\!=\!\boldsymbol{\rho}_l{\boldsymbol{\rho}_l}^T$ is rank-1. Thus ${\cal F}^{opt}_l$ is the only non-zero eigenvalue of ${\boldsymbol{\cal B}_1}^{-1}|\boldsymbol{\mu}_l|{|\boldsymbol{\mu}_l|}^T$ and ${\boldsymbol{s}}^{opt}_l$ is the corresponding eigenvector. 
%
%\vspace{-.2cm}
%\begin{subequations}\label{J_opt and s_opt in terms of P_l and v_l}
%\begin{align} 
%{\cal F}^{opt}_l&\!=\!{|\boldsymbol{\mu}_l|}^T{(\frac{\sigma_{v_{l}}^2\boldsymbol{R}_{t_l}}{{\cal V}_l\!-\!P_l}\!+\!({|\hat{h}_l|}^2\!+\!\zeta_l^2)\boldsymbol{\Sigma}_{q_l}\!+\!P_l\boldsymbol{B}_l)}^{-1}\!|\boldsymbol{\mu}_l|,\label{J_opt after rank-1 prop}\\
%{\boldsymbol{s}}^{opt}_l&\!=\!{(\frac{\sigma_{v_{l}}^2\boldsymbol{R}_{t_l}}{{\cal V}_l\!-\!P_l}\!+\!({|\hat{h}_l|}^2\!+\!\zeta_l^2)\boldsymbol{\Sigma}_{q_l}\!+\!P_l\boldsymbol{B}_l)}^{-1}\!|\boldsymbol{\mu}_l|,\label{s_opt after rank-1 prop}
%\end{align}
%\end{subequations}
%
Proposition \ref{simplified w_l and F_l in term of P_l} gives expressions for ${\boldsymbol{w}}^{opt}_l$ and ${\cal F}^{opt}_l$ in terms of $P_l$.
\vspace{-.2cm}
\begin{prop} \label{simplified w_l and F_l in term of P_l}
\textup{
Considering problem \eqref{max problem to find P_l,w_l}, the optimal fusion vector ${\boldsymbol{w}}^{opt}_l$ and the maximum value of the objective function ${\cal F}^{opt}_l$ in terms of $P_l$ are:
\vspace{-.25cm}
\begin{equation} \label{opt w_l and f_l as a FN of P_l}
{\boldsymbol{w}}^{opt}_l\!=\!\sqrt{\frac{{\cal V}_l\!-\!P_l}{\tau_l}}{\boldsymbol{R}_{t_l}}^{-1}\boldsymbol{\rho}_l,\ \ {\cal F}^{opt}_l\!=\!\frac{{|\hat{h}_l|}^2\beta_lP_l\tau_l}{\sigma_{v_{l}}^2(1+\frac{\beta_l}{{\cal V}_l-P_l})},
\end{equation}
where $\tau_l\!=\!{\boldsymbol{\rho}_l}^T{\boldsymbol{R}_{t_l}}^{-1}\boldsymbol{\rho}_l,\ \beta_l\!=\!\frac{\sigma_{v_{l}}^2}{{|\hat{h}_l|}^2(1-\sigma^2_{\theta}P_l\tau_l)+\zeta_l^2}$.
\vspace{-.1cm}
}
\end{prop}
\vspace{-.3cm}
\begin{proof}
	See Appendix \ref{finding w_opt and J_opt}.
	%\vspace{-.1cm}
\end{proof}
%$\!\!\!\!\!\!${\it Proof}. See Appendix \ref{finding w_opt and J_opt}.\\
%
%\begin{remark} \label{w_opt is LMMSE estimate}
%\textup{
For our system model $\boldsymbol{R}_{t_l\theta}\!=\!\mathbb{E}\{\theta{\boldsymbol{t}_l}\}\!=\!\sigma^2_{\theta}\sqrt{P_l}\boldsymbol{\rho}_l$. Hence, we can rewrite ${\boldsymbol{w}}^{opt}_l$ in \eqref{opt w_l and f_l as a FN of P_l} as:
\vspace{-.2cm}
\begin{equation} \label{w_l_opt is LMMSE est of theta}
{\boldsymbol{w}}^{opt}_l=\underbrace{\sigma^{-2}_{\theta}\sqrt{\frac{{\cal V}_l-P_l}{P_l\tau_l}}}_{=\chi_l}(\boldsymbol{R}_{t_l}^{-1}\boldsymbol{R}_{t_l\theta}).
\vspace{-.1cm}
\end{equation}
%
%One can immediately see that the local message sent by CH$_l$ is exactly the LMMSE estimate $\boldsymbol{R}_{\theta t_l}\boldsymbol{R}_{t_l}^{-1}\boldsymbol{t}_l$ multiplied by a scalar $\sigma^{-2}_{\theta}\!\sqrt{\frac{{\cal V}_l-P_l}{P_l\tau_l}}$. This means that LMMSE estimation followed by an amplification factor is optimal in a power-distortion sense.
Since $\boldsymbol{R}_{t_l}^{-1}\boldsymbol{R}_{t_l\theta}$ is the linear operator corresponding to the LMMSE estimator, \eqref{w_l_opt is LMMSE est of theta} implies that the optimal linear fusion rule at CH$_l$ is {\blue equal to the linear operator corresponding to} the LMMSE estimation of $\theta$ based on $\boldsymbol{t}_l$, {\blue multiplied} by the amplification factor $\chi_l$.  
%}
%\end{remark}
%
%%%%%%%%%%%%%%%%%%%%% 

%\vspace{-.1cm}
$\bullet$ {\red Computing} $P^{opt}_l$ given $\boldsymbol{w}_l$:
Note that \eqref{eq:KKT_P6_der_eta} results in $\eta\!=\!0$ for active clusters with $P_l>0$. Letting $\eta\!=\!0$ in \eqref{eq:KKT_P6_der_P} and solving for $\gamma$ we find:
%Then from \eqref{eq:KKT_P6_der_P} we get:
%
\vspace{-.4cm}
\begin{equation}\label{second equality for gamma}
\gamma=\frac{1+\boldsymbol{w}_l^T\boldsymbol{\Omega}_l\boldsymbol{w}_l}{\boldsymbol{w}_l^T({|\hat{h}_l|}^2\boldsymbol{\Pi}_l\!-\!{\cal U}_l\boldsymbol{B}_l)\boldsymbol{w}_l}.
\vspace{-.1cm}
\end{equation}
Equating \eqref{second equality for gamma} with \eqref{gamma} and solving for ${\cal U}_l$ we get:
\vspace{-.2cm}
\begin{equation} \label{u using gammas}
{\cal U}_l=\frac{({\cal V}_l-P_l){|\hat{h}_l|}^2\boldsymbol{w}_l^T\boldsymbol{\Pi}_l\boldsymbol{w}_l}{\sigma_{v_{l}}^2(1+\boldsymbol{w}_l^T\boldsymbol{\Omega}_l\boldsymbol{w}_l)+({\cal V}_l-P_l)\boldsymbol{w}_l^T\boldsymbol{B}_l\boldsymbol{w}_l}.
\vspace{-.1cm}
\end{equation}
On the other hand, solving \eqref{active constraint P6 2} for ${\cal U}_l$ results in:
\vspace{-.2cm}
\begin{equation} \label{Reforming active constraint P6 2}
{\cal U}_l=\frac{P_l{|\hat{h}_l|}^2\boldsymbol{w}_l^T\boldsymbol{\Pi}_l\boldsymbol{w}_l}{\sigma_{v_{l}}^2+({|\hat{h}_l|}^2\!+\!\zeta_l^2)\boldsymbol{w}_l^T\boldsymbol{\Sigma}_{q_l}\boldsymbol{w}_l+P_l\boldsymbol{w}_l^T\boldsymbol{B}_l\boldsymbol{w}_l}.
\vspace{-.1cm}
\end{equation}
Combining \eqref{u using gammas} and \eqref{Reforming active constraint P6 2}, we obtain $P^{opt}_l$ in terms of ${\boldsymbol{w}}_l$ as the following:
\vspace{-.45cm}
\begin{equation}\label{P_opt in terms of w_opt and v_l}
P^{opt}_l=\frac{{\cal V}_l(\sigma_{v_{l}}^2+({|\hat{h}_l|}^2+\zeta_l^2){\boldsymbol{w}_l}^T\boldsymbol{\Sigma}_{q_l}\boldsymbol{w}_l)}{\sigma_{v_{l}}^2(2+\boldsymbol{w}_l^T\boldsymbol{\Omega}_l\boldsymbol{w}_l)+({|\hat{h}_l|}^2+\zeta_l^2){\boldsymbol{w}_l}^T\boldsymbol{\Sigma}_{q_l}\boldsymbol{w}_l}.
\vspace{-.1cm}
\end{equation}
At this point, we have obtained two equations: \eqref{opt w_l and f_l as a FN of P_l} provides ${\boldsymbol{w}}^{opt}_l$ in terms of $P_l$, and \eqref{P_opt in terms of w_opt and v_l} provides $P^{opt}_l$ in terms of $\boldsymbol{w}_l$. Substituting ${\boldsymbol{w}}^{opt}_l$ from \eqref{opt w_l and f_l as a FN of P_l} in \eqref{P_opt in terms of w_opt and v_l} yields in:
{\small
\vspace{-.2cm}
\begin{align} \label{eqn to solve to obtain P_l}
&\!\!\!\sigma_{v_{l}}^2\!({\cal V}_l\!-\!2P^{opt}_l)\tau_l\!+\!({|\hat{h}_l|}^2\!\!+\!\zeta_l^2)({\cal V}_l\!-\!P^{opt}_l)^2{\boldsymbol{\rho}_l}^T\!{\boldsymbol{R}_{t_l}}^{-1}\boldsymbol{\Sigma}_{q_l}{\boldsymbol{R}_{t_l}}^{-1}\!\boldsymbol{\rho}_l\nonumber\\
&\!-\!P^{opt}_l\sigma_{v_{l}}^2\!({\cal V}_l\!-\!P^{opt}_l){\boldsymbol{\rho}_l}^T\!{\boldsymbol{R}_{t_l}}^{-1}\boldsymbol{\Omega}_l{\boldsymbol{R}_{t_l}}^{-1}\!\boldsymbol{\rho}_l\!=\!0.
\vspace{-.1cm}
\end{align}
}
%in which $\tau^{'}_l\!=\!{\boldsymbol{\rho}_l}^T\!{\boldsymbol{R}_{t_l}}^{-1}\boldsymbol{\Sigma}_{q_l}{\boldsymbol{R}_{t_l}}^{-1}\!\boldsymbol{\rho}_l$ and $\tau^{''}_l\!=\!{\boldsymbol{\rho}_l}^T\!{\boldsymbol{R}_{t_l}}^{-1}\boldsymbol{\Omega}_l{\boldsymbol{R}_{t_l}}^{-1}\!\boldsymbol{\rho}_l$. 
Note that $\tau_l,\!\boldsymbol{R}_{t_l}$ in \eqref{eqn to solve to obtain P_l} depend on $P^{opt}_l\!\!$, and thus, a closed-form solution for $P^{opt}_l$ remains elusive. $\!$One can employ a line search method (e.g., the Golden section method \cite[p. 216]{lin_and_nonlin_prog}) to solve \eqref{eqn to solve to obtain P_l} in the interval $(0,{\cal V}_l)$. 
%\footnote{We need to say something about the uniqueness of solution for $P^{opt}_l$. As it seems, the second derivative of left hand side w.r.t. $P^{opt}_l$ does not give us any clue! Maybe we resort to numerical methods to show the uniqueness.}. 
%After finding $P^{opt}_l$, we substitute it in \eqref{opt w_l and f_l as a FN of P_l} to find ${\boldsymbol{w}}^{opt}_l$ and therefore the optimal solution of the constrained maximization problem in \eqref{max problem to find P_l,w_l} is found.  
Having $P^{opt}_l\!$ we find ${\boldsymbol{w}}^{opt}_l\!$ using \eqref{opt w_l and f_l as a FN of P_l}. 
\subsubsection{Solving Optimization Problem \eqref{max problem to find V_l}} \label{Solving P2-2}
%\vspace{-.1cm}
By substituting ${\cal F}^{opt}_l$ from \eqref{opt w_l and f_l as a FN of P_l} in the objective function, problem \eqref{max problem to find V_l} becomes:
%\leqnomode
{\small
\vspace{-.4cm}
\begin{align} \label{simplified max problem P5-2}
&\text{given}\ P_{trn},\{P_l, \boldsymbol{w}_l\}_{l=1}^L,\ \mathop{\text{max}}_{\{{\cal V}_l\}_{l=1}^{L}}\ \ \sum_{l=1}^{L}\!\frac{{|\hat{h}_l|}^2\beta_lP_l\tau_l}{\sigma_{v_{l}}^2(1+\frac{\beta_l}{{\cal V}_l-P_l})}\nonumber\\%\tag{P4}\\
&\text{s.t.}\ \ \ \sum_{l=1}^{L}{\cal V}_l\!\leq\!\sigma P_{tot},{\cal V}_l\in\mathbb{R}^{+}, \forall l.
\end{align}
}
%\reqnomode
{\blue The maximization problem in \eqref{simplified max problem P5-2} is concave and its solution can be found via solving the KKT conditions}. In particular, we find
%can be solved analytically using Lagrangian function and KKT conditions.
%, which leads to a water filling type power allocation scheme. In particular, by using the procedure described in Appendix \ref{finding delta_opt}, the optimal value ${\cal V}_l^{opt}$, the Lagrange multiplier $\lambda$, and the set of active clusters $\cal A$ can be uniquely derived as 
(see Appendix \ref{finding delta_opt} for derivations):
\vspace{-.2 cm}
\begin{subequations}\label{opt_delta_l and lambda}
\begin{eqnarray} 
{\cal V}_l^{opt}&=&\big{[}\beta_l(\frac{|\hat{h}_l|}{\sigma_{v_{l}}}\sqrt{\frac{P_l\tau_l}{\lambda}}-1)\big{]}^{+}+P_l,\label{opt_delta_l}\\
\lambda&=&(\frac{\sum_{l\in{\cal A}}\frac{|\hat{h}_l|\beta_l\sqrt{P_l\tau_l}}{\sigma_{v_{l}}}}{\sigma P_{tot}\!-\!\sum_{l\in{\cal A}}P_l\!+\!\sum_{l\in{\cal A}}\beta_l})^2.\label{opt lambda}
\end{eqnarray}
\end{subequations}
%
%The detailed derivations on the solutions in \eqref{opt_delta_l} and \eqref{opt lambda} are given in Appendix \ref{finding delta_opt}.\\
%Having $\delta_l^{opt}$, the optimal value ${\cal V}_l^{opt}$ is computed as $\!{\cal V}_l^{opt}\!=\delta_l^{opt}+P_l$.
% -------------------------------------------------------------
%\subsection{Proposed Algorithm for Solving \eqref{max problem to minimize D}} \label{proposed algorithm}
%Note that given $\lambda$, \eqref{opt_delta_l} implies that the decision of CH$_l$ whether to transmit its fused signal or become silent depends on the observation qualities of sensors within cluster $l$ as well as physical layer parameters of the channel between CH$_l$ and the FC.
Note that the first term of the right side of the equality in \eqref{opt_delta_l} is ${\cal P}_l$ introduced in Section \ref{Power Constraints and Problem Statement}. Given $\lambda, |\hat{h}_l|, \sigma_{v_{l}}$ in \eqref{opt_delta_l} and the easy-to-prove fact that $\tau_l\!+\!P_l\frac{\partial \tau_l}{\partial P_l}\!>\!0$, it is straightforward to show that $\frac{\partial {\cal P}_l}{\partial P_l}\!>\!0$ for active clusters, i.e., increasing $P_l$ increases ${\cal P}_l$. 
%However, $P_l\!+\!{\cal P}_l\!=\!{\cal V}_l$. This implies that there is an optimal $P_l$ which can be found via solving \eqref{max problem to find P_l,w_l}. 
Having the solutions to problems \eqref{max problem to find P_l,w_l} and \eqref{max problem to find V_l}, Algorithm \ref{solution to problem P1} summarizes our proposed solution to problem \eqref{max problem to find P_l,w_l all l}. {\blue Essentially, this algorithm iteratively solves \eqref{max problem to find P_l,w_l} and \eqref{max problem to find V_l} in a block-coordinate ascent manner until the convergence is reached. 
%Since the solutions of \eqref{max problem to find P_l,w_l} and \eqref{max problem to find V_l} are unique, the output of this algorithm, denoted as $\{P_l^{opt},w_l^{opt}\}_{l=1}^L$, converges to a stationary point 
In Section \ref{conv of alg to solve P_A}, we argue that the algorithm output converges to a stationary point of \eqref{max problem to find P_l,w_l all l}.}
\vspace{-.2cm}
\begin{algorithm}[h!]
{\blue 
\caption{proposed solution of \eqref{max problem to find P_l,w_l all l}}
\label{solution to problem P1}
 {\bf Input:} $P_{tot}, P_{trn}, \{\hat{h}_l\}_{l=1}^L, \epsilon$, and system parameters defined in Section \ref{System Model and Problem Formulation}\\
 {\bf Output:} optimal optimization variables $\{\!P_l^{opt}\!, \boldsymbol{w}_l^{opt}\!\}_{l=1}^L$
  \vspace{-0.1cm}
  \hbox to \hsize{\dashfill\hfil}
% % % % % % % % % % % %
 \vspace{-0.1cm}
 Let $i$ indicate the iteration index, ${\cal V}_l^{(i)},P_l^{(i)},\boldsymbol{w}_l^{(i)},{\cal A}^{(i)}$ denote ${\cal V}_l,P_l,\boldsymbol{w}_l,{\cal A}$ values and ${\cal F}^{(i)}\!=\!\sum_{l\in{\cal A}^{(i)}}{\cal F}_l(P_l^{(i)},\boldsymbol{w}_l^{(i)})$ at iteration $i$.\\
 - Given the channel estimates, sort the clusters as described in Appendix \ref{finding delta_opt}.\\
 - Initialization: $i\!=\!1$, ${\cal A}^{(0)}\!=\{1, ..., L\}$, randomly choose $\{P_l^{(0)},\boldsymbol{w}_l^{(0)}\}_{l=1}^L$ such that $0\!<\!P_l^{(0)}\!<\!{\cal V}_l^{(0)}\!=\!\frac{P_{tot}\!-\!P_{trn}}{L}$ and \eqref{max problem to find P_l,w_l,V_l} holds with active constraints, and compute ${\cal F}^{(0)}$.\\
 - Iterate between solving \eqref{max problem to find P_l,w_l} and \eqref{max problem to find V_l} until convergence. At iteration $i$ do below:\\
$\ \ \!$1: Obtain $P_l^{(i)}\!\in\!(0,{\cal V}_l^{(i-1)})$ via solving \eqref{eqn to solve to obtain P_l}, substitute $P_l^{(i)}$ into \eqref{opt w_l and f_l as a FN of P_l} to obtain $\boldsymbol{w}_l^{(i)}\!$, compute ${\cal F}^{(i)}\!$.\\
$\ \ \!$2: If $|\frac{{\cal F}^{(i)}-{\cal F}^{(i-1)}}{{\cal F}^{(i-1)}}|\!\leq\!\epsilon$, terminate the iteration and return the optimal solution $\{P_l^{opt}\!=\!P^{(i)}_l,\boldsymbol{w}_l^{opt}\!=\!\boldsymbol{w}^{(i)}_l\}_{\forall l\in{\cal A}^{(i)}}$ and $\{P_l^{opt}\!=\!0,\boldsymbol{w}_l^{opt}\!=\!\boldsymbol{0}\}_{\forall l\notin{\cal A}^{(i)}}$.\\
$\ \ \!$3: Increase $i$, update ${\cal A}^{(i)}$, and find $\{{\cal V}_l^{(i+1)}\}_{\forall l\in{\cal A}^{(i)}}$ using \eqref{opt_delta_l}, \eqref{opt lambda}.\\
- Continue the iteration until the stopping criteria in step 2 is met.
%\uIf{$|\frac{{\cal F}^{(i)}-{\cal F}^{(i-1)}}{{\cal F}^{(i-1)}}|\!\leq\!\epsilon$}
 %{Return the optimal solution $\{P_l^{opt}\!=\!P^{(i)}_l,\boldsymbol{w}_l^{opt}\!=\!\boldsymbol{w}^{(i)}_l\}_{\forall l\in{\cal A}^{(i)}}$ and $\{P_l^{opt}\!=\!0,\boldsymbol{w}_l^{opt}\!=\!\boldsymbol{0}\}_{\forall l\notin{\cal A}^{(i)}}$.}
 %\Else
 %{Increase $i$, update ${\cal A}^{(i)}$, and find $\{{\cal V}_l^{(i+1)}\}_{\forall l\in{\cal A}^{(i)}}$ using \eqref{opt_delta_l}, \eqref{opt lambda} until the stopping criteria is met.}
 % % % % % % % % % % % %
 % % % % % % % % % % % %
 }
 \end{algorithm}
\vspace{-.6cm}
\subsection{Finding Optimal {\blue Total Training Power and its Distribution Among CHs}} \label{find opt P_trn and psi_l}
\vspace{-.1cm}
%Since multiplying $P_l$ and $\boldsymbol{w}_l$ by a scalar $\alpha>1$ (strictly) increases both $\cal J$ and ${\cal P}_d$ (and for $\alpha<1$, strictly decreases them), the optimal solution of \eqref{max problem to minimize D} holds with active constraint ${\cal P}_d(\{P_l,\boldsymbol{w}_l\}_{l=1}^{L})+P_{trn}=P_{tot}$. 
%The optimal solution of \eqref{max problem to minimize D} holds with active constraint $P_{trn}+\sum_{l=1}^{L}C_l(P_l,\boldsymbol{w}_l)=P_{tot}$. The argument is as follows. If a set of $\{P_l,\boldsymbol{w}_l\}_{l=1}^{L}$ is such that strict inequality holds, we can equally scale up each $\{P_l,\boldsymbol{w}_l\}$ so that equality holds. In addition, if we equally scale up each $\{P_l,\boldsymbol{w}_l\}$, we obtain a larger function value of $\cal J$. Consequently, with optimal $\{P_l,\boldsymbol{w}_l\}$, the inequality constraint must be active. 
%In section \ref{finding P_{trn}^{*}}, we propose an algorithm to find optimal total training power $P_{trn}^{*}$. Then having $P_{trn}^{*}$, we propose our approach to obtain $\{\psi_l\}_{l=1}^L$ in section \ref{finding psi_l}.
In this section, we focus on (P$_B$) and find $P_{trn}$ as well as training power distribution $\{\psi_l\}_{l=1}^L$ among the CHs such that $\sum_{l=1}^L\psi_l\!=\!P_{trn}$. 
%%%%%%%%%%%%%%%%%%%%%%%%%%%%%%%%%%%%
\begin{comment}
\subsubsection{Finding $P_{trn}^{opt}$} \label{finding P_{trn}^{*}}
Considering the objective function in \eqref{max problem to minimize D} we note that ${\cal J}_l(P_{trn},P_l,\boldsymbol{w}_l)$ depends on the channel estimate $\hat{h}_l$, as shown in the definition \eqref{def of J_l and C_l}, through ${|\hat{h}_l|}^2$ in the numerator and $\boldsymbol{\Lambda}_{1_l}$ in the denominator. Using this objective function directly to find the optimal training power $P_{trn}$ would lead into an answer that depends on the channel estimates, and thus cannot be implemented in practice for training power allocation. To obtain a solution that does not depend on the channel estimates, and instead depends on the statistics of these estimates, one can find $P_{trn}^{opt}$ from the constrained maximization of the average 
\end{comment}
As we mentioned earlier, to find $P_{trn}$ we consider a modified objective function, i.e., instead of $\sum_{l=1}^{L}{\cal J}_l$ in \eqref{max problem to minimize D} we consider $\sum_{l=1}^{L}\mathbb{E}\{{\cal J}_l\}$, where the expectation is taken over the channel estimates ${|\hat{h}_l|}^2$. Since solving this problem analytically is still intractable, we use the Jensen's inequality for concave functions \cite[pp. 77-78]{boyd_convex_opt_book}, to establish a lower bound on $\mathbb{E}\{{\cal J}_l(P_{trn},P_l,\boldsymbol{w}_l)\}$:
\vspace{-.15cm}
\begin{equation*} %\label{J_l using Jensens}
\mathbb{E}\{{\cal J}_l(P_{trn},P_l,\boldsymbol{w}_l)\}\!\leq\!{\cal G}_l(P_{trn},P_l,\boldsymbol{w}_l),
\vspace{-.15cm}
\end{equation*}
where ${\cal G}_l(P_{trn},P_l,\boldsymbol{w}_l)$ is obtained from ${\cal J}_l(P_{trn},P_l,\boldsymbol{w}_l)$, after replacing ${|\hat{h}_l|}^2$ with $\mathbb{E}\{{|\hat{h}_l|}^2\}$. To find $\mathbb{E}\{{|\hat{h}_l|}^2\}$ needed for ${\cal G}_l(P_{trn},P_l,\boldsymbol{w}_l)$ we revisit the error corresponding to the LMMSE channel estimation in \eqref{channel estimate and channel estimation error variance}. Note that $\hat{h}_l$ is a zero-mean complex Gaussian.
%Considering \eqref{pilot symbol at the FC}, we note that $h_l$ and $\nu_l$ are zero-mean independent complex Gaussian, and hence from \eqref{channel estimate and channel estimation error variance} we find that $\hat{h}_l$ is also a zero-mean complex Gaussian. 
Let $2\sigma_{\hat{h}_l}^2$ denote the variance of $\hat{h}_l$. For the model $h_l\!=\!\hat{h}_l\!+\!\tilde{h}_l$, we invoke the orthogonality principle from the linear estimation theory \cite{Kay_book_est_theory}, that states $var({\hat{h}_l})\!=\!var({h_l})-var({\tilde{h}_l})\!=\!2\sigma_{h_l}^2-\zeta_l^2$, where $\zeta_l^2$ in \eqref{channel estimate and channel estimation error variance} depends on $\psi_l$. Since $\hat{h}_l$ is zero-mean, we have $\mathbb{E}\{{|\hat{h}_l|}^2\}\!=\!var({\hat{h}_l})$. Thus, ${\cal G}_l(P_{trn},P_l,\boldsymbol{w}_l)\!=\!\frac{(2\sigma_{h_l}^2-\zeta_l^2)P_l{\boldsymbol{w}_l}^T\boldsymbol{\Pi}_l\boldsymbol{w}_l}{\sigma_{v_{l}}^2+{\boldsymbol{w}_l}^T\boldsymbol{\Lambda}_{1_l}\boldsymbol{w}_l}$, where $\boldsymbol{\Lambda}_{1_l}\!=\!\sigma^2_{\theta}\zeta_l^2P_l\boldsymbol{\Pi}_l+2\sigma_{h_l}^2(\boldsymbol{\Sigma}_{q_l}+P_l\boldsymbol{\Delta}_l)$. Notice that ${\cal G}_l$ depends on the optimization variable $P_{trn}$ through $\zeta_l^2$ in the numerator and $\boldsymbol{\Lambda}_{1_l}$ in the denominator.
%To find $P_{trn}^{opt}$ we consider the following constrained maximization problem:
%\leqnomode
%
%\begin{align} \label{max problem to obtain P_trn}
%~~~~~~&\mathop{\text{max}}_{P_{trn},\{P_l, \boldsymbol{w}_l\}_{l=1}^L}\ \ \sum_{l=1}^{L}{\cal G}_l(P_{trn},P_l,\boldsymbol{w}_l)\tag{P3}\\
%\!\text{s.t.}\ \ P_{trn}+\!\!&\sum_{l=1}^{L}\!C_l(\!P_l,\boldsymbol{w}_l\!)\!\leq\! P_{tot},P_{trn}\!\in\!\mathbb{R}^{+}\!,P_l\!\in\! \mathbb{R}^{+}\!,\boldsymbol{w}_l\!\in\! \mathbb{R}^{K_l}\!, \forall l.\nonumber
%\end{align}
% 
%\reqnomode
%which is equivalent to \eqref{max problem to minimize D} if $\sigma_{\hat{h}_l}^2$ is substituted by ${|\hat{h}_l|}^2$. In the following, we argue that $P_{trn}^{*}$ given $\{P_l,\boldsymbol{w}_l\}_{l=1}^{L}$ 
%Suppose (P$^{'}_{B}$) is the second sub-problem obtained from the decomposition of \eqref{max problem to obtain P_trn}, where the goal is to find $P_{trn}$ given $\{P_l, \boldsymbol{w}_l\}_{l=1}^L$ (obtained from solving the first sub-problem using Algorithm 1 described in Section \ref{find opt P_l and w_l}). In the following, we argue that the solution to (P$^{'}_{B}$) can be found via a simple one-dimensional optimization method. Recall from section \ref{find opt P_l and w_l} that $P_{trn}\!=\!(1\!-\!\sigma)P_{tot}$ such that $\sigma\!\in\!(0,1)$. 
We reconsider \eqref{max problem to minimize D} in which ${\cal J}_l$ is now replaced with ${\cal G}_l$: 
%Given $\{P_l,\boldsymbol{w}_l\}_{l=1}^{L}$ the optimal $P_{trn}$ can be found via solving the following problem:
%
\vspace{-.3cm}
{\blue
	\leqnomode
\begin{align} \label{max problem to obtain P_trn}
%&\text{given}\ \{P_l, \boldsymbol{w}_l\}_{l=1}^L,\ 
~~~~~~&\mathop{\text{max}}_{P_{trn}, \{{P}_l, {\boldsymbol{w}}_l\}_{l=1}^L}\ \ \sum_{l=1}^{L}{\cal G}_l(P_{trn}, {P}_l, {\boldsymbol{w}}_l)\tag{P$_{B'}$}\\
%&\text{s.t.}\ \ P_{trn}\leq (1-\sigma)P_{tot},\ P_{trn}\in\mathbb{R}^{+}.\nonumber
\text{s.t.}\ \ &P_{trn}\!+\!\sum_{l=1}^{L}\!C_l(\!{P}_l,{\boldsymbol{w}}_l\!)\!\leq \!P_{tot}, P_{trn},{P}_l\!\in\! \mathbb{R}^{+}\!,{\boldsymbol{w}}_l\!\in\! \mathbb{R}^{K_l}\!, \forall l.\nonumber
\vspace{-.2cm}
\end{align}
\reqnomode
{\blue Examining (\ref{max problem to obtain P_trn}), we realize that solving it for $P_{trn}$ provides an answer that depends on  ${P}_l, {\boldsymbol{w}}_l$ (which is undesirable). To circumvent this problem we propose a method to find $P_{trn}$ based on the following observation. 
We observe that, although (\ref{max problem to obtain P_trn}) is a non-concave maximization problem,  given $\{{P}_l, {\boldsymbol{w}}_l\}_{l=1}^L$ and letting $\sigma=1-\frac{P_{trn}}{P_{tot}}$, 
the problem (\ref{max problem to obtain P_trn}) under these conditions becomes 
strictly concave with respect to the variable $\sigma$ over the interval $(0,1)$, and hence, the objective function has a unique global maximum in this interval. Let ${\sigma}^{opt}$ denote the solution to this problem, which can be efficiently found using numerical line search methods (e.g., the Golden section{\blue \footnote{{\blue  Let $x^{opt}$ denote the maximum value that a concave function $f(x)$  attains over a search interval $x\!\in\!(x_b,x_e)$. This numerical method finds $x^{opt}$ via successively narrowing the range of the search interval. 
Let $i$ be the iteration index, $({\cal I}_b^{(i)}, {\cal I}_e^{(i)})$ be the starting and ending points of the search interval at iteration $i$, $\alpha_b^{(i)}\!=\!0.382({\cal I}_e^{(i)}-{\cal I}_b^{(i)})+ {\cal I}_b^{(i)}$ and $\alpha_e^{(i)}\!=\!0.618({\cal I}_e^{(i)}-{\cal I}_b^{(i)})+ {\cal I}_b^{(i)}$ be the evaluating points. Also let $f_b^{(i)}, f_e^{(i)}$, denote the values of the function $f(x)$ when it is evaluated at the evaluating points $\alpha_b^{(i)}, \alpha_e^{(i)}$, respectively. 
For initialization,  we let $i\!=\!0, {\cal I}_b^{(0)}\!=\!x_b, {\cal I}_e^{(0)}\!=\!x_e$. At iteration $i$, we compute $f_b^{(i)}$ and $f_e^{(i)}$ and then update the search interval to find $x^{opt}$ as the following: 
if $f_b^{(i)}\!>\!f_e^{(i)}$, then ${\cal I}_b^{(i+1)}\!=\!{\cal I}_b^{(i)}, {\cal I}_e^{(i+1)}\!=\!\alpha_e^{(i)}$, if $f_b^{(i)}\!=\!f_e^{(i)}$, then ${\cal I}_b^{(i+1)}\!=\!\alpha_b^{(i)}, {\cal I}_e^{(i+1)}\!=\!\alpha_e^{(i)}$, and if $f_b^{(i)}\!<\!f_e^{(i)}$, then ${\cal I}_b^{(i+1)}\!=\!\alpha_b^{(i)}, {\cal I}_e^{(i+1)}\!=\!{\cal I}_e^{(i)}$. As the stopping criterion, we check whether the length of the search interval exceeds a pre-determined threshold $\epsilon$. If the stopping criterion is met at iteration $j$, the algorithm returns the optimal solution $x^{opt}\!=\!{\cal I}_b^{(j)}$. Otherwise, we update the search interval and continue the iterations until the stopping criterion is met.}}} method \cite[p. 216]{lin_and_nonlin_prog})}. 
%In the absence of a closed-form solution, we resort to numerical line search methods to find ${\sigma}^{opt}$. 
%
Since this problem is concave over  $(0,1)$, the convergence of Golden section method to ${\sigma}^{opt}$ is guaranteed.

Based on the above observation, we propose the method described in Algorithm \ref{solution to find P_trn} to solve (\ref{max problem to obtain P_trn}) and find $P_{trn}^{opt}$. The proposed method is basically Golden section method, where in each iteration we apply Algorithm \ref{solution to problem P1} to find $\{{P}_l, {\boldsymbol{w}}_l\}_{l=1}^L$, only for the purpose of successively narrowing the search interval for $\sigma$. The output of Algorithm \ref{solution to find P_trn} converges to ${\sigma}^{opt}$ and thus $P_{trn}^{opt}=(1-{\sigma}^{opt}) P_{tot}$.}
\begin{algorithm}[h!]
{\blue
\caption{proposed solution of \eqref{max problem to obtain P_trn}}
\label{solution to find P_trn}
 {\bf Input:} $P_{tot}, \epsilon$, system parameters defined in Section \ref{System Model and Problem Formulation}
 {\bf Output:} optimal optimization variable ${\sigma}^{opt}$
  \vspace{-0.1cm}
  \hbox to \hsize{\dashfill\hfil}
% % % % % % % % % % % %
 \vspace{-0.1cm}
Apply the iterative Golden section method to find ${\sigma}^{opt}\!\in\!(0,1)$ \\
- Initialization: $i\!=\!0, {\cal \sigma}_b^{(0)}\!=\!0, {\cal \sigma}_e^{(0)}\!=\!1$.\\
- At iteration $i$ of Golden section method, do below:\\
$\ \ \!$1: Compute two evaluating points $\alpha_b^{(i)}$ and $\alpha_e^{(i)}$ using the starting and the ending points of the search interval $({\cal \sigma}_b^{(i)}, {\cal \sigma}_e^{(i)})$.\\
$\ \ \!$2: For each evaluating point, use Algorithm \ref{solution to problem P1} to obtain $\{P_l,\boldsymbol{w}_l\}_{l=1}^L$ and compute the objective function $\sum_{l=1}^{L}{\cal G}_l$. Suppose ${\cal G}_b^{(i)},{\cal G}_e^{(i)}$ denote the values of $\sum_{l=1}^{L}{\cal G}_l$ when it is evaluated at  $\alpha_b^{(i)}$ and $\alpha_e^{(i)}$, respectively.\\
 % % % % % % % % % % % %
- Depending on the values of ${\cal G}_b^{(i)},{\cal G}_e^{(i)}$ update the search interval $({\cal \sigma}_b^{(i)},{\cal \sigma}_e^{(i)})$.

- Continue the iteration until ${\cal \sigma}_e^{(i)}-{\cal \sigma}_b^{(i)}\leq \epsilon$. 
 % \hbox to \hsize{\dashfill\hfil}
%Return ${\sigma}^{opt}$.
}
 \end{algorithm}
%
%\vspace{-.3cm}
%\end{comment}
% -------------------------------------------------------------
%\subsubsection{Finding $\{\psi_l\}_{l=1}^L$} \label{finding psi_l}
%According to \eqref{channel estimate and channel estimation error variance}, the problem of minimizing the average of $\zeta_l^2$ across clusters, under the training power constraint $\sum_{l=1}^{L}\psi_l\leq P_{trn}$ can be formulated as:
Given  $P_{trn}^{opt}$, 
%obtained from solving \eqref{max problem to find sigma}, 
%
we find $\{\psi_l\}_{l=1}^L$, via minimizing the MSE of the LMMSE channel estimates for all clusters:
\vspace{-.1cm}
\begin{align} \label{min problem in training phase}
&\text{given}\ P_{trn}^{opt}~~~~\mathop{\text{min}}_{\{\psi_l\}_{l=1}^{L}}\ \ \ \sum_{l=1}^{L}\zeta_l^2\\
&\text{s.t.}\ \ \ \ \sum_{l=1}^{L}\psi_l\leq P_{trn}^{opt},\ \psi_l\in \mathbb{R}^{+},\ \forall l\nonumber.
%\vspace{-.15cm}
\end{align}
{\blue The above is a convex minimization problem.} Solving the associated KKT conditions, we obtain:
\vspace{-.1cm}
%\begin{subequations}\label{final formulas for channel estimation}
\begin{equation} \label{final formulas for channel estimation}
\psi_l=\big{[}\frac{\sigma_{v_{l}}^2}{\sigma_{h_{l}}^2}(\frac{\sigma_{h_{l}}^2}{\kappa\sigma_{v_{l}}}-1)\big{]}^{+},\ \ \kappa=\frac{\sum_{l=1}^L\sigma_{v_{l}}}{P_{trn}^{opt}+\sum_{l=1}^L\frac{\sigma_{v_{l}}^2}{\sigma_{h_{l}}^2}}.
%\hat{h}_l&=&\frac{\kappa\sqrt{\psi_l}\hat{z}_l}{\sigma_{v_{l}}},\label{final formula h_hat_l}\\
%\zeta_l^2&=&2\sigma_{v_{l}}\kappa,\label{final formula zeta_sq_l}
%\vspace{-.15cm}
\end{equation}
%\end{subequations}
%
%where 
%
%\begin{equation} \label{kappa}
%\kappa=\frac{\sum_{l=1}^{S}\sigma_{v_{l}}}{P_{trn}+\sum_{l=1}^{S}\frac{\sigma_{v_{l}}^2}{\sigma_{h_{l}}^2}}, 
%\end{equation}
%
%in which ${\cal T}\!=\!\{l: \psi_l\!>\!0, l=1,...,L\}$. This implies that in this approach, according to the value of $P_{trn}$, only a subset of the clusters might get active at each training period which is not desirable. In particular, \eqref{final formula psi_l} infers $\psi_l\!=\!\frac{\sigma_{v_{l}}^2}{\sigma_{h_{l}}^2}(\frac{\sigma_{h_{l}}^2}{\kappa\sigma_{v_{l}}}-1)$, and thus, if 
%Suppose the clusters are sorted such that $\frac{\sigma_{h_{1}}^2}{\sigma_{v_{1}}}\!\geq\!\frac{\sigma_{h_{2}}^2}{\sigma_{v_{2}}}\!\geq\!...\!\geq\!\frac{\sigma_{h_{L}}^2}{\sigma_{v_{L}}}$. When 
The solution in \eqref{final formulas for channel estimation} is based on the assumption that all CHs participate in pilot transmission and $P_{trn}^{opt}$ satisfies the inequality  $P_{trn}^{opt}\!\geq\!\frac{\sigma_{v_{L}}}{\sigma_{h_{L}}^2}\sum_{l=1}^{L}\sigma_{v_{l}}\!-\!\sum_{l=1}^{L}\frac{\sigma_{v_{l}}^2}{\sigma_{h_{l}}^2}\!=\!\Upsilon$. However, when $P_{trn}^{opt}\!<\!\Upsilon$, the solutions in \eqref{final formulas for channel estimation} imply that $\psi_l\!=\!0$ for some clusters. 
%To ensure that all CHs participate in pilot transmission, regardless of $\Upsilon$ value, 
In this case, we propose to choose $\psi_l\!=\!a\frac{\sigma_{h_{l}}^2}{\sigma_{v_{l}}}$, in which $a$ is a common factor. Imposing the constraint $\sum_{l=1}^{L}\psi_l\!=\!P_{trn}^{opt}$ results in:
\vspace{-.1cm}
\begin{equation} \label{proposed psi_l}
\psi_l\!=\!\frac{\sigma_{h_{l}}^2P_{trn}^{opt}}{\sigma_{v_{l}}\sum_{l=1}^{L}\frac{\sigma_{h_{l}}^2}{\sigma_{v_{l}}}},\ l=1, ..., L, \text{when}\ P_{trn}^{opt}<\Upsilon. 
%\vspace{-.15cm}
\end{equation}
\begin{figure*}[t]
	\centering
	\hspace*{-.4cm}
	\includegraphics[width=7.1in,height=3.1in]{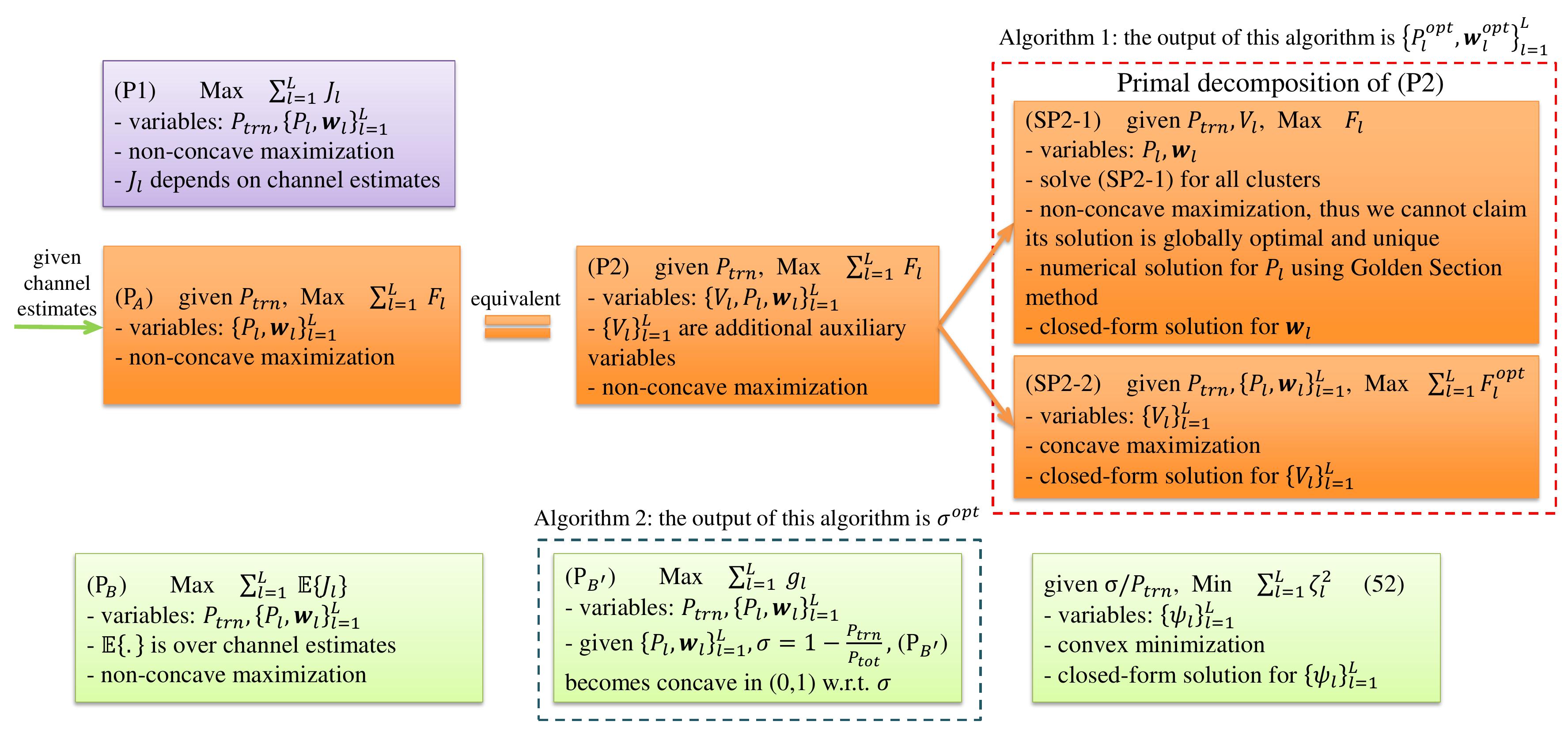}
	\vspace{-0.1cm}
	\caption{\blue 
	This block diagram is the pictorial narrative of our  approach  to  solve  the  original  constrained  optimization  problem  \eqref{max problem to minimize D}.}
	\label{diagram for P1)}
	\vspace{-.5cm}
\end{figure*}
{\blue Fig.~\ref{diagram for P1)} shows a block diagram that summarizes  our  approach  to  solve  the  original  constrained  optimization  problem  \eqref{max problem to minimize D}.} 
{\blue  Overall,  the  sequence  of  algorithm  implementations
and  network  operation  follow. 
The FC implements Algorithm \ref{solution to find P_trn} to obtain $P_{trn}^{opt}$, and consequently to find $\{\psi_l\}_{l=1}^L$ given in (\ref{final formulas for channel estimation}). The FC feeds back this information to CHs (all the obtained $\{{P}_l, {\boldsymbol{w}}_l\}_{l=1}^L$ values during the execution of Algorithm \ref{solution to find P_trn} are discarded at this point).
CHs send their pilot symbols to the FC and the FC estimates the channels $\{\hat{h}_l\}_{l=1}^L$. Now, given $P_{trn}^{opt}, \{\hat{h}_l\}_{l=1}^L$, the FC implements Algorithm \ref{solution to problem P1}, finds $\{P_l^{opt}, \boldsymbol{w}_l^{opt}\}_{l=1}^L$, feeds back{\blue \footnote{{\blue Similar to  \cite{Cihan_TSP_2008,Wu_Elsevier_2011} we assume that the FC energy resource is much larger than those of the sensors/CHs. Therefore, the overhead required for feeding back the necessary information from the FC to the sensors/CHs is neglected.}}}
this new information to CHs, and feeds back $P_{l,k}=\frac{P_l^{opt}}{K_l}$ to sensors. Sensors send their amplified measurements to their CHs. CHs send their fused signals to the FC.  Finally, the FC estimates $\theta$.}
%%%%%%%%%%%%%%%%%%%%%%%%%%%%%%%%%%%%%
\vspace{-.3cm}
\subsection{Minimizing Lower Bounds on MSE $D$} \label{min D1 and D2}
\vspace{-.05cm}
This section discusses constrained minimization of the lower bounds $D_1,D_2$ we derived in Section \ref{3 lower bounds}. The lower bound $D_1$ depends on $\{P_l, \boldsymbol{w}_l\}_{l=1}^L$ and hence its constrained minimization becomes:
\leqnomode
{\small
\vspace{-.2cm}
\begin{align} \label{max problem to minimize D1}
\mathop{\text{max}}_{\{P_l, \boldsymbol{w}_l\}_{l=1}^L}\ \ \sum_{l=1}^{L}\frac{P_l{|h_l|}^2{\boldsymbol{w}_l}^T\boldsymbol{\Pi}_l\boldsymbol{w}_l}{\sigma_{v_{l}}^2+{|h_l|}^2{\boldsymbol{w}_l}^T(\boldsymbol{\Sigma}_{q_l}\!+\!P_l\boldsymbol{\Delta}_l)\boldsymbol{w}_l}&\tag{P3}\\
\!\!\text{s.t.}\ \sum_{l=1}^{L}\!{\boldsymbol{w}_l}^T\!\boldsymbol{\Sigma}_{q_l}\boldsymbol{w}_l\!+\!P_l(\!1\!+\!{\boldsymbol{w}_l}^T\!\boldsymbol{\Omega}_l\boldsymbol{w}_l\!)\!\leq\! P_{tot},P_l\!\in\! \mathbb{R}^{+}\!,\boldsymbol{w}_l&\!\in\! \mathbb{R}^{K_l}\!,\forall l.\nonumber
\vspace{-.1cm}
\end{align}
}
\reqnomode
%A similar procedure described in algorithm 2 can be followed to solve \eqref{max problem to minimize D1}. Particularly at iteration $i$, given ${\cal V}_l^{(i)}$, we compute $P_l^{(i)}$ and $\boldsymbol{w}_l^{(i)}$ from \eqref{eqn to solve to obtain P_l} and \eqref{opt w_l and f_l as a FN of P_l}, respectively, in which ${|\hat{h}_l|}^2\!+\zeta_l^2$ is substituted by ${|h_l|}^2$. Then we update ${\cal A}$ and find $\{{\cal V}_l^{(i+1)}\}_{\forall l\in{\cal A}}$ according to section \ref{Solving P2-2} where:
%
%\begin{align} 
%{\cal V}_l^{opt}&=\left[\beta_l^{'}(\frac{|h_l|}{\sigma_{v_{l}}}\sqrt{\frac{P_l\tau_l}{\lambda}}\!-\!1)\right]^{+}\!+\!P_l,\ \beta_l^{'}\!=\!\frac{\sigma_{v_{l}}^2}{{|h_l|}^2(1-\sigma^2_{\theta}P_l\tau_l)},\nonumber\\ 
%\lambda&=(\frac{\sum_{l\in{\cal A}}\frac{|h_l|\beta_l^{'}\sqrt{P_l\tau_l}}{\sigma_{v_{l}}}}{P_{tot}-\sum_{l\in{\cal A}}P_l+\sum_{l\in{\cal A}}\beta_l^{'}})^2.\nonumber
%\end{align}
%
This is similar to \eqref{max problem to find P_l,w_l,V_l}, with the difference that $P_{trn}\!=\!0$, and hence in \eqref{opt_delta_l} and \eqref{opt lambda} expressions we let $\zeta_l^2\!=\!0, {|\hat{h}_l|}^2\!=\!{|h_l|}^2, \sigma\!=\!1$. Algorithm \ref{solution to problem P1} can be followed to find the solution to \eqref{max problem to minimize D1}, using $\boldsymbol{w}_l^{opt}$ in \eqref{opt w_l and f_l as a FN of P_l}.
%The problem of minimizing $D_2$ subject to total network power constraint is:
The lower bound $D_2$ depends on $\{\boldsymbol{w}_l\}_{l=1}^{L}$ and hence its constrained minimization becomes:
\leqnomode
{\small
\vspace{-.2cm}
\begin{align} \label{max problem to minimize D2}
~~~~~~\mathop{\text{max}}_{\{\boldsymbol{w}_l\}_{l=1}^{L}}\ \ &\sum_{l=1}^{L}\frac{{|h_l|}^2{\boldsymbol{w}_l}^T\boldsymbol{\Sigma}_l\boldsymbol{w}_l}{\sigma_{v_{l}}^2+{|h_l|}^2{\boldsymbol{w}_l}^T\boldsymbol{\Sigma}_{n_l}\boldsymbol{w}_l}\tag{P4}\\
\text{s.t.}\ \ \ &\sum_{l=1}^{L}\!{\boldsymbol{w}_l}^T(\sigma^2_{\theta}\boldsymbol{\Sigma}_l\!+\!\boldsymbol{\Sigma}_{n_l})\boldsymbol{w}_l\!\leq\! P_{tot},\boldsymbol{w}_l\!\in\! \mathbb{R}^{K_l}\!,\forall l.\nonumber
\vspace{-.1cm}
\end{align}
}
\reqnomode
%To solve \eqref{max problem to minimize D2}, we first obtain the optimal ${\cal A}$ and $\{{\cal V}_l\}_{\forall l\in{\cal A}}$ as follows:
This is similar to \eqref{max problem to find P_l,w_l,V_l}, with the differences that $P_{trn}\!=\!0$ and $P_l\!=\!0, \forall l$. Following similar steps we took in Section \ref{find opt P_l and w_l} to solve \eqref{max problem to find P_l,w_l,V_l}, we find that \eqref{opt_delta_l} and \eqref{opt lambda} become:
\vspace{-.2cm}
\begin{align} 
{\cal V}_l^{opt}&\!=\!\big{[}\beta_l^{''}(\frac{|h_l|}{\sigma_{v_{l}}}\sqrt{\frac{\tau_l^{'}}{\lambda}}\!-\!1)\big{]}^{+}\!,\ \ \beta_l^{''}\!=\!\frac{\sigma_{v_{l}}^2}{{|h_l|}^2(1\!-\!\sigma^2_{\theta}\tau_l^{'})},\nonumber\\ 
\lambda&\!=\!(\frac{\sum_{l\in{\cal A}}\frac{|h_l|\beta_l^{''}\sqrt{\tau_l^{'}}}{\sigma_{v_{l}}}}{P_{tot}\!+\!\sum_{l\in{\cal A}}\beta_l^{''}})^2,\ \ \tau_l^{'}\!=\!{\boldsymbol{1}_l}^T{(\sigma^2_{\theta}\boldsymbol{\Sigma}_l\!+\!\boldsymbol{\Sigma}_{n_l})}^{-1}\boldsymbol{1}_l.\nonumber 
\end{align}
%\vspace{-.15cm}
%
The optimal weight vector $\boldsymbol{w}_l^{opt}$ corresponding to the solution of \eqref{max problem to minimize D2} is computed as ${\boldsymbol{w}}^{opt}_l\!=\!\sqrt{\frac{{\cal V}_l}{\tau_l^{'}}}{(\sigma^2_{\theta}\boldsymbol{\Sigma}_l\!+\!\boldsymbol{\Sigma}_{n_l})}^{-1}\boldsymbol{1}_l$.
%
%\begin{equation*}
%{\boldsymbol{w}}^{opt}_l=\sqrt{\frac{{\cal V}_l}{\tau_l^{'}}}{\boldsymbol{R}_{t_l}}^{-1}\boldsymbol{1}_l.
%\end{equation*}
%%%%%%%%%%%%%%%%%%%%%%%%%%%%%%%
\vspace{-.2cm}
\section{{\blue Solving the Special Cases of the Original Problem}} \label{Special Cases of P1}
\vspace{-.1cm}
The original problem \eqref{max problem to minimize D} aims at constrained minimization of $D$, with respect to three sets of optimization variables: $P_{trn}$ total training power, $P_l$ power allocated to sensors in cluster $l$ to send their measurements to CH$_l$, and ${\cal P}_l$ power allocated to CH$_l$ to transmit its signal to the FC. To untangle the performance gain that optimizing each set of these optimization variables provides, we consider the following three special cases of \eqref{max problem to minimize D}. In problem (P1-SC1) assuming $P_{trn}$ is given and $\psi_l\!=\!P_{trn}/L$, we optimize $\{P_l, {\cal P}_l\}_{l=1}^L$. In problem \eqref{max problem to minimize D_c} assuming $P_l\!=\!P, \forall l$, we optimize $P_{trn}, P, \{{\mathcal P}_l\}_{l=1}^L$. In problem \eqref{max problem to minimize D_d} assuming ${\mathcal P}_l\!=\!{\mathcal P}, \forall l$, we optimize $P_{trn},{\mathcal P},\{P_l\}_{l=1}^L$. Note that problem (P1-SC1) is the same as problem \eqref{max problem to find P_l,w_l all l} addressed in Section \ref{find opt P_l and w_l}. In the following we address problems \eqref{max problem to minimize D_c} and \eqref{max problem to minimize D_d}. 
\vspace{-.35cm}
\subsection{{\blue Solving Special Case \eqref{max problem to minimize D_c}: When Intra-Cluster Powers of all Clusters are Equal}} \label{solving P1-SC2}
\vspace{-.05cm}
Problem \eqref{max problem to minimize D_c} becomes: 
\leqnomode
\vspace{-.2cm}
\begin{align} \label{max problem to minimize D_c}
\mathop{\text{max}}_{P_{trn},P,\{\!\boldsymbol{w}_l\!\}_{l=1}^{L}}\ \ \sum_{l=1}^{L}\frac{P{|\hat{h}_l|}^2{\boldsymbol{w}_l}^T\boldsymbol{\Pi}_l\boldsymbol{w}_l}{\sigma_{v_{l}}^2+{\boldsymbol{w}_l}^T\boldsymbol{\Lambda}_{1_l}\boldsymbol{w}_l}&\tag{P1-SC2}\\
\!\!\!\!\!\!\text{s.t.}\ P_{trn}\!+\!\!\sum_{l=1}^{L}\!{\boldsymbol{w}_l}^T\boldsymbol{\Sigma}_{q_l}\boldsymbol{w}_l\!+\!P(\!1\!+\!{\boldsymbol{w}_l}^T\boldsymbol{\Omega}_l\boldsymbol{w}_l\!)\!\leq\!P_{tot}, &P_{trn}, P\!\in\!\mathbb{R}^{+}\!\!,\nonumber
\vspace{-.15cm}
\end{align}
\reqnomode
where $\boldsymbol{\Lambda}_{1_l}\!=\!\sigma^2_{\theta}\zeta_l^2P\boldsymbol{\Pi}_l+({|\hat{h}_l|}^2+\zeta_l^2)(\boldsymbol{\Sigma}_{q_l}+P\boldsymbol{\Delta}_l)$. 
%To solve \eqref{max problem to minimize D_c}, we take a similar approach as we did to solve \eqref{max problem to minimize D}. We need to deal with three sub-problems:
To address \eqref{max problem to minimize D_c} we consider the following two sub-problems: 
($a$) finding $P^{*}, \{\boldsymbol{w}_l^{*}\}_{l=1}^L$ given $P_{trn}$, ($b$) finding $P_{trn}^{*}$ as well as $\{\psi_l^{*}\}_{l=1}^L$ such that $\sum_{l=1}^L \psi_l^{*}\!=\!P_{trn}^{*}$. Sub-problem ($a$) is a special case of \eqref{max problem to find P_l,w_l all l} in which, for finding $P^{*}$, we use Golden section method, and sub-problem ($b$) is similar to (P$_B$). Recall that $P_{trn}\!=\!(1-\sigma)P_{tot}$ and thus $\sum_{l=1}^{L}(P\!+\!{\cal P}_l)\!=\!\sigma P_{tot}$. We let $\sigma_c\in(0,1)$ such that $P\!=\!(1-\sigma_c)\sigma P_{tot}$. It is simple to show that 
%finding $P^{*}, P_{trn}^{*}$ in sub-problems ($a$) and ($b$), respectively, are both concave problems. This proves that 
sub-problems ($a$) and ($b$) are both concave and hence $P^{*}$ and $P_{trn}^{*}$ are unique. 
Next, we summarize our proposed solutions for solving sub-problems ($a$) and ($b$) in Algorithms 3-a and 3-b, respectively.

{\bf Description of Algorithm 3-a}: Let $P^{*}\!=\!(1-\sigma_c^{*})\sigma P_{tot}$ denote the optimal $P$. We apply Golden section method to find $\sigma_c^{*}\in(0,1)$ and thus $P^{*}$ that maximizes the objective function in \eqref{max problem to minimize D_c}, denoted as ${\cal F}(\sigma_c)$. At iteration $i$, for each evaluating point we first compute the optimal ${\cal V}_l^{(i)}$, denoted as $\{\bar{{\cal V}}_l^{(i)}\}_{l=1}^L$ using \eqref{opt_delta_l}, and substitute $\bar{{\cal V}}_l^{(i)}$ into \eqref{opt w_l and f_l as a FN of P_l} to obtain $\{\bar{\boldsymbol{w}}_l^{(i)}\}_{l=1}^L$. Next we compute ${\cal F}_{b}^{(i)}$ and ${\cal F}_{e}^{(i)}$. The stopping criterion is similar to Algorithm \ref{solution to find P_trn}. Algorithm 3-a returns the optimal $\sigma_c^{*}, \{\boldsymbol{w}_l^{*}\}_{l=1}^L$.

{\bf Description of Algorithm 3-b}: We address sub-problem 
%\eqref{max problem to minimize D_c}-$(b)$ using an approach similar to the one we took in Section \ref{find opt P_trn and psi_l} to solve (P$_B$).
$(b)$ similar to problem (P$_B$) in Section \ref{find opt P_trn and psi_l}. 
More specifically, we consider problem \eqref{max problem to obtain P_trn}, where $P_l$ is substituted by $P$, 
%\leqnomode
%
%\begin{align} \label{max problem to obtain P_trn^{*}}
%~~~~~~\mathop{\text{max}}_{P_{trn},P,\{\boldsymbol{w}_l\}_{l=1}^L}\ \ &\sum_{l=1}^{L}{\cal G}_l(P_{trn},P,\boldsymbol{w}_l)\\
%\text{s.t.}\ \ \ \ \ \ \ &P_{trn}+\sum_{l=1}^{L}C_l(P,\boldsymbol{w}_l)\leq P_{tot},\nonumber\\
%&P_{trn}\in\mathbb{R}^{+},P\in \mathbb{R}^{+},\boldsymbol{w}_l\in \mathbb{R}^{K_l},\ \forall l.\nonumber
%\end{align}
% 
%\reqnomode
and apply a modified version of Algorithm \ref{solution to find P_trn} to solve it.  
In particular, at iteration $i$ of Algorithm \ref{solution to find P_trn}, we use Algorithm 3-a to obtain the optimal variables $\bar{P}^{(i)}, \{\bar{\boldsymbol{w}}_l^{(i)}\}_{l=1}^L$, and then compute ${\cal R}_b^{(i)}$ and ${\cal R}_e^{(i)}$. The rest is similar to Algorithm \ref{solution to find P_trn}. 
Algorithm 3-b returns the optimal $P_{trn}^{*}, \{\psi_l^{*}\}_{l=1}^L$.

%%%%%%%%%%%%%%%%%%%%%%%%%%%
\vspace{-.1cm}
\subsection{{\blue Solving Special Case \eqref{max problem to minimize D_d}: When Powers of all CHs for Their Data Transmission to the FC are Equal}} \label{solving P1-SC3}
%In problem \eqref{max problem to minimize D_d} we assumed ${\mathcal P}_l\!=\!{\mathcal P}$. To impose this constraint in a meaningful way, we note that as discussed in section \ref{Solving P2-1}, $\boldsymbol{w}_L^{opt}$ is exactly 
To incorporate the constraint ${\mathcal P}_l\!=\!{\mathcal P}$ in the cost function of problem \eqref{max problem to minimize D_d}, from Section \ref{Solving P2-1} we recall that $\boldsymbol{w}_l^{opt}\!=\!\chi_l({\boldsymbol{R}_{t_l}}^{-1}\sigma^2_{\theta}\sqrt{P_l}\boldsymbol{\rho}_l)$. Therefore from ${\mathcal P}_l\!=\!{\boldsymbol{w}_l}^T\boldsymbol{R}_{t_l}\boldsymbol{w}_l$ in \eqref{transmission power of CH l} and ${\mathcal P}_l\!=\!{\mathcal P}$, we conclude $\chi_l^2\!=\!{\mathcal P}/{\sigma^4_{\theta}P_l\tau_l}$. Substituting for $\boldsymbol{w}_l$ in \eqref{max problem to minimize D}, problem \eqref{max problem to minimize D_d} becomes: 
\leqnomode
\vspace{-0.1cm}
{\small
\begin{align} \label{max problem to minimize D_d}
~~~~~\mathop{\text{max}}_{P_{trn},{\mathcal P}\!,\{\!P_l\!\}_{l=1}^{L}} \sum_{l=1}^{L}\!\!\frac{P_l{|\hat{h}_l|}^2\tau_l}{\frac{\sigma_{v_{l}}^2}{{\mathcal P}}\!+\!\zeta_l^2\!+\!\frac{{|\hat{h}_l|}^2}{\tau_l}{\boldsymbol{\rho}_l}^T\!{\boldsymbol{R}_{t_l}^{-1}}\boldsymbol{\Sigma}_{q_l}\!{\boldsymbol{R}_{t_l}^{-1}}\!\boldsymbol{\rho}_l}&\tag{\small P1-SC3}\\
\text{s.t.}\ \ P_{trn}\!+\!\sum_{l=1}^{L}(P_l\!+\!{\mathcal P})\leq P_{tot},P_{trn}, {\mathcal P}\in \mathbb{R}^{+},P_l&\in \mathbb{R}^{+},\forall l.\nonumber
\end{align}
}
\reqnomode
To address \eqref{max problem to minimize D_d} we consider the following two sub-problems: 
($a$) finding ${\cal P}^{*}, \{P_l^{*}\}_{l=1}^L$ given $P_{trn}$, ($b$) finding $P_{trn}^{*}$ as well as $\{\psi_l^{*}\}_{l=1}^L$ such that $\sum_{l=1}^L \psi_l^{*}\!=\!P_{trn}^{*}$. Sub-problem ($a$) is a special case of \eqref{max problem to find P_l,w_l all l} in which, for finding ${\cal P}^{*}$, we use Golden section method, and sub-problem ($b$) is similar to (P$_B$). 
We let $\sigma_d\in(0,1)$ such that ${\cal P}\!=\!(1-\sigma_d)\sigma P_{tot}$. It is easy to show that finding ${\cal P}^{*}, P_{trn}^{*}$ in sub-problems ($a$) and ($b$), respectively, are concave problems, and hence ${\cal P}^{*}$ and $P_{trn}^{*}$ are unique. In Appendix \ref{prove P13_b is concave}, we prove that finding $\{P_l^{*}\}_{l=1}^L$ in sub-problem ($a$) is jointly concave over $P_l$'s and therefore its solution is unique. In the absence of a closed form expression we use gradient-ascent algorithm to find the solution. 
%In the following, we summarize our proposed solutions for solving sub-problems ($a$) and ($b$) in Algorithm 4-a and 4-b, respectively.
Algorithms 4-a and 4-b summarize how we solve sub-problems ($a$) and ($b$), respectively. 

{\bf Description of Algorithm 4-a}: Let ${\cal P}^{*}\!=\!(1-\sigma_d^{*})\sigma P_{tot}$ denote the optimal ${\cal P}$. We apply Golden section method to find $\sigma_d^{*}\in(0,1)$ and thus ${\cal P}^{*}$ that maximizes the objective function in \eqref{max problem to minimize D_d}, denoted as ${\cal F}(\sigma_d)$. At iteration $i$, for each evaluating point we compute the optimal $P_l^{(i)}$, denoted as $\{\bar{P}_l^{(i)}\}_{l=1}^L$ using gradient-ascent algorithm, and substitute them in \eqref{max problem to minimize D_d} to compute ${\cal F}_{b}^{(i)}$ and ${\cal F}_{e}^{(i)}$. The stopping criterion is similar to Algorithm \ref{solution to find P_trn}. Algorithm 4-a returns the optimal $\sigma_d^{*}, \{P_l^{*}\}_{l=1}^L$.

{\bf Description of Algorithm 4-b}: We address sub-problem 
%\eqref{max problem to minimize D_d}-$(b)$ using an approach similar to the one we took in Section \ref{find opt P_trn and psi_l} to solve (P$_B$). More specifically, we solve the constrained optimization problem in 
$(b)$ similar to problem (P$_B$) in Section \ref{find opt P_trn and psi_l}. Specifically, we consider problem \eqref{max problem to obtain P_trn}, where ${\cal P}_l$ is substituted by ${\cal P}$ and apply a modified version of Algorithm \ref{solution to find P_trn} to solve it. In particular, at iteration $i$ of Algorithm \ref{solution to find P_trn}, we use Algorithm 4-a to obtain the optimal variables $\bar{\cal P}^{(i)}, \{\bar{P}_l^{(i)}\}_{l=1}^L$, and then compute ${\cal R}_b^{(i)}$ and ${\cal R}_e^{(i)}$. The rest is similar to Algorithm \ref{solution to find P_trn}. Algorithm 4-b returns the optimal $P_{trn}^{*}, \{\psi_l^{*}\}_{l=1}^L$.
%%%%%%%%%%%%%%%%%%%%%%%%%%%%%%%%%%%%
%%%%%%%%%%%%%%%%%%%%%%%%%%%%%%%%%%%
\vspace{-.2cm}
\section{Complexity of Algorithms} \label{comp complexity}
\vspace{-.1cm}
We discuss the computational complexity of Golden section method as well as Algorithms 1, 2, 3-a, 3-b, 4-a, 4-b, which allows us to compare the {\red computational} complexity of solving \eqref{max problem to minimize D} versus those of (P1-SC1), \eqref{max problem to minimize D_c}, \eqref{max problem to minimize D_d}.\\
$\bullet$ Golden section method: This method includes a one-dimensional search to find the optimal point. If no matrix inversion is required, its complexity order for convergence to an $\epsilon$-accurate solution is $\bar{\epsilon}$, where $\bar{\epsilon}\!=\!\log(1/\epsilon)$ \cite[p. 217]{lin_and_nonlin_prog}. We use this method for solving \eqref{eqn to solve to obtain P_l}. In each iteration, to compute the left side of \eqref{eqn to solve to obtain P_l} we employ the matrix inversion algorithm in \cite{mtx_inv_alg} to calculate ${\boldsymbol{R}_{t_l}}^{-1}$ with complexity order of ${\cal O}(K_l^{2.37})$. Therefore, the overall complexity order of finding $P_l^{opt}\!\in\!(0,{\cal V}_l)$ becomes ${\cal O}(\bar{\epsilon}K_l^{2.37})$.\\
$\bullet$ Algorithm \ref{solution to problem P1} for solving \eqref{max problem to find P_l,w_l all l}: We switch between solving \eqref{max problem to find P_l,w_l} and \eqref{max problem to find V_l} until the stopping criteria is met. In each iteration, we need to (i) find $\{P_l\}_{l=1}^L$ using Golden section method, with the overall complexity order of ${\cal O}(\bar{\epsilon}\bar{K})$, where $\bar{K}\!=\!\sum_{l=1}^{L}K_l^{2.37}$, and (ii) calculate $\{{\cal V}_l\}_{l=1}^L$ using \eqref{opt_delta_l and lambda}, which needs $\tau_l, \beta_l$ that are found in (i) and hence, the complexity order of finding $\{{\cal V}_l\}_{l=1}^L$ is ${\cal O}(L)$. The overall complexity order of Algorithm \ref{solution to problem P1} becomes ${\cal O}(\bar{\epsilon}(L\!+\!\bar{\epsilon}\bar{K}))$.\\
$\bullet$ Algorithm \ref{solution to find P_trn} for solving {\blue \eqref{max problem to obtain P_trn}}: In each iteration, for each evaluating point we use Algorithm \ref{solution to problem P1} to obtain $\{P_l, \boldsymbol{w}_l\}_{l=1}^L$. Therefore, the overall complexity order of Algorithm \ref{solution to find P_trn} becomes ${\cal O}({\bar{\epsilon}}^2(L\!+\!\bar{\epsilon}\bar{K}))$.\\
$\bullet$ Algorithm 3-a for solving sub-problem ($a$) of \eqref{max problem to minimize D_c}: In each iteration, for each evaluating point computing $\tau_l$ in \eqref{opt_delta_l and lambda}, \eqref{opt w_l and f_l as a FN of P_l} involves the matrix inversion ${\boldsymbol{R}_{t_l}}^{-1}$, and thus, the complexity order of finding $\{{\cal V}_l\}_{l=1}^L$ and then $\{\boldsymbol{w}_l\}_{l=1}^L$ is ${\cal O}(\bar{K})$. Therefore, the overall complexity order of Algorithm 3-a is ${\cal O}(\bar{\epsilon}\bar{K})$.\\
$\bullet$ Algorithm 3-b for solving sub-problem ($b$) of \eqref{max problem to minimize D_c}: In each iteration, for each evaluating point we use Algorithm 3-a to obtain $P, \{\boldsymbol{w}_l\}_{l=1}^L$. Therefore, the overall complexity order of Algorithm 3-b is ${\cal O}({\bar{\epsilon}}^2\bar{K})$.\\
$\bullet$ Algorithm 4-a for solving sub-problem ($a$) of \eqref{max problem to minimize D_d}: Note that the complexity order of the gradient-ascent algorithm to maximize a {\blue general non-smooth} convex function $f(x)$ and converge to an $\epsilon$-accurate solution is {\blue ${\cal O}(1/\epsilon)$}, if no matrix inversion is required for finding $f(x)$ and its gradient $\triangledown f(x)$ \cite[p. 232]{lin_and_nonlin_prog}.
%strongly convex function $f(x)$ and converge to an $\epsilon$-accurate solution is ${\cal O}(\bar{\epsilon})$, if no matrix inversion is required for finding $f(x)$ and its gradient $\triangledown f(x)$ \cite[p. 232]{lin_and_nonlin_prog}. 
In each iteration of Algorithm 4-a, for each evaluating point, since computing the objective function in \eqref{max problem to minimize D_d} and its derivative with respect to $P_l$ involves the matrix inversion ${\boldsymbol{R}_{t_l}}^{-1}$, the complexity order of finding $\{P_l\}_{l=1}^L$ using the gradient-ascent algorithm is {\blue ${\cal O}(\bar{K}/\epsilon)$}. Therefore, the overall complexity order of Algorithm 4-a becomes {\blue ${\cal O}({\bar{\epsilon}}\bar{K}/\epsilon)$}.\\
$\bullet$ Algorithm 4-b for solving sub-problem ($b$) of \eqref{max problem to minimize D_d}: In each iteration, for each evaluating point we use Algorithm 4-a to obtain ${\cal P}, \{P_l\}_{l=1}^L$. Therefore, the overall complexity order of Algorithm 4-b is {\blue ${\cal O}({\bar{\epsilon}}^2\bar{K}/\epsilon)$}. 

To solve \eqref{max problem to minimize D} we need to solve \eqref{max problem to find P_l,w_l all l}, {\blue \eqref{max problem to obtain P_trn}}. Therefore, the complexity order of solving \eqref{max problem to minimize D} is $e_0\!=\!{\cal O}(\bar{\epsilon}(1\!+\!\bar{\epsilon})(L\!+\!\bar{\epsilon}\bar{K}))$. 
To solve (P1-SC1) we need to solve \eqref{max problem to find P_l,w_l all l}. Therefore, the complexity order of solving (P1-SC1) is $e_1\!=\!{\cal O}(\bar{\epsilon}(L\!+\!\bar{\epsilon}\bar{K}))$. 
To solve \eqref{max problem to minimize D_c} we need to solve sub-problems ($a$) and ($b$) of \eqref{max problem to minimize D_c}. Therefore, the complexity order of solving \eqref{max problem to minimize D_c} is $e_2\!=\!{\cal O}(\bar{\epsilon}(1\!+\!\bar{\epsilon})\bar{K})$. 
To solve \eqref{max problem to minimize D_d} we need to solve sub-problems ($a$) and ($b$) of \eqref{max problem to minimize D_d}. Therefore, the complexity order of solving \eqref{max problem to minimize D_d} is %$e_3\!=\!{\cal O}({\bar{\epsilon}}^2(1\!+\!\bar{\epsilon})\bar{K})$. 
{\blue $e_3\!=\!{\cal O}({\bar{\epsilon}(1\!+\!\bar{\epsilon})}\bar{K}/\epsilon)$}
It is clear that {\blue $e_1\!<\!e_2\!<\!e_0\!<\!e_3$}.
%=======================================
%=======================================
%=======================================
{\blue
\vspace{-.2cm}
\section{Convergence Analysis} \label{conv of alg to solve P_A}
\vspace{-.1cm}
We discuss the convergence analysis of Algorithms \ref{solution to problem P1} and \ref{solution to find P_trn} which solve problems \eqref{max problem to find P_l,w_l all l} and \eqref{max problem to obtain P_trn}, respectively. 

$\bullet$ {\bf Convergence of Algorithm \ref{solution to problem P1}}: Problems \eqref{max problem to find P_l,w_l all l} and \eqref{max problem to find P_l,w_l,V_l} are equivalent. In  \eqref{max problem to find P_l,w_l,V_l}, the cost function is non-concave and the constraint is a closed convex set w.r.t. the optimization variables $\{{\cal V}_l, P_l, \boldsymbol{w}_l\}_{l=1}^L$. Algorithm \ref{solution to problem P1} is indeed a block-coordinate ascent type algorithm with two blocks. 
The first block solves  \eqref{max problem to find P_l,w_l} for all clusters to obtain $\{P_l, \boldsymbol{w}_l\}_{l=1}^L$.  \eqref{max problem to find P_l,w_l} is a non-concave maximization problem for which we have a numerical solution for $P_l$ using Golden Section method and a closed-form solution for $\boldsymbol{w}_l$. Since \eqref{max problem to find P_l,w_l} is a non-concave maximization problem, we cannot claim that our proposed solution for $P_l, \boldsymbol{w}_l$ is globally optimal and unique. 
The second block solves \eqref{max problem to find V_l} to obtain $\{{\cal V}_l\}_{l=1}^L$.  \eqref{max problem to find V_l} is a concave maximization problem for which we have a closed-form solution for ${\cal V}_l$. Since \eqref{max problem to find V_l} is a concave maximization problem, its solution is globally optimal and unique.

The authors in \cite{Sciandrone_OMS_1999} proved that in a block-coordinate descent algorithm with only two blocks, which solves the unconstrained minimization problem 
\vspace{-.2cm}
\begin{align}\label{min f unconst}
	\mathop{\text{min}}_{(\boldsymbol{x}_1,\boldsymbol{x}_2)\in\mathbb{R}^{n_1}\times\mathbb{R}^{n_2}}\ \ f(\boldsymbol{x}_1,\boldsymbol{x}_2),
	\vspace{-.1cm}
\end{align}
given we have the global minimizer $\boldsymbol{x}_1^{(k+1)}\!\!=\!\underset{\boldsymbol{x}_1}{\text{argmin}}\ f(\boldsymbol{x}_1,\boldsymbol{x}_2^{(k)}),\forall k$, the algorithm converges to a stationary point if we can find a point $\boldsymbol{x}_2^{(k+1)}$ such that $f(\boldsymbol{x}_1^{(k+1)}\!,\boldsymbol{x}_2^{(k+1)})\!\leq\! f(\boldsymbol{x}_1^{(k+1)}\!,\boldsymbol{x}_2^{(k)})$ and $\nabla_2f(\boldsymbol{x}_1^{(k+1)}\!,\boldsymbol{x}_2^{(k+1)})\!=\!0,\forall k$.
%
%\begin{equation*}
%f(\boldsymbol{x}_1^{(k+1)},\boldsymbol{x}_2^{(k+1)})\leq f(\boldsymbol{x}_1^{(k+1)},\boldsymbol{x}_2^{(k)})\ \ \textnormal{and}\ \ \nabla_2\ f(\boldsymbol{x}_1^{(k+1)},\boldsymbol{x}_2^{(k+1)})=0,\ \forall k.
%\end{equation*}
%
%
%Therefore, the uniqueness assumption in Proposition 2.7.1. in \cite{nonlinear_prog_Bertsekas} is not necessary for convergence to a stationary point, \underline{when there are only two blocks}. The authors in \cite{Sciandrone_OMS_1999} furthered their work in \cite{Sciandrone_Elsevier_2000} and proved the convergence when $f$ in \eqref{min f unconst} is minimized subject to a convex constraint set (see Corollary 1 and Section 4 in \cite{Sciandrone_Elsevier_2000}).
%
The authors also proved the convergence when $f$ in \eqref{min f unconst} is minimized subject to a convex constraint set (see Corollary 1 and Section 4 in \cite{Sciandrone_Elsevier_2000}). 
Equipped with this result from \cite{Sciandrone_OMS_1999,Sciandrone_Elsevier_2000}, we return to our own problem. Let $f\!=\!-\sum_{l=1}^L{\cal F}_l,\ \boldsymbol{x}_1\!=\!\{{\cal V}_l\}_{l=1}^L,\ \boldsymbol{x}_2\!=\!\{P_l,\boldsymbol{w}_l\}_{l=1}^L$ in problem \eqref{max problem to find P_l,w_l,V_l}. When solving problem \eqref{max problem to find P_l,w_l,V_l} using the block-coordinate method with two blocks, we note that  \eqref{max problem to find V_l} has a globally optimal solution and thus $\boldsymbol{x}_1^{(k+1)}\!\!=\!\underset{\boldsymbol{x}_1}{\text{argmin}}\ f(\boldsymbol{x}_1,\boldsymbol{x}_2^{(k)})$ is completely known. Also, our proposed solution for  \eqref{max problem to find P_l,w_l} satisfies the condition $\nabla_2\ f(\boldsymbol{x}_1^{(k+1)},\boldsymbol{x}_2^{(k+1)})\!=\!0$ (because it is the solution of KKT conditions for \eqref{max problem to find P_l,w_l}). Furthermore, our extensive simulations indicate that the condition $f(\boldsymbol{x}_1^{(k+1)},\boldsymbol{x}_2^{(k+1)})\leq f(\boldsymbol{x}_1^{(k+1)},\boldsymbol{x}_2^{(k)})$ is always satisfied $\forall k$. Hence, we conclude that the output of the block-coordinate ascent method between  \eqref{max problem to find P_l,w_l} and \eqref{max problem to find V_l} converges to a stationary point. 

Regarding the convergence speed of the block-coordinate descent method, few works have obtained a convergence rate under special conditions on $f$ in (\ref{min f unconst}).
%
%and $\Omega$ in \eqref{min f st const} (e.g.,  see Theorem 6.3 in \cite{Tetruashvili_SIAM_2013}, which proved a sublinear rate of convergence for \eqref{min f st const}, when $f, \Omega_i,\ i=1, ..., m$ are convex and \eqref{min f st const} is solved using the block-coordinate gradient projection method). 
%
However, for the general case of non-convex $f$, even under convex constraints, no convergence rate is established in the literature.
%for the block-coordinate descent method. 
Our extensive simulations indicate that the average number of iterations needed for Algorithm \ref{solution to problem P1} to converge to an $\epsilon$-accurate solution for $\{P_l^{opt},\boldsymbol{w}_l^{opt}\}_{l=1}^L$ is 30. 

$\bullet$ {\bf Convergence of Algorithm \ref{solution to find P_trn}}: 
In this algorithm, we employ Golden section method to obtain $P_{trn}^{opt}$, where in each iteration we apply Algorithm \ref{solution to problem P1} to find $\{{P}_l, {\boldsymbol{w}}_l\}_{l=1}^L$, only for the purpose of successively narrowing the search interval of Golden section method. 
Consider solving the following non-convex minimization problem under convex constraints:
\vspace{-.2cm}
\begin{align}\label{min f(x,y)}
	\mathop{\text{min}}_{x,\boldsymbol{y}}&\ \ f(x,\boldsymbol{y})\nonumber\\
	\text{s.t.}&\ \ x\in\Omega_1\subseteq\mathbb{R},\ \boldsymbol{y}\in\Omega_2\subseteq\mathbb{R}^{n_2},    
	\vspace{-.1cm}
\end{align}
where Golden section method is used to obtain $x^{opt}$.  If $\boldsymbol{y}^{(k)}=\underset{\boldsymbol{y}}{\text{argmin}}\ f(x^{(k)},\boldsymbol{y})$ given $x^{(k)},\ k=0, 1, ...$ is known instantly,  Golden section method converges linearly, and the rate of convergence is approximately 0.62 \cite[p. 217]{lin_and_nonlin_prog}. 
Equipped  with  this result  from  \cite{lin_and_nonlin_prog},  we  return  to  our  own  problem. 
Let $f\!=\!-\sum_{l=1}^L{\cal G}_l,\ x=P_{trn},\ \boldsymbol{y}\!=\!\{P_l,\boldsymbol{w}_l\}_{l=1}^L$ in problem \eqref{max problem to obtain P_trn}. We obtain $x^{opt}$ using Golden section method, where in each iteration Algorithm \ref{solution to problem P1} is applied to obtain $\boldsymbol{y}^{opt}$. Note that in Section \ref{find opt P_trn and psi_l}, we proved that $f$ is strictly convex w.r.t. $x$ and thus, the convergence of Algorithm \ref{solution to find P_trn} to $x^{opt}$ is guaranteed. Since Algorithm \ref{solution to problem P1} is an iterative algorithm with an unknown convergence rate, the exact convergence rate of Algorithm \ref{solution to find P_trn} is unknown. We only know that convergence rate of Algorithm \ref{solution to find P_trn} is less than 0.62. Our extensive simulations indicate that the average number of iterations needed for Algorithm \ref{solution to find P_trn} to converge to an $\epsilon$-accurate solution for $x^{opt}$ is 15.
}
%%%%%%%%%%%%%%%%%%%%%%%%%%%%%%%%%%%%%%%%
%%%%%%%%%%%%%%%%%%%%%%%%%%%%%%%%%%%%
%%%%%%%%%%%%%%%%%%%%%%%%%%%%%%%%%%%%%%%%
\vspace{-.2cm}
\section{Numerical and Simulation Results} \label{simulation}
\vspace{-.1cm}
%We compare the effectiveness of the proposed power allocation schemes in section \ref{comp of diff PA algs}. In Section \ref{Behavior of PA across clusters} we study the behavior of the optimal power allocation across clusters to solve \eqref{original min problem} as $P_{tot}$ increases.
In this section, we corroborate our analytical results with numerical simulations, compare the effectiveness of different proposed power optimization schemes in acheiveing an MSE distorion-power tradeoff which is close to the Bayesian CRB, and investigate how the allocated power across clusters vary as signal-to-noise ratio (SNR) changes. 
%%%%%%%%%%%%%%%%%%%%%%%%%
\vspace{-.2cm}
\subsection{Comparing $D$ and its Lower Bounds}\label{D, D_1, D_2, D_3 vs P_tot}
\vspace{-.1cm}
\begin{figure}[t]
	\centering
	\includegraphics[width=3.3in,height=1.3in]{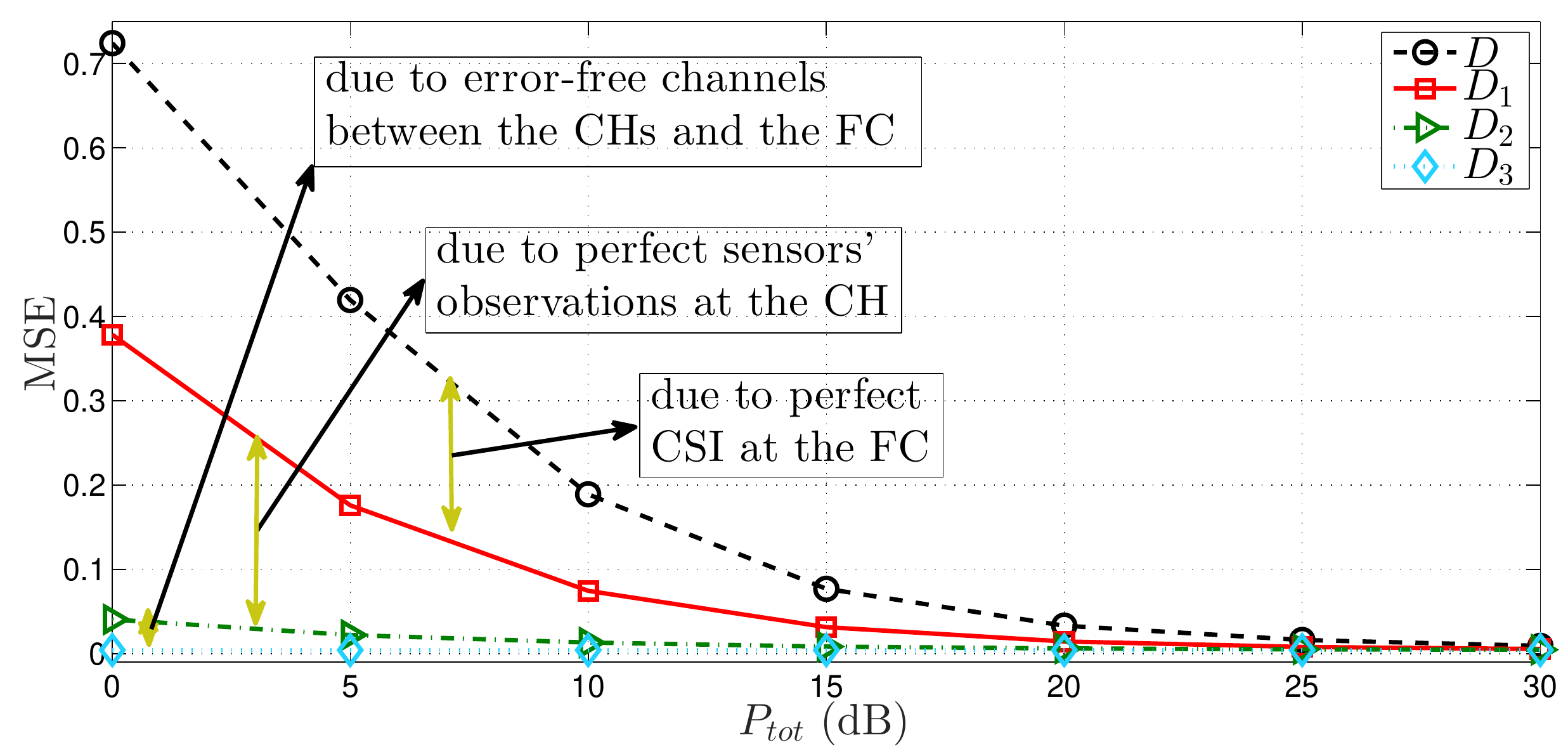}
	%\vspace{-0.4cm}
	\caption{$D, D_1, D_2$, and $D_3$ versus $P_{tot}$ (dB).}
	\label{Diff_D}
	%\vspace{-.45cm}
\end{figure}
Suppose $\theta$ is zero-mean with $\sigma^2_{\theta}\!=\!1$ and $L\!=\!10$ clusters. To enforce the heterogeneity in the network, we randomly choose $\sigma_{h_l},\sigma_{v_l},\sigma_{n_{l,k}},\sigma_{q_{l,k}}\!\in\!(0,1)$, and $K_l\!\in\!\{1,2,...,10\}, l=1,..., L, k=1,..., K_l$. To capture the effect of randomness in flat fading channel coefficients and communication noise, the numerical results are computed based on $10^{6}$ Monte-Carlo trials, where in each trial, one realization of $|h_l|, \nu_l, \forall l$ are generated. {\blue We also assume $\epsilon\!=\!10^{-3}$.}
%Then we save these as fixed parameters for all simulations in this subsection.
In Section \ref{3 lower bounds} we derived three lower bounds on $D$, of which we optimized $D_1, D_2$ in problems \eqref{max problem to minimize D1}, \eqref{max problem to minimize D2}, respectively. Fig.~\ref{Diff_D} plots optimized $D$, optimized $D_1$, optimized $D_2$ versus $P_{tot}$. Note that $D_3\!=\!0.0043$ is constant. Clearly, $D_3\!<\!D_2\!<\!D_1\!<\!D\!<\!\sigma^2_{\theta}$. Also, $D_2, D_1, D$ decrease as $P_{tot}$ increases.
%%%%%%%%%%%%%%%%%%%%%%%%%%%%%%%%
\vspace{-.35cm}
\subsection{Comparing Different Power Allocation Schemes} \label{comp of diff PA algs}
\begin{figure}[t]
	\centering
	\includegraphics[width=3.3in,height=1.3in]{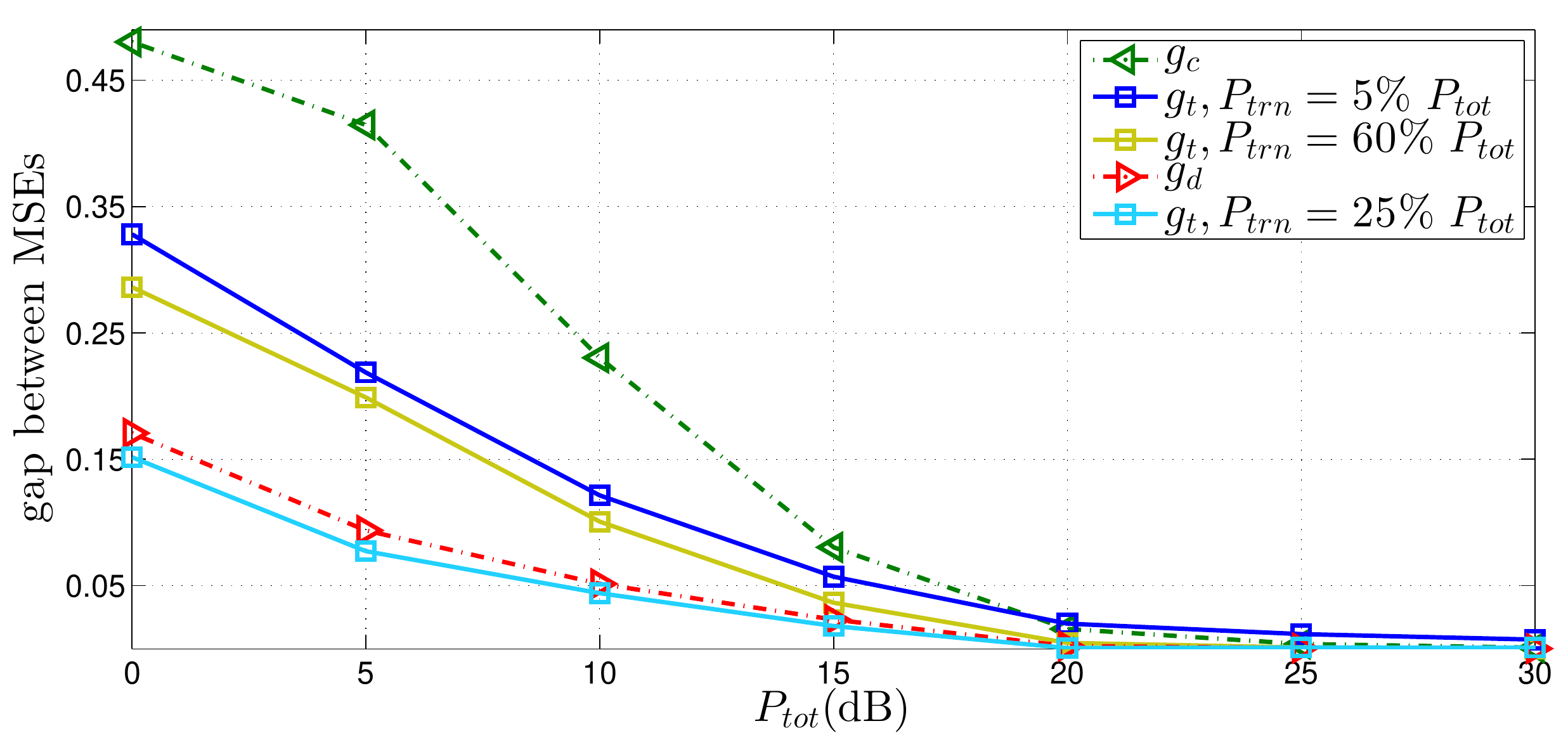}
	%\vspace{-0.4cm}
	\caption{$g_t, g_c, g_d$ versus $P_{tot}$ (dB) {\blue for the first set of system parameters}.}
	\label{gaps}
	%\vspace{-.45cm}
\end{figure}

\begin{figure}[t]
	\centering
	\includegraphics[width=3.3in,height=1.3in]{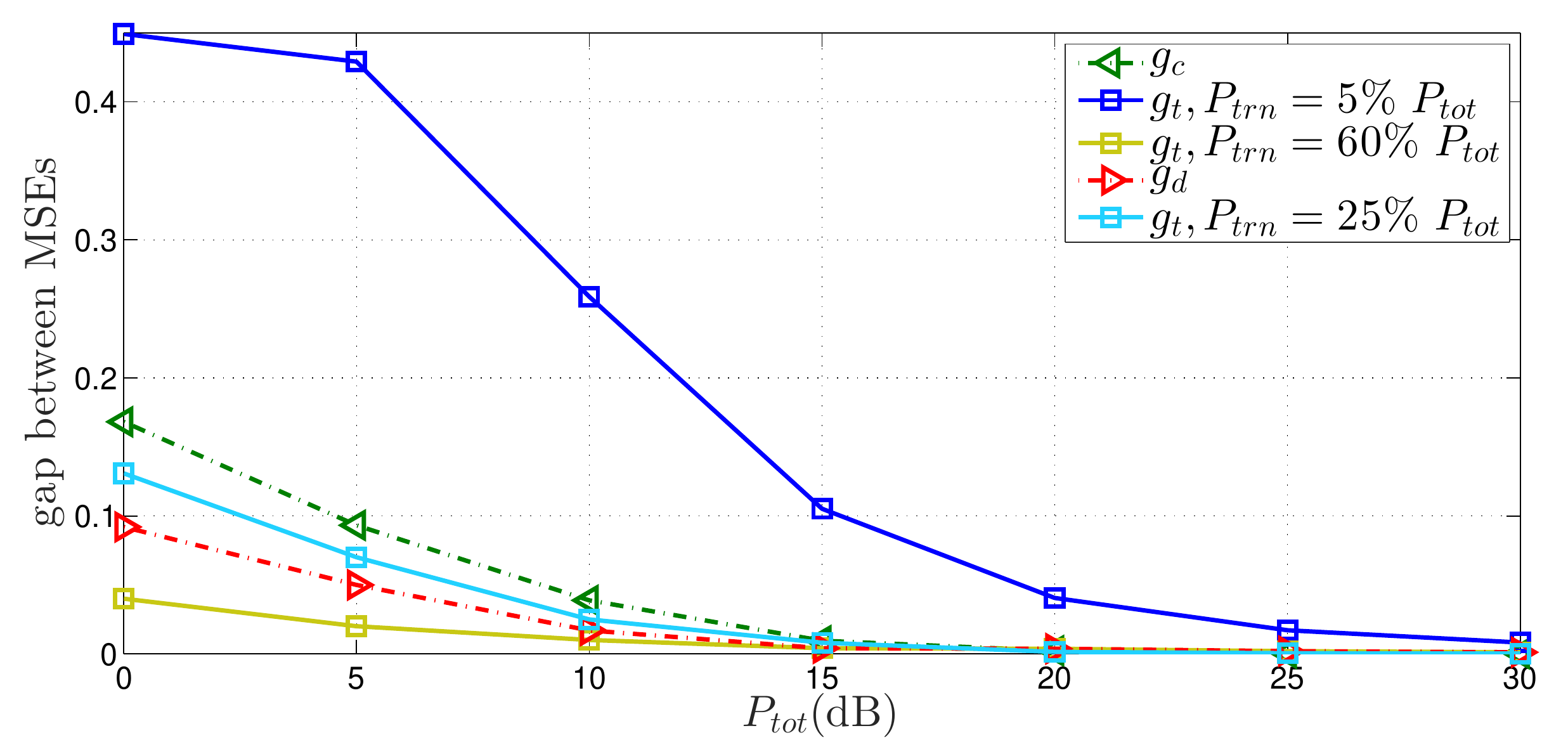}
	%\vspace{-0.4cm}
	\caption{$g_t, g_c, g_d$ versus $P_{tot}$ (dB) {\blue for the second set of system parameters}.}
	\label{gaps new setup parameters}
	%\vspace{-.45cm}
\end{figure}

\vspace{-.1cm}
\begin{comment}
\begin{figure}[t]
	\centering
	\includegraphics[width=3.3in,height=1.3in]{MSEs_and_gaps.eps}
	%\vspace{-0.4cm}
	\caption{$D, D_t, D_c, D_d$ and Bayesian CRB versus $P_{tot}$ (dB). 
	%and a schematic representation of the gains $g_t, g_c, g_d$.
    }
	\label{MSEs_and_gaps}
	%\vspace{-.45cm}
\end{figure}
\end{comment}
%
We compare the effectiveness of power optimization schemes, obtained from solving \eqref{max problem to minimize D} and its special cases (P1-SC1), \eqref{max problem to minimize D_c}, \eqref{max problem to minimize D_d}, in decreasing the MSE of the LMMSE estimator. We also compare the optimized MSE with the Bayesian CRB $G^{-1}$ derived in Section \ref{BCRB}. 
Let $D_t,D_c,D_d$ denote the MSE corresponding to the optimal solutions of (P1-SC1), \eqref{max problem to minimize D_c}, \eqref{max problem to minimize D_d}, respectively. 
%Recall that $D$ and $D_3$ denote the MSEs corresponding to the optimal solution of the problem \eqref{original min problem} and that of clairvoyant estimator, respectively. We can write:
%
%\begin{equation} \label{diff distortions to define MSE gain}
%D_3\leq D\leq D_t\ \text{or}\ D_c\ \text{or}\ D_d\leq\sigma_{\theta}^2,
%\end{equation}
%
We know $D_3\!< \!G^{-1}\!<\! D\!<\! D_t, D_c, D_d\!<\!\sigma_{\theta}^2$. 
%where $\sigma_{\theta}^2$ is the worst case distortion that corresponds to the prior information only. 
%Equation \eqref{diff distortions to define MSE gain} as well as Bayesian CRB are illustrated in Fig.~\ref{MSEs_and_gaps}, where a typical operational region is depicted along with the power-distortion tradeoff. We observe that all metrics 
%
%
%Fig.~\ref{MSEs_and_gaps} plots $D_t,D_c,D_d,D,G^{-1}$ versus $P_{tot}$, showing that all decrease as $P_{tot}$ increases. 
%
%
%The goal of any estimation application is to close as much of the performance gap $(\sigma_{\theta}^2-D_3)$ as possible using efficient allocation of the limited network power. We define the following metrics (gains): 
To quantify the efficacy of different power allocation (w.r.t three sets of optimization variables $P_{trn}$, $P_l$'s, ${\mathcal P}_l$'s) in closing the MSE performance gap $\sigma_{\theta}^2-G^{-1}$, we define three factors as the following:
\vspace{-.15cm}
\begin{equation} \label{def of gains}
g_t=\frac{D_t-D}{\sigma_{\theta}^2-G^{-1}},\ g_c=\frac{D_c-D}{\sigma_{\theta}^2-G^{-1}},\ g_d=\frac{D_d-D}{\sigma_{\theta}^2-G^{-1}},
\end{equation}
%
%that, respectively, quantify the efficacy of power allocation with respect to optimization variables $P_{trn}$, $P_l$'s, and ${\mathcal P}_l$'s. These gains are depicted in Fig.~\ref{MSEs_and_gaps}. Note that
where $0\!\leq g_t, g_c, g_d\leq\!1$. A {\blue larger} factor $g$ means that the particular power allocation is more effective in reducing the MSE performance gap (closing the MSE performance gap). 
Fig.~\ref{gaps} {\blue and Fig. \ref{gaps new setup parameters}} plot $g_t, g_c, g_d$ versus $P_{tot}$
{\blue for two sets of noise variances (in Fig. ~\ref{gaps new setup parameters}   $\sigma_{h_l},\sigma_{q_{l,k}}$ are chosen from a smaller interval $(0,0.5)$)}. For $g_t$ we plot three curves corresponding to $P_{trn}\!=\!5\%, 25\%, 60\% P_{tot}$. 
%%%%%%%%%%%%%%%%%%%%%%%%%%%%%%
%delete due to lack of space
%As expected, when $P_{tot}$ increases, all metrics decrease and approach zero in high-region of $P_{tot}$, because the schemes converge to uniform power allocation among clusters. 
%%%%%%%%%%%%%%%%%%%%%%%%%%%%%%%%%%%%%%%%
{\blue Fig.~\ref{gaps} shows $g_c \!>\! g_t(P_{trn} \!=\!5\% P_{tot}) \!>\! g_t(P_{trn}\!=\!60\% P_{tot}) \!>\! g_d  \!>\! g_t(P_{trn}\!=\!25\% P_{tot})$. Whereas Fig. \ref{gaps new setup parameters} shows $g_t(P_{trn} \!=\!5\% P_{tot}) \!>\! g_c \!>\! g_t(P_{trn}\!=\!25\% P_{tot})\!>\! g_d \!>\! g_t(P_{trn}\!=\!60\% P_{tot})$. 
Evidently, a more accurate channel estimation does not necessarily lead into a smaller $D_t$. Two takeaway messages are: (1) $g_t,g_c,g_d \!>\!0$, i.e., the solution obtained from solving \eqref{max problem to minimize D} always leads into an MSE improvement, (2) the actual values of $g_t,g_c,g_d$ depend on the system parameters and $P_{tot}$.} 
Note {\blue that in Fig.~\ref{gaps}} at $P_{tot}\!=\!0\text{dB}$, $g_t\!=\!0.15$ (for $P_{trn}\!=\!25\% P_{tot}$), $g_d\!=\!0.17, g_c\!=\!0.48$, meaning that power allocation among CHs for training and ${\cal P}_l$, and among clusters for obtaining $P_l$ reduce the MSE performance gap to $15\%, 17\%, 48\%$, respectively. 
%Moreover, at $P_{tot}\!=\!10\text{dB}$, $g_t\!=\!0.04$ (for $P_{trn}\!=\!25\% P_{tot}$), $g_d\!=\!0.05, g_c\!=\!0.23$, meaning that power allocation among CHs for training and ${\cal P}_l$, and among clusters for obtaining $P_l$ reduce the MSE performance gap to $4\%, 5\%, 23\%$, respectively. 
%
{\blue Combining the information given by $g_t,g_c,g_d, G$ with the computational complexity analysis in Section \ref{comp complexity}
provides the system designer with quantitative  complexity-versus-MSE improvement tradeoffs offered by different power optimization schemes.}

%These observations imply that power allocation among CHs for training as well as ${\cal P}_l$ is always beneficial for low-region of $P_{tot}$, and power allocation among clusters for obtaining $P_l$ is beneficial for low-region to moderate-region of $P_{tot}$. 
% 

%
%%%%%%%%%%%%%%%%%%%%%%%%%%%%%%
\vspace{-.4cm}
\subsection{Behavior of Power Allocation Across Clusters} \label{Behavior of PA across clusters}
\vspace{-.15cm}
We study the effect of heterogeneous clusters on the behavior of our proposed power allocation scheme to solve \eqref{max problem to minimize D} as $P_{tot}$ increases. Consider a network consisting $L\!=\!3$ clusters with $K_l\!=\!6, \sigma_{n_{l,k}}\!=\!\sigma_{n_l}, \sigma_{q_{l,k}}\!=\!\sigma_{q_l}, \forall l,k$. We define 
${\gamma}_l^o\!=\!\frac{\sigma^2_{\theta}}{\sigma_{n_l}^2}$ as observation SNR of sensors within cluster $l$, ${\gamma}_l^c\!=\!\frac{1}{\sigma_{q_l}^2}$ as channel-to-noise ratio (CNR) corresponding to sensors-CH$_l$ links, and ${\gamma}_l^d\!=\!\frac{\sigma^2_{h_l}}{\sigma_{v_l}^2}$ as CNR corresponding to CH$_l$-FC link. 
%denoting the observation SNR for the sensors within cluster $l$, average SNR for the channels between the sensors within cluster $l$ and CH$_l$, and average SNR for the channel between the CH$_l$ and the FC. 
%To be able to interpret the results in this section, we plot the variables $\psi_l, P_l, {\cal P}_l$ as well as the auxilary variables ${\cal V}_l\!=\!P_l\!+\!{\cal P}_l, \forall l$ in each figure. Recall that ${\cal V}_l$ represents the amount of power allocated to $l$-th cluster excluding the training power $\psi_l$.
Let $\psi_l\ (\text{dB})\!=\!10\text{log}_{10}(\psi_l), P_l\ (\text{dB})\!=\!10\text{log}_{10}(P_l), {\cal P}_l\ (\text{dB})\!=\!10\text{log}_{10}({\cal P}_l), {\cal V}_l\ (\text{dB})\!=\!10\text{log}_{10}({\cal V}_l)$, where ${\cal V}_l\!=\!P_l\!+\!{\cal P}_l$ represents the allocated power to cluster $l$, excluding its training power $\psi_l$. In the following we consider three scenarios: (i) when observation SNR ${\gamma}_l^o$ and CNR ${\gamma}_l^c$ are equal and CNR ${\gamma}_l^d$ are different across clusters, (ii) when observation SNR ${\gamma}_l^o$ and CNR ${\gamma}_l^d$ are equal and CNR ${\gamma}_l^c$ are different across clusters, (iii) when CNRs ${\gamma}_l^c$ and ${\gamma}_l^d$ are equal and observation SNR ${\gamma}_l^o$ are different across clusters.
 
\begin{figure}[t]
	
	\centering
	\hspace{-.5cm}
	\begin{subfigure}[b]{0.25\textwidth}
		
		\centering
		%\hspace{-.2cm}
		\subcaptionbox{\label{P_trn_l_vs_P_tot_diff_SNR_d}}{\vspace{-.2 cm}\includegraphics[width=1.8in,height=.85in]{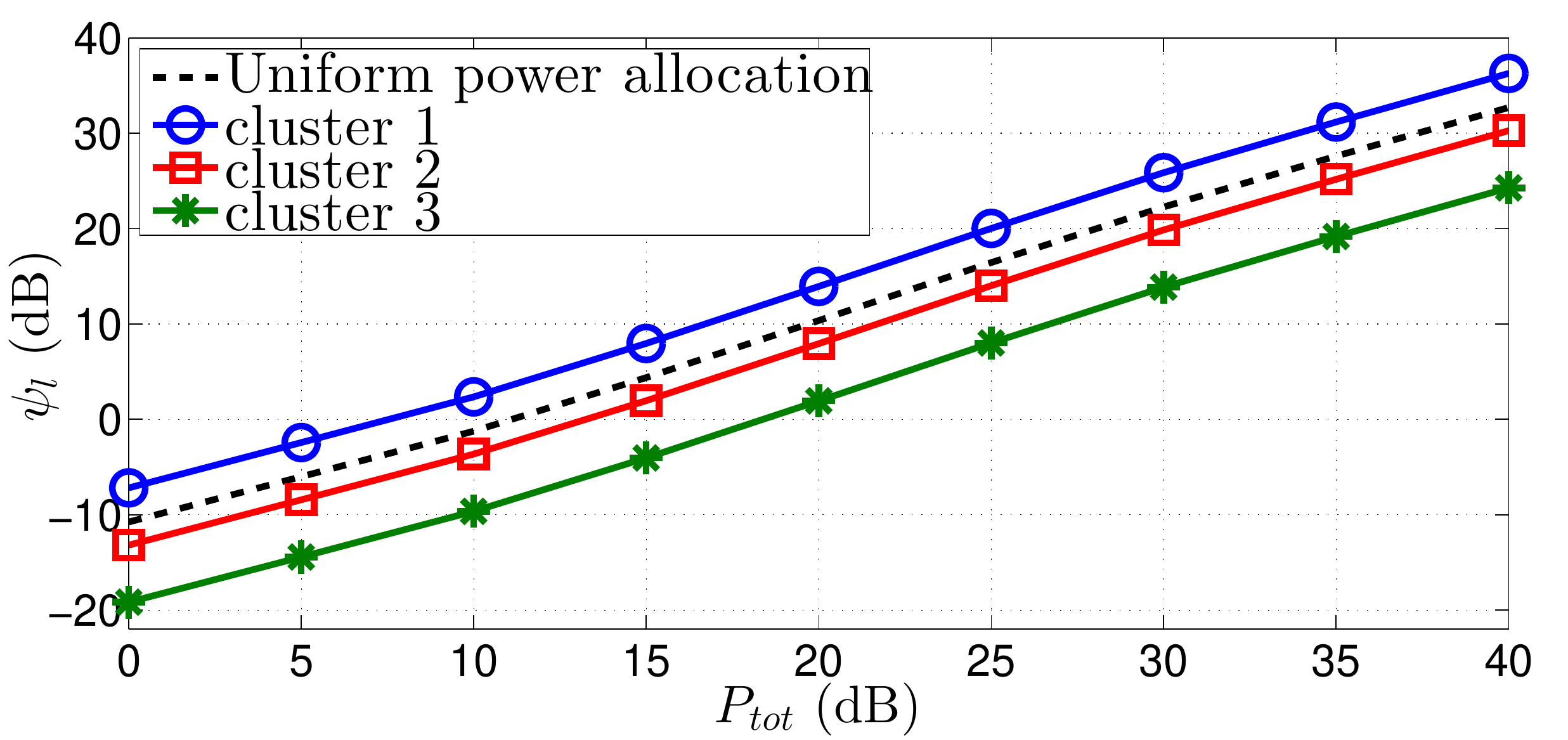}}
		
	\end{subfigure}%	
	\begin{subfigure}[b]{0.25\textwidth}
		
		\centering
		%\hspace{-.2cm}
		\subcaptionbox{\label{P_prime_l_vs_P_tot_diff_SNR_d}}{\vspace{-.2 cm}\includegraphics[width=1.8in,height=.85in]{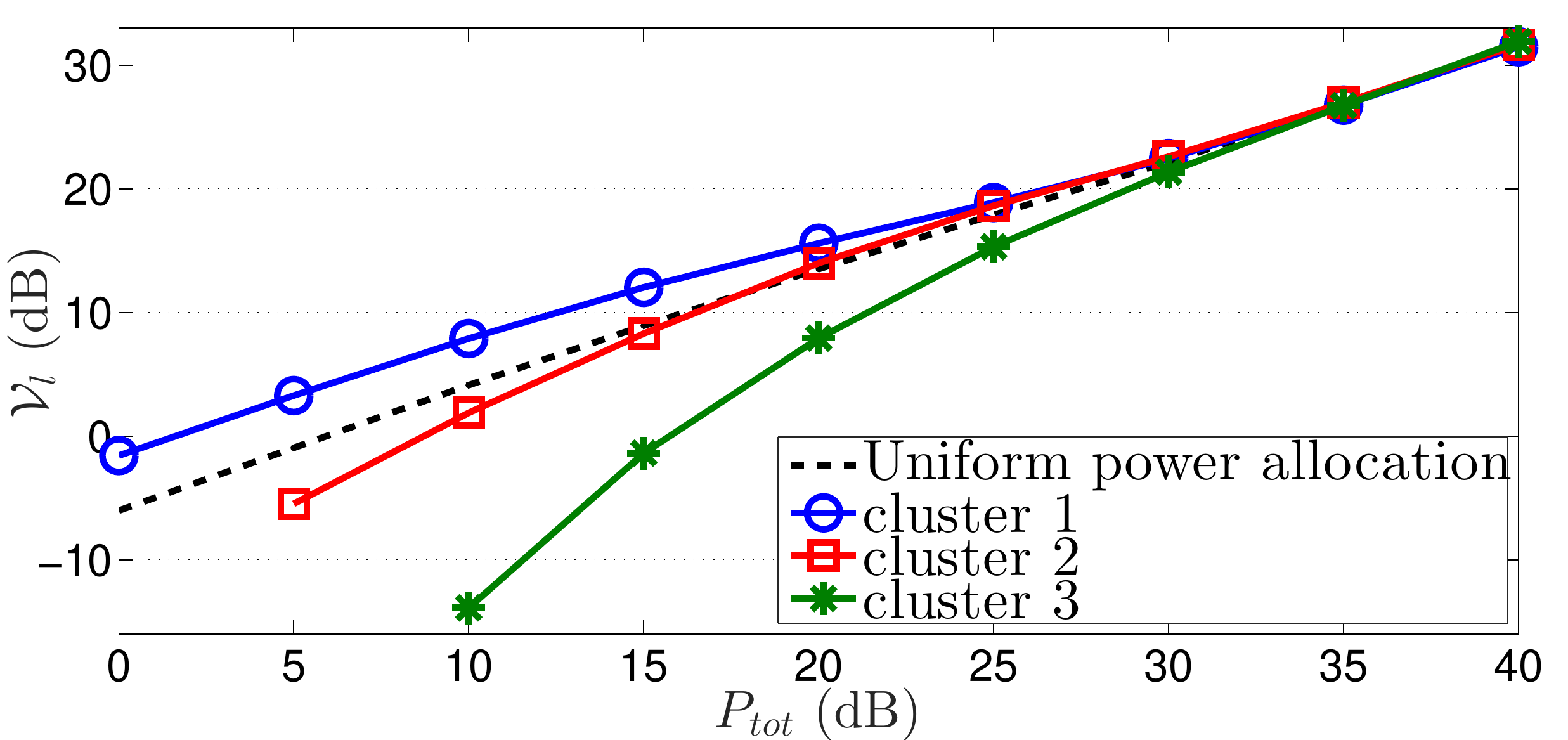}}
		
	\end{subfigure} \\
\centering
\hspace{-.5cm}
\begin{subfigure}[b]{0.25\textwidth}
	
	\centering
	%\hspace{-.2cm}
	\subcaptionbox{\label{P_d_l_vs_P_tot_diff_SNR_d}}{\vspace{-.2 cm}\includegraphics[width=1.8in,height=.85in]{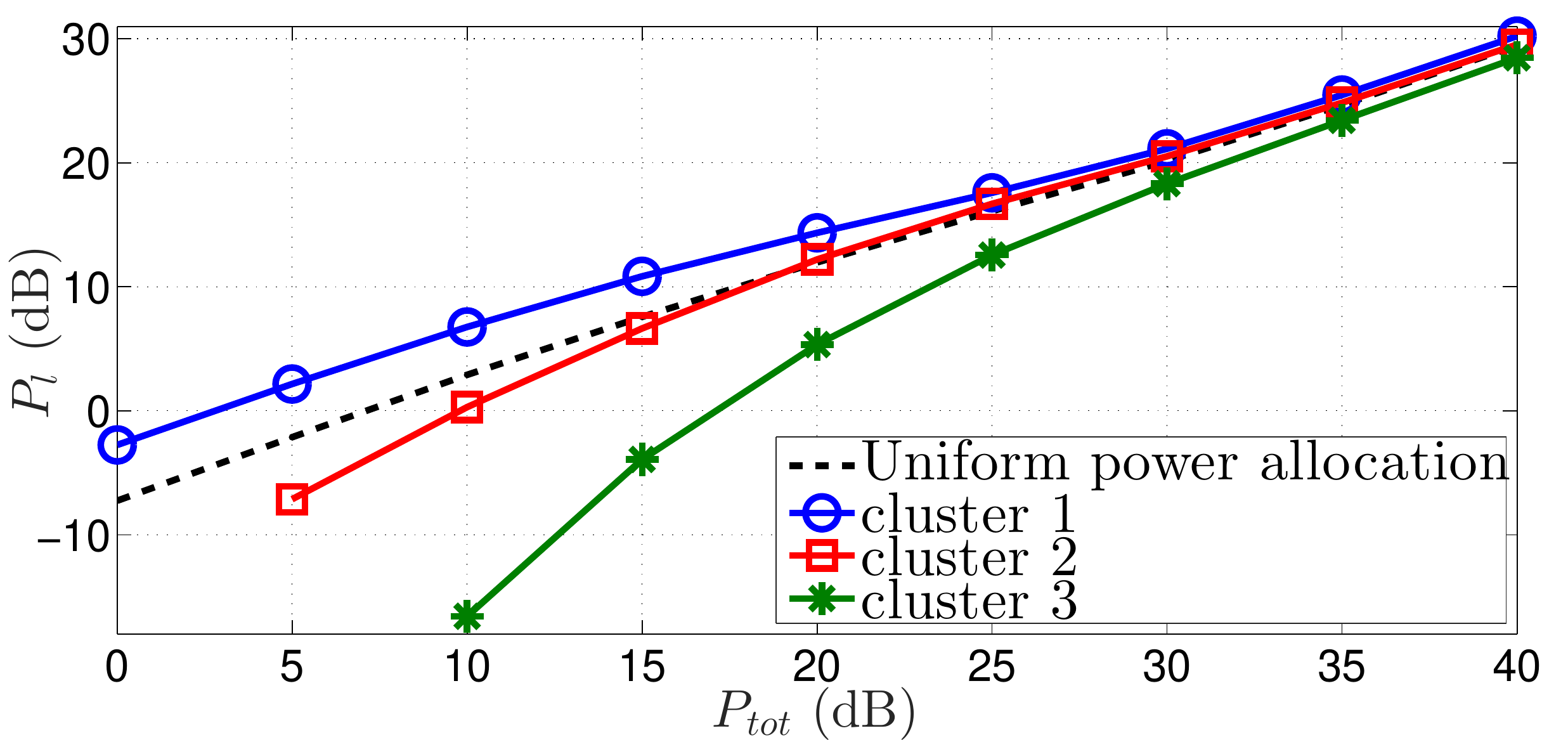}}
	
\end{subfigure}%	
\begin{subfigure}[b]{0.25\textwidth}
	
	\centering
	%\hspace{-.2cm}
	\subcaptionbox{\label{P_cal_l_vs_P_tot_diff_SNR_d}}{\vspace{-.2 cm}\includegraphics[width=1.8in,height=.85in]{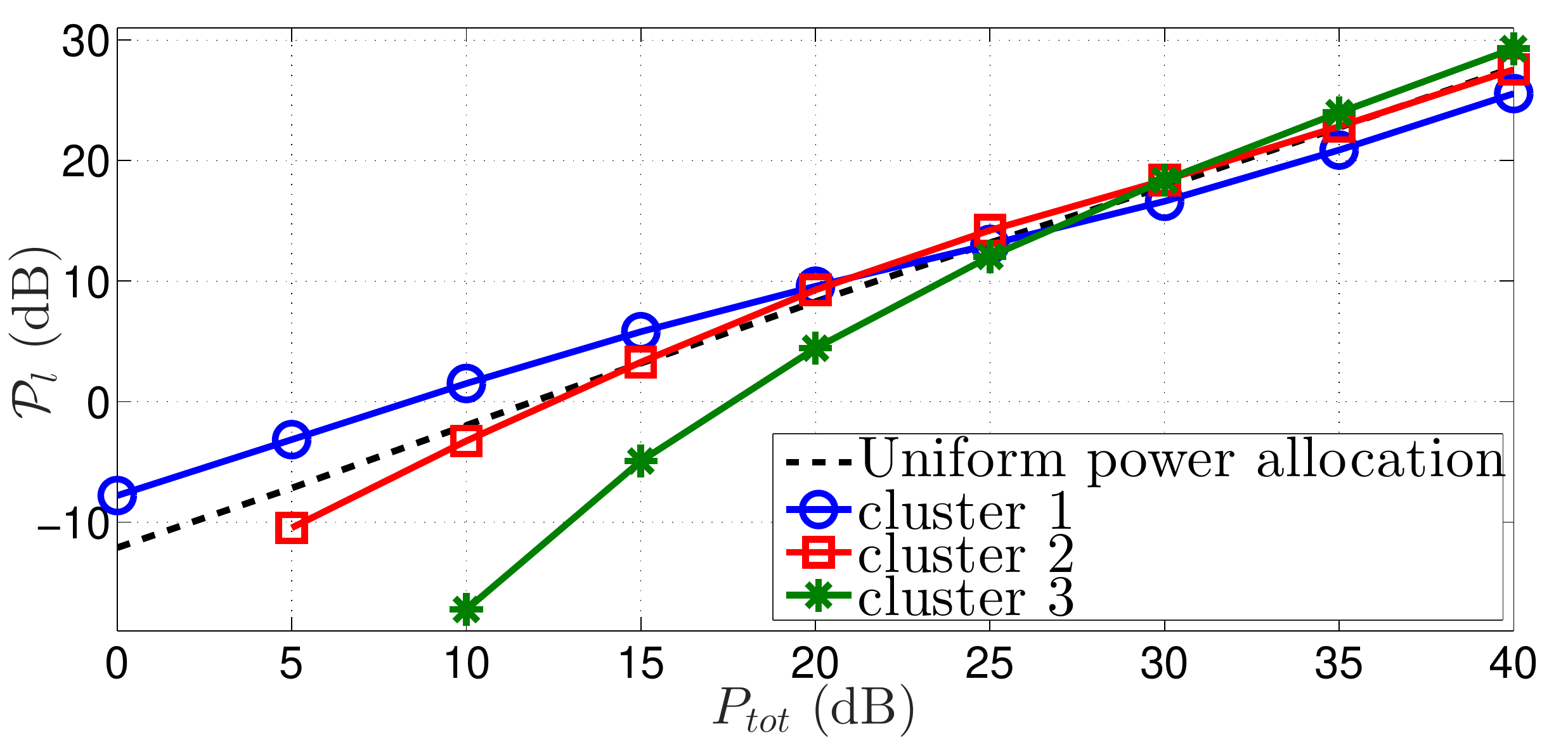}}
	
\end{subfigure}\\
	
	\caption{
		%behavior of (a) $\{10\text{log}_{10}(\psi_l)\}_{l=1}^3$ (dB), (b) $\{10\text{log}_{10}({\cal V}_l)\}_{l=1}^3$ (dB), (c) $\{10\text{log}_{10}(P_l)\}_{l=1}^3$ (dB), (d) $\{10\text{log}_{10}({\cal P}_l)\}_{l=1}^3$ (dB) versus $P_{tot}$ (dB) to solve \eqref{max problem to minimize D}, when 
		$\{{\gamma}_l^o\!=\!5\ \text{dB}, {\gamma}_l^c\!=\!5\ \text{dB}\}_{l=1}^3$ and ${\gamma}_1^d\!>\!{\gamma}_2^d\!>\!{\gamma}_3^d$.}   
	\label{heterog wrt gamma_d}
	
\end{figure}

Figs.~\ref{P_trn_l_vs_P_tot_diff_SNR_d}, \ref{P_prime_l_vs_P_tot_diff_SNR_d}, \ref{P_d_l_vs_P_tot_diff_SNR_d}, \ref{P_cal_l_vs_P_tot_diff_SNR_d}, respectively, depict $\psi_l\!$ (dB),${\cal V}_l\!$ (dB),$P_l\!$ (dB),${\cal P}_l\!\ (\text{dB}), \forall l$, versus $P_{tot}$ for ${\gamma}_l^o\!=\!5\!\ \text{dB}, {\gamma}_l^c\!=\!5\!\ \text{dB}, \forall l$ and ${\gamma}_1^d\!=\!14\!\ \text{dB}, {\gamma}_2^d\!=\!8\!\ \text{dB}, {\gamma}_3^d\!=\!2\!\ \text{dB}$.
%, where $P_{trn}, \{P_l, {\cal P}_l\}_{l=1}^L$ are the solutions of the constrained minimization problem in \eqref{original min problem} and $\psi_l$ is computed using \eqref{final formula psi_l}. 
Regarding Fig.~\ref{heterog wrt gamma_d} we make the following observations: 1) all powers increase as $P_{tot}$ increases, 2) when $P_{tot}$ is small, only cluster 1 is active, and as $P_{tot}$ increases, clusters 2 and 3 become active in a sequential order, 3) in all regions of $P_{tot}$, a cluster with a larger ${\gamma}_l^d$ is allotted a larger $\psi_l$ (water filling), 4) in low-region to moderate-region of $P_{tot}$, a cluster with a larger ${\gamma}_l^d$ is allocated a larger ${\cal V}_l$ (water filling), and in high-region of $P_{tot}$, ${\cal V}_l$ of all clusters converge (uniform power allocation), 5) in all regions of $P_{tot}$, a cluster with a larger ${\gamma}_l^d$ is assigned a larger $P_l$ (water filling), 6) in low-region of $P_{tot}$, a cluster with a larger ${\gamma}_l^d$ is allocated a larger ${\cal P}_l$ (water filling), and in high-region of $P_{tot}$, a cluster with a larger ${\gamma}_l^d$ is allotted a smaller ${\cal P}_l$ (inverse of water filling). The behavior of $P_l$ and ${\cal P}_l$ in high-region of $P_{tot}$ can be explained by examining the behavior of ${\cal V}_l$. 
%In particular, since the clusters differ in the average channel SNR between the CHs and the FC, this difference can be compensated by increasing $P_{tot}$, such that ${\cal V}_l$ converges to uniform power allocation scheme among all clusters. 
Note that, although CNRs ${\gamma}_1^d, {\gamma}_2^d, {\gamma}_3^d$ are different, the differences are compensated as $P_{tot}$ increases and ${\cal V}_l$ of all clusters converge. 
This fact implies the behaviors of $P_l$ and ${\cal P}_l$ in high-region of $P_{tot}$ are opposite, i.e., water filling and inverse of water filling power allocation for $P_l$ and ${\cal P}_l$, respectively. 
\begin{figure}[t]
	
	\centering
	\hspace{-.5cm}
	\begin{subfigure}[b]{0.25\textwidth}
		
		\centering
		%\hspace{-.2cm}
		\subcaptionbox{\label{P_trn_l_vs_P_tot_diff_SNR_c}}{\vspace{-.2 cm}\includegraphics[width=1.8in,height=.85in]{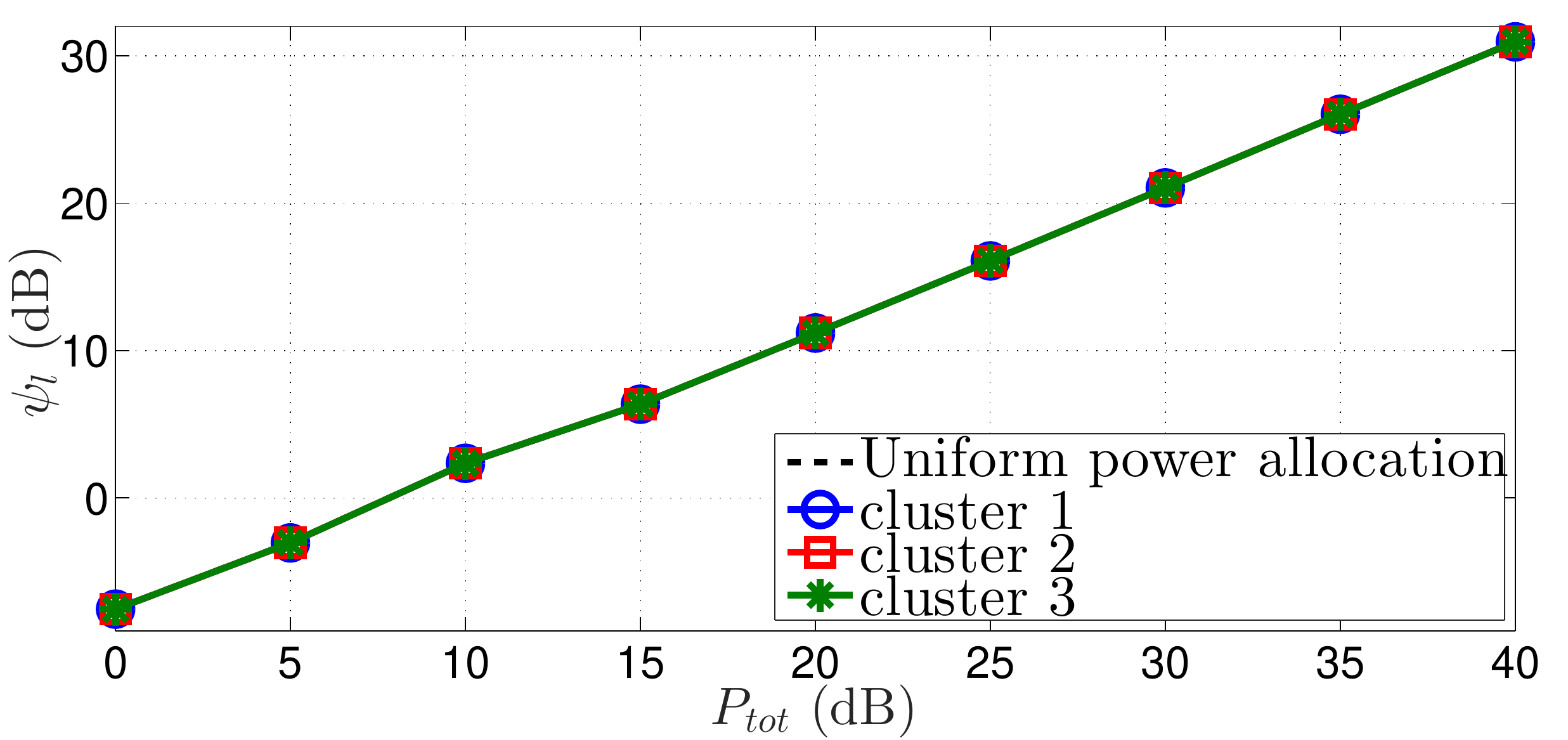}}
		
	\end{subfigure}%	
	\begin{subfigure}[b]{0.25\textwidth}
		
		\centering
		%\hspace{-.2cm}
		\subcaptionbox{\label{P_prime_l_vs_P_tot_diff_SNR_c}}{\vspace{-.2 cm}\includegraphics[width=1.8in,height=.85in]{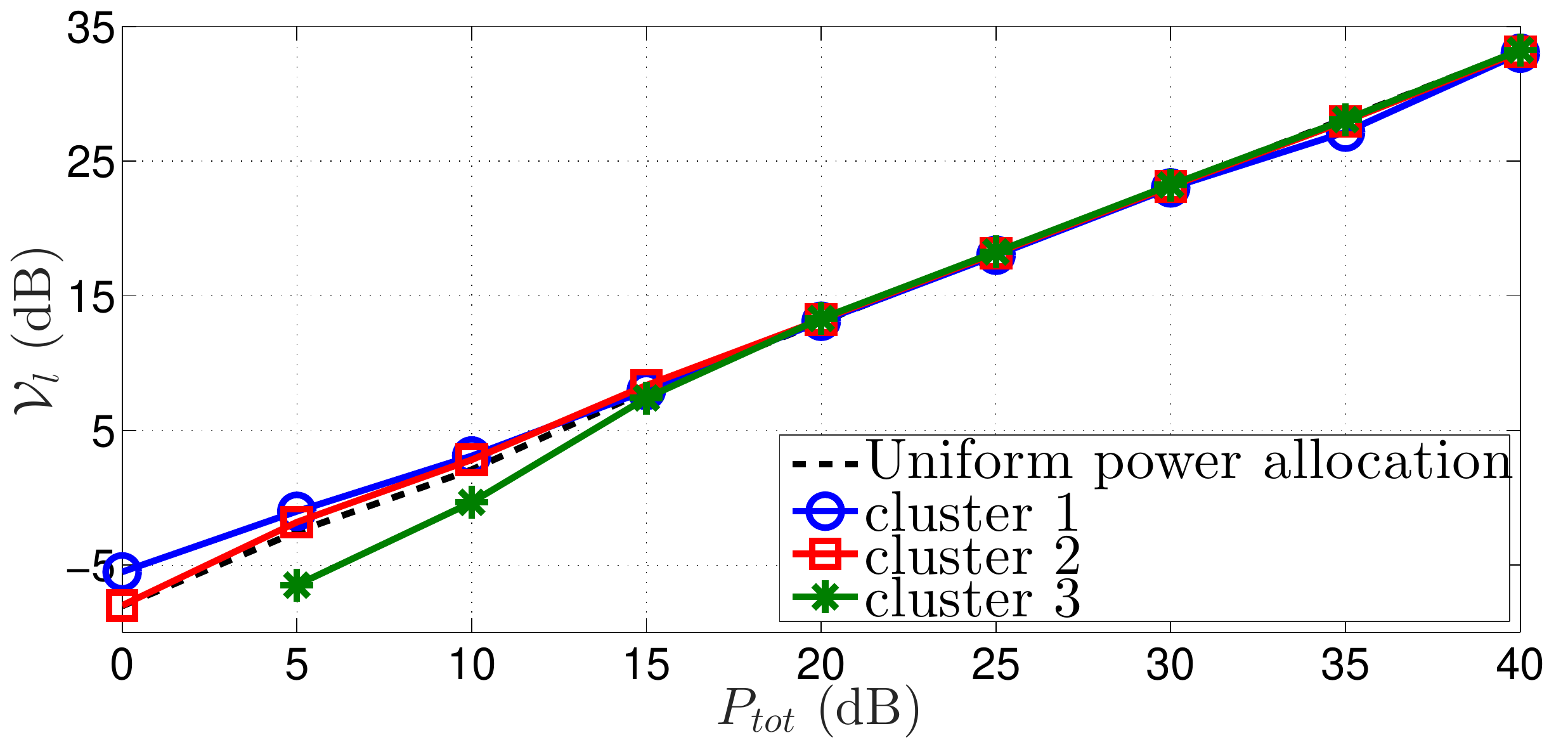}}
		
	\end{subfigure} \\
	\centering
	\hspace{-.5cm}
	\begin{subfigure}[b]{0.25\textwidth}
		
		\centering
		%\hspace{-.2cm}
		\subcaptionbox{\label{P_d_l_vs_P_tot_diff_SNR_c}}{\vspace{-.2 cm}\includegraphics[width=1.8in,height=.85in]{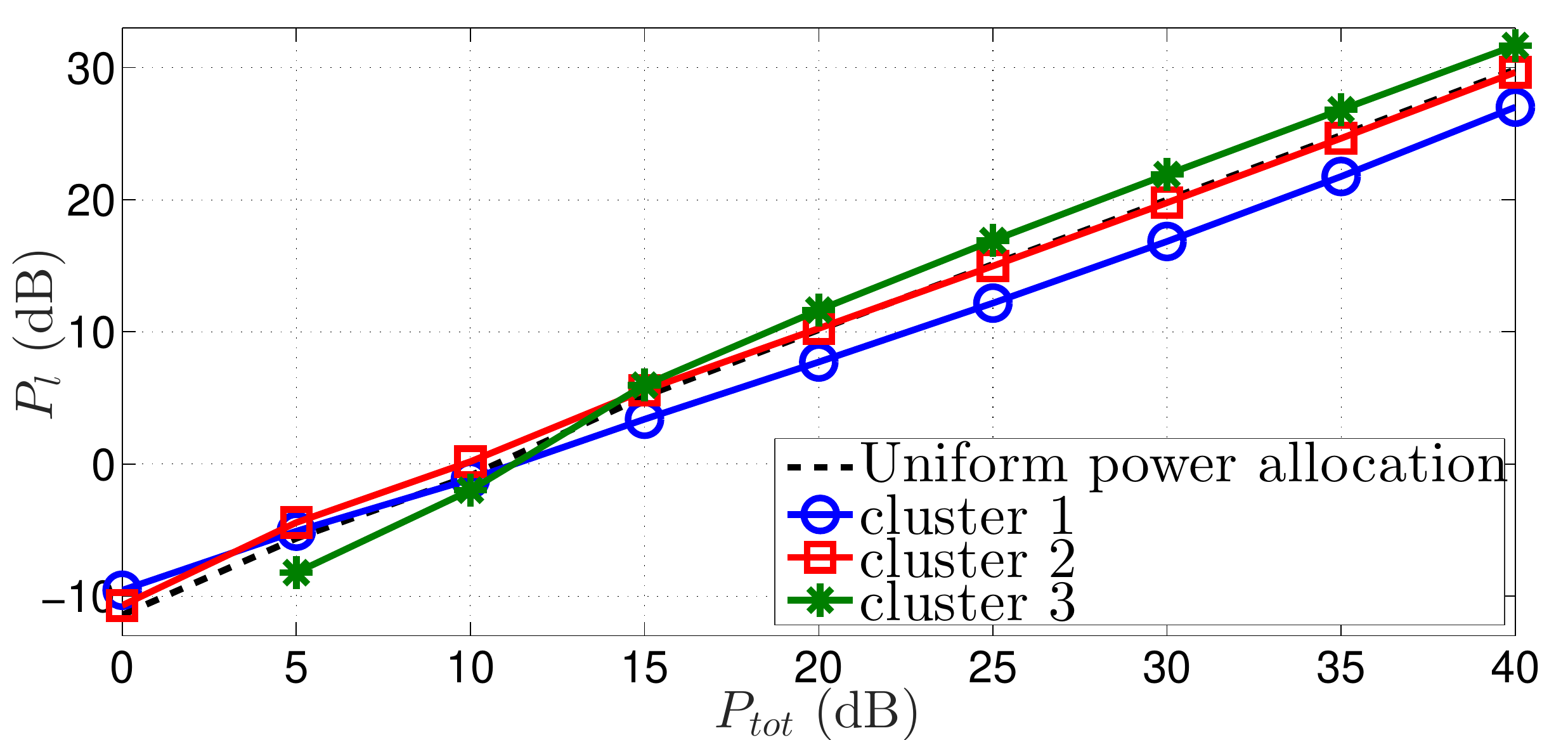}}
		
	\end{subfigure}%	
	\begin{subfigure}[b]{0.25\textwidth}
		
		\centering
		%\hspace{-.2cm}
		\subcaptionbox{\label{P_cal_l_vs_P_tot_diff_SNR_c}}{\vspace{-.2 cm}\includegraphics[width=1.8in,height=.85in]{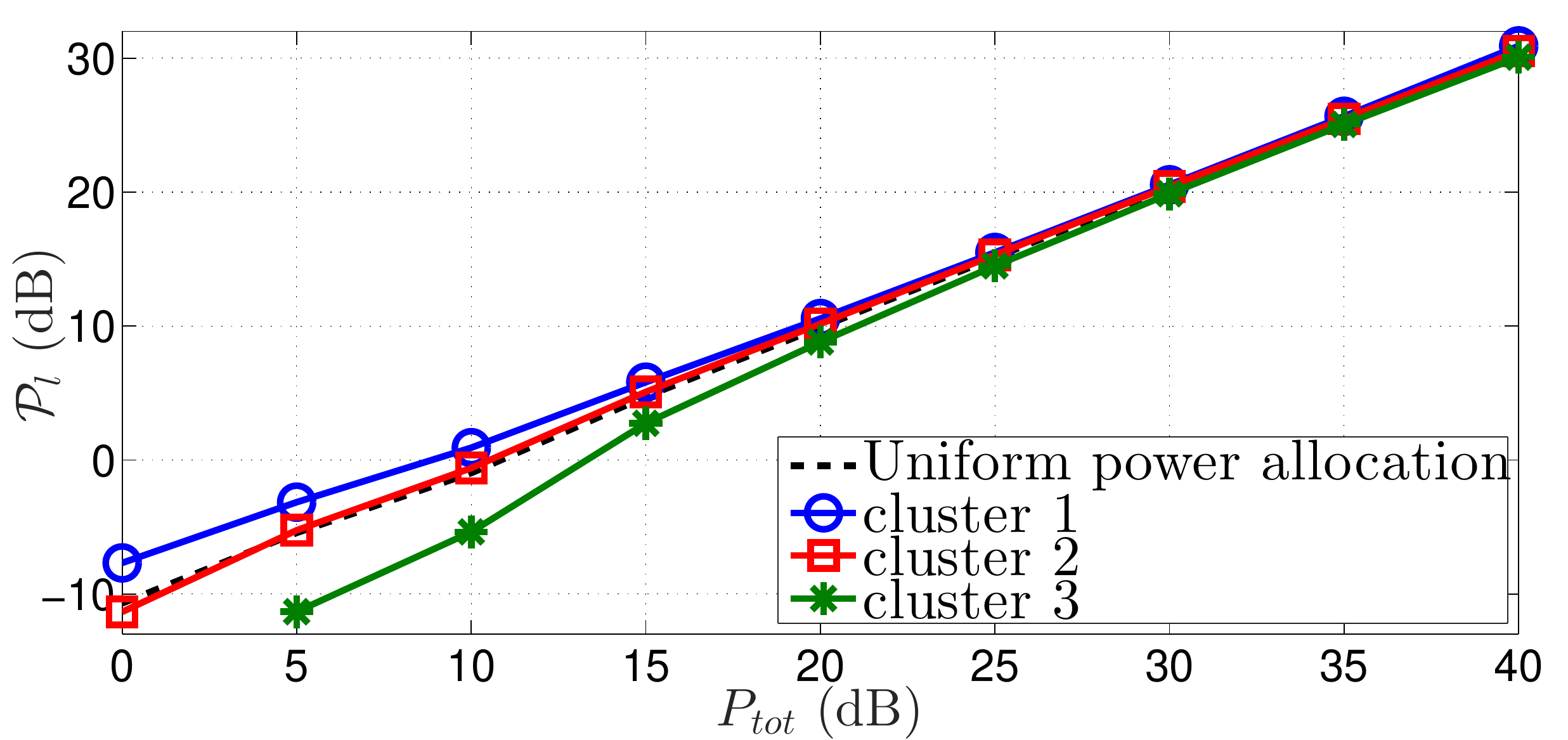}}
		
	\end{subfigure}\\
	
	\caption{
		%behavior of (a) $\{10\text{log}_{10}(\psi_l)\}_{l=1}^3$ (dB), (b) $\{10\text{log}_{10}({\cal V}_l)\}_{l=1}^3$ (dB), (c) $\{10\text{log}_{10}(P_l)\}_{l=1}^3$ (dB), (d) $\{10\text{log}_{10}({\cal P}_l)\}_{l=1}^3$ (dB) versus $P_{tot}$ (dB) to solve \eqref{max problem to minimize D}, when 
		$\{{\gamma}_l^o\!=\!5\ \text{dB}, {\gamma}_l^d\!=\!5\ \text{dB}\}_{l=1}^3$ and ${\gamma}_1^c\!>\!{\gamma}_2^c\!>\!{\gamma}_3^c$.}   
	\label{heterog wrt gamma_c}
	
\end{figure}
\begin{figure}[h!]
	
	\centering
	\hspace{-.5cm}
	\begin{subfigure}[b]{0.25\textwidth}
		
		\centering
		%\hspace{-.2cm}
		\subcaptionbox{\label{P_trn_l_vs_P_tot_diff_SNR_o}}{\vspace{-.2 cm}\includegraphics[width=1.8in,height=.85in]{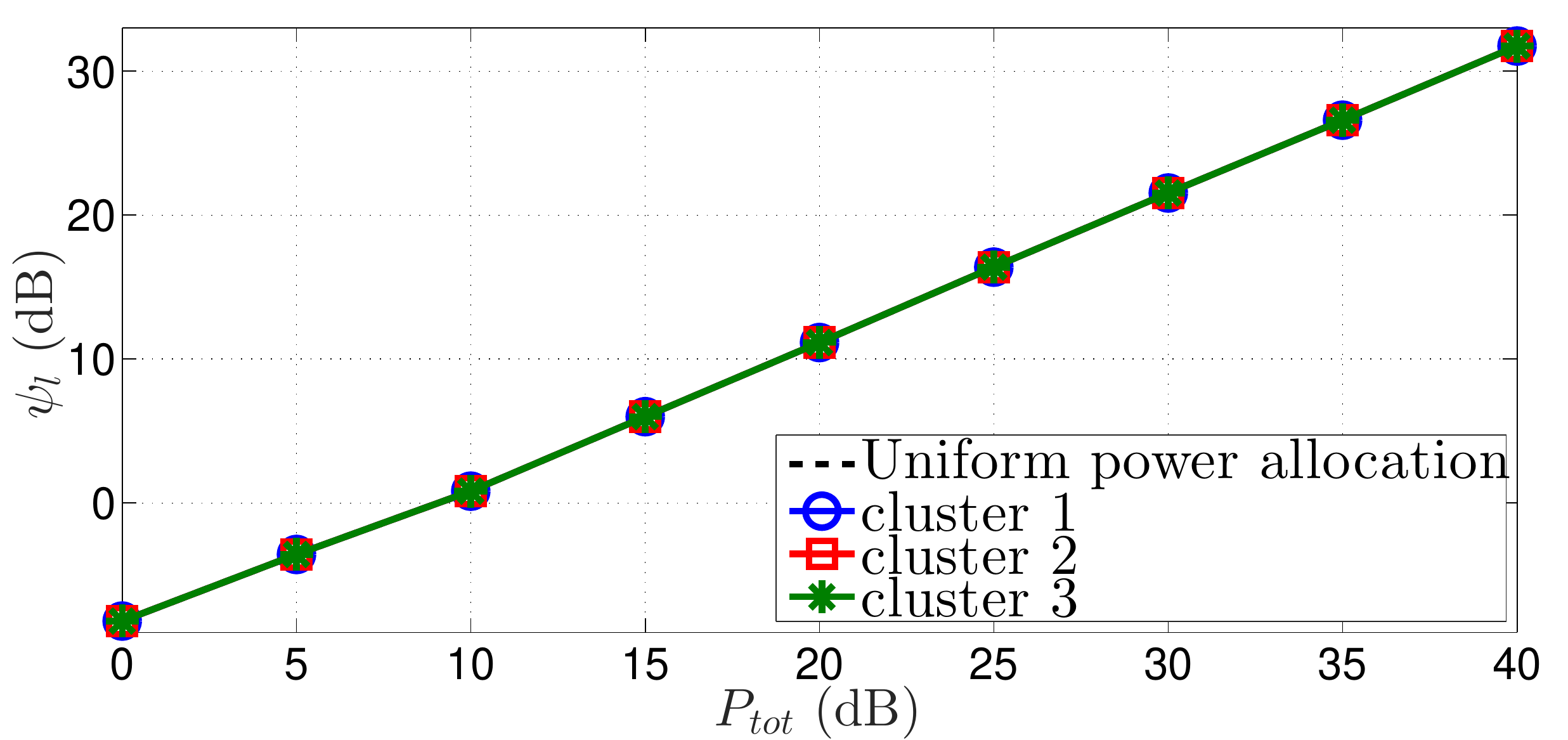}}
		
	\end{subfigure}%	
	\begin{subfigure}[b]{0.25\textwidth}
		
		\centering
		%\hspace{-.2cm}
		\subcaptionbox{\label{P_prime_l_vs_P_tot_diff_SNR_o}}{\vspace{-.2 cm}\includegraphics[width=1.8in,height=.85in]{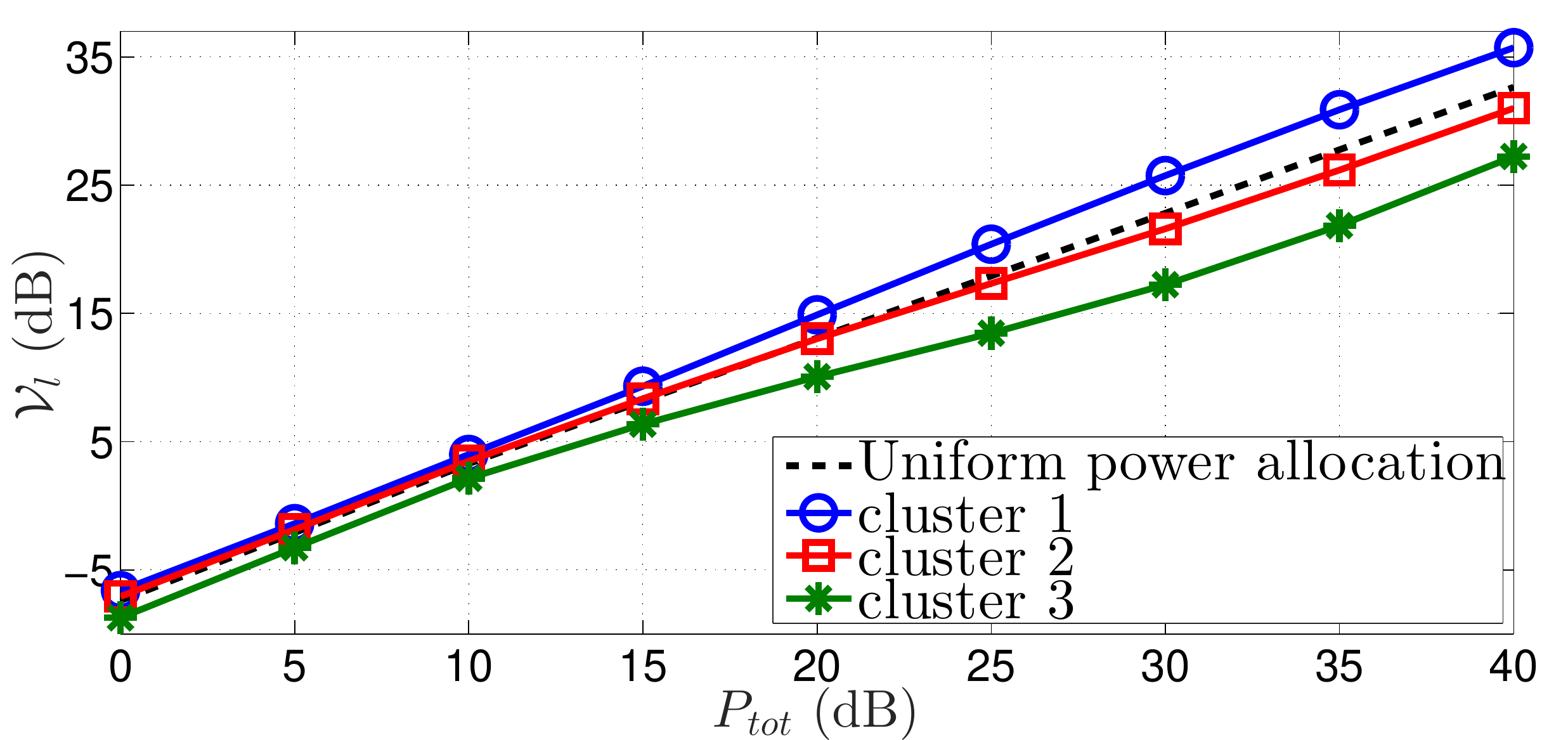}}
		
	\end{subfigure} \\
	\centering
	\hspace{-.5cm}
	\begin{subfigure}[b]{0.25\textwidth}
		
		\centering
		%\hspace{-.2cm}
		\subcaptionbox{\label{P_d_l_vs_P_tot_diff_SNR_o}}{\vspace{-.2 cm}\includegraphics[width=1.8in,height=.85in]{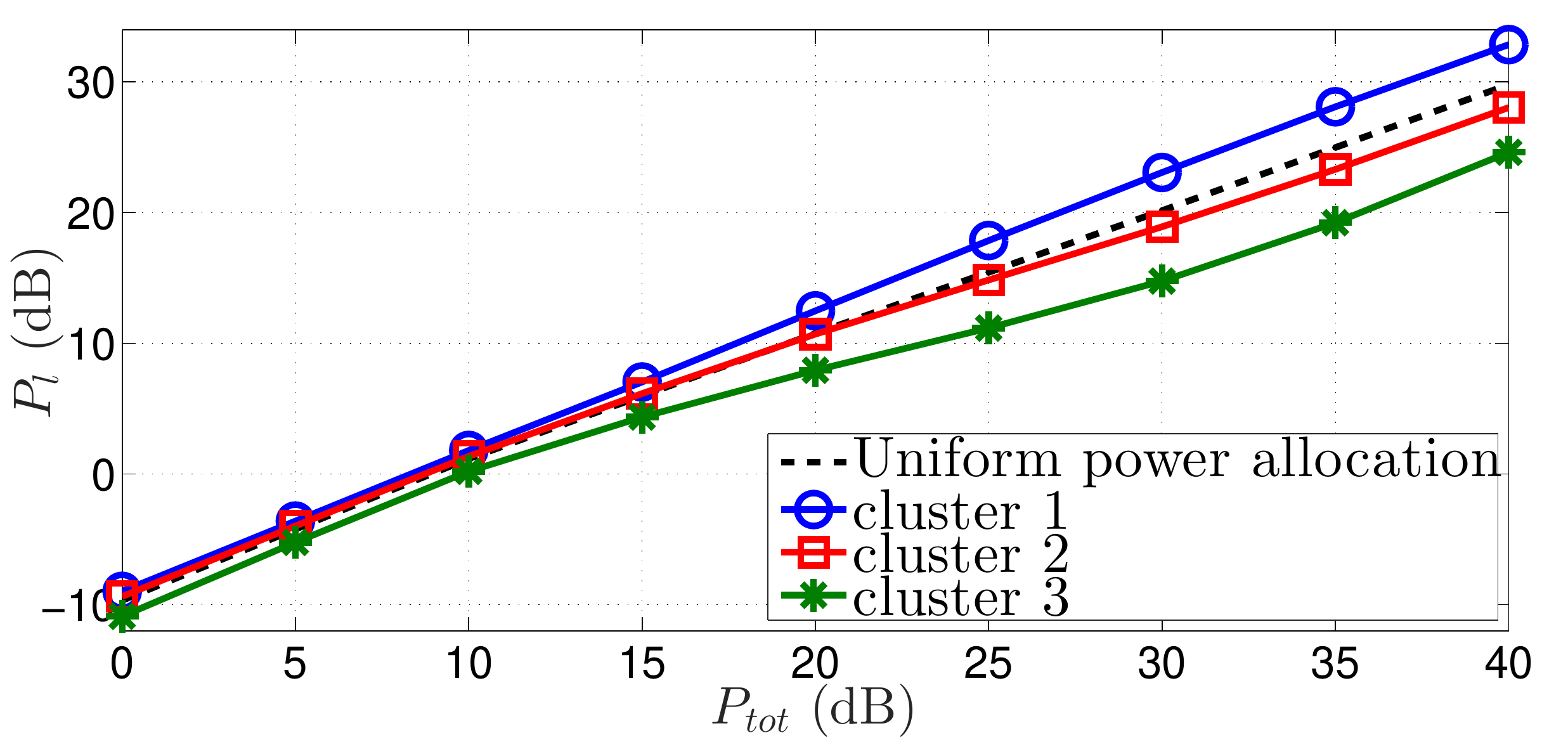}}
		
	\end{subfigure}%	
	\begin{subfigure}[b]{0.25\textwidth}
		
		\centering
		%\hspace{-.2cm}
		\subcaptionbox{\label{P_cal_l_vs_P_tot_diff_SNR_o}}{\vspace{-.2 cm}\includegraphics[width=1.8in,height=.85in]{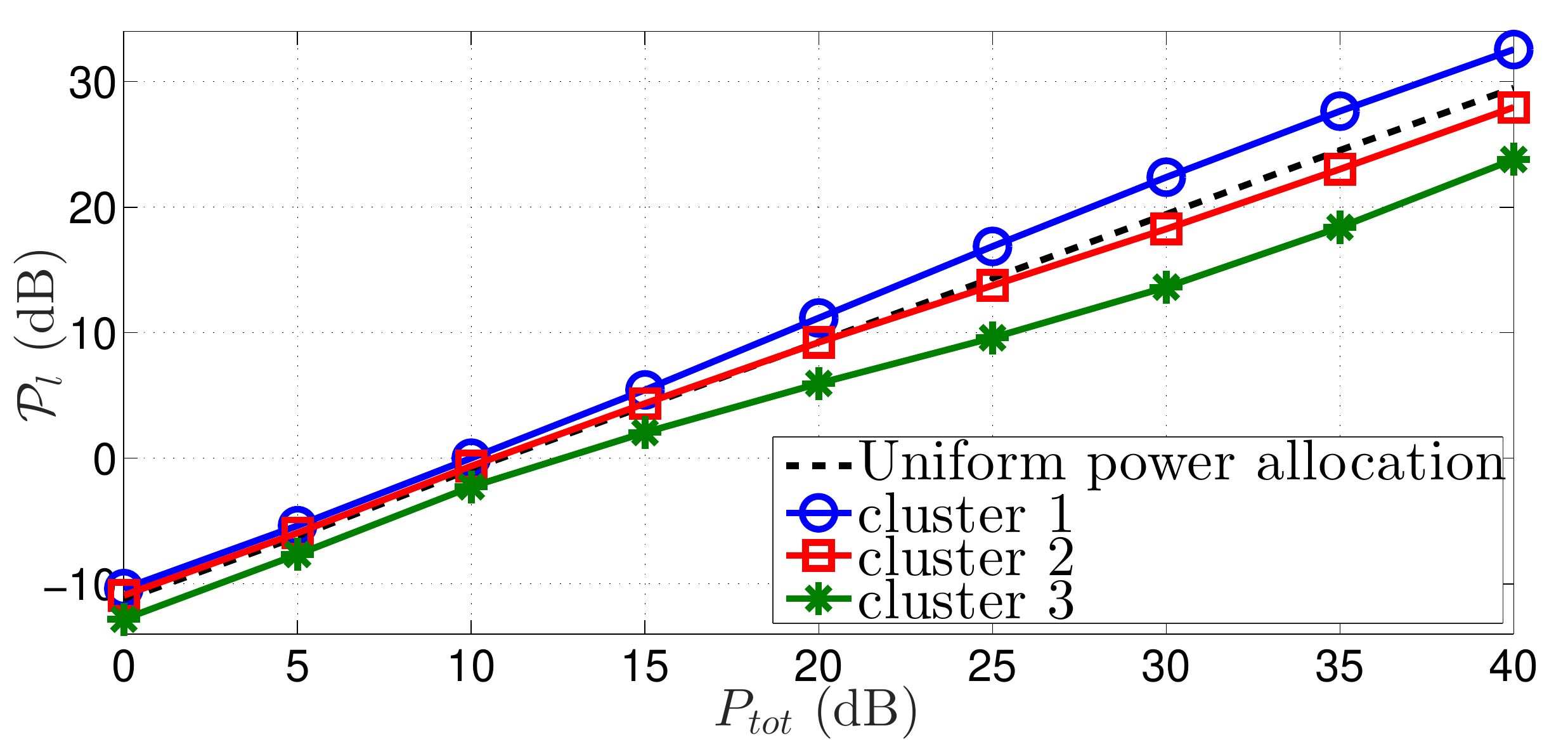}}
		
	\end{subfigure}\\
	
	\caption{
		%behavior of (a) $\{10\text{log}_{10}(\psi_l)\}_{l=1}^3$ (dB), (b) $\{10\text{log}_{10}({\cal V}_l)\}_{l=1}^3$ (dB), (c) $\{10\text{log}_{10}(P_l)\}_{l=1}^3$ (dB), (d) $\{10\text{log}_{10}({\cal P}_l)\}_{l=1}^3$ (dB) versus $P_{tot}$ (dB) to solve \eqref{max problem to minimize D}, when 
		$\{{\gamma}_l^c\!=\!5\ \text{dB}, {\gamma}_l^d\!=\!5\ \text{dB}\}_{l=1}^3$ and ${\gamma}_1^o\!>\!{\gamma}_2^o\!>\!{\gamma}_3^o$.}   
	\label{heterog wrt gamma_o}
	
\end{figure}

Figs.~\ref{P_trn_l_vs_P_tot_diff_SNR_c}, \ref{P_prime_l_vs_P_tot_diff_SNR_c}, \ref{P_d_l_vs_P_tot_diff_SNR_c}, \ref{P_cal_l_vs_P_tot_diff_SNR_c}, respectively, depict $\psi_l$ (dB), ${\cal V}_l$ (dB), $P_l$ (dB), ${\cal P}_l\ (\text{dB}), \forall l$, versus $P_{tot}$ for ${\gamma}_l^o\!=\!5\ \text{dB}, {\gamma}_l^d\!=\!5\ \text{dB}, \forall l$ and ${\gamma}_1^c\!=\!14\ \text{dB}, {\gamma}_2^c\!=\!8\ \text{dB}, {\gamma}_3^c\!=\!2\ \text{dB}$. 
The following observations can be made for Fig.~\ref{heterog wrt gamma_c}: comments 1) and 2) for Fig.~\ref{heterog wrt gamma_d} also hold for Fig.~\ref{heterog wrt gamma_c}, 3) in all regions of $P_{tot}$, $\psi_l$ of all clusters are equal (uniform power allocation) since ${\gamma}_l^d$'s are equal, 4) behavior of ${\cal V}_l$ in Fig.~\ref{P_prime_l_vs_P_tot_diff_SNR_c} is the same as that of Fig.~\ref{P_prime_l_vs_P_tot_diff_SNR_d}, 5) in low-region of $P_{tot}$, a cluster with a larger ${\gamma}_l^c$ is allocated a larger $P_l$ (water filling), and in high-region of $P_{tot}$, a cluster with a larger ${\gamma}_l^c$ is allocated a smaller $P_l$ (inverese of water filling), 6) in all regions of $P_{tot}$, a cluster with a larger ${\gamma}_l^c$ is allocated a larger ${\cal P}_l$ (water filling). 
%Regarding the behavior of $P_l$ and ${\cal P}_l$ in high-region of $P_{tot}$, similar discussion to that of Fig.~\ref{heterog wrt gamma_d} can be made for Fig.~\ref{heterog wrt gamma_c}. In particular, since the clusters differ in the average SNR for the channels between the sensors within each cluster and their associated CH, this difference can be compensated by increasing $P_{tot}$, such that ${\cal V}_l$ converges to uniform power allocation scheme among all clusters. 
Note that, although CNRs ${\gamma}_1^c, {\gamma}_2^c, {\gamma}_3^c$ are different, the differences are compensated as $P_{tot}$ increases and ${\cal V}_l$ of all clusters converge. 
This fact implies the behaviors of $P_l$ and ${\cal P}_l$ in high-region of $P_{tot}$ are opposite, i.e., inverse of water filling and water filling power allocation for $P_l$ and ${\cal P}_l$, respectively.
 
Figs.~\ref{P_trn_l_vs_P_tot_diff_SNR_o}, \ref{P_prime_l_vs_P_tot_diff_SNR_o}, \ref{P_d_l_vs_P_tot_diff_SNR_o}, \ref{P_cal_l_vs_P_tot_diff_SNR_o}, respectively, depict $\psi_l$ (dB), ${\cal V}_l$ (dB), $P_l$ (dB), ${\cal P}_l\ (\text{dB}), \forall l$, versus $P_{tot}$ for ${\gamma}_l^c\!=\!5\ \text{dB}, {\gamma}_l^d\!=\!5\ \text{dB}, \forall l$ and ${\gamma}_1^o\!=\!14\ \text{dB}, {\gamma}_2^o\!=\!8\ \text{dB}, {\gamma}_3^o\!=\!2\ \text{dB}$. 
The following observations can be made for Fig.~\ref{heterog wrt gamma_o}: comments 1) and 2) for Figs.~\ref{heterog wrt gamma_d} and \ref{heterog wrt gamma_c} also hold for Fig.~\ref{heterog wrt gamma_o}, 3) in all regions of $P_{tot}$, $\psi_l$ of all clusters are equal (uniform power allocation) since ${\gamma}_l^d$'s are equal, 4) in all regions of $P_{tot}$ a cluster with a larger ${\gamma}_l^o$ is allocated a larger ${\cal V}_l$, a larger $P_l$, and a larger ${\cal P}_l$ (water filling). The behaviors of ${\cal V}_l$, $P_l$, ${\cal P}_l$ in high-region of $P_{tot}$ are different from the two previous scenarios (CNRs across clusters were different), in which ${\cal V}_l$ of all clusters converge as $P_{tot}$ increases. Here the difference in observation SNR across clusters cannot be compensated as $P_{tot}$ increases. Hence, ${\cal V}_l$ of clusters are different, such that a cluster with a larger (smaller) ${\gamma}_l^o$ is allocated a larger (smaller) ${\cal V}_l$.
%4) behavior of ${\cal V}_l$, $P_l$ and ${\cal P}_l$ in Fig.~\ref{P_prime_l_vs_P_tot_diff_SNR_o}, Fig.~\ref{P_d_l_vs_P_tot_diff_SNR_o} and Fig.~\ref{P_cal_l_vs_P_tot_diff_SNR_o}, respectively: they follows water filling power allocation scheme in all regions of $P_{tot}$, and they diverge from uniform power allocation scheme as $P_{tot}$ increases. The behaviors of ${\cal V}_l$, $P_l$ and ${\cal P}_l$ can be justified by the fact that the clusters differ in the observation SNR of the sensors within the clusters. This difference, however, can not be compensated by increasing $P_{tot}$ and thus, it is always more beneficial in terms of minimizing the MSE to allocate a bigger share of $P_{tot}$ to the cluster with larger ${\gamma}_l^o$ (better observation quality) and less share of $P_{tot}$ to the cluster with smaller ${\gamma}_l^o$ (worse observation quality). 
%%%%%%%%%%%%%%%%%%%%%%%%%%%%%%%%%%%%%%%%%%%%%%%%
%
\begin{figure*}[t]
	\centering
	%\hspace{-.8cm}
	\begin{subfigure}[b]{0.25\textwidth}
		\hspace{-2.8cm}
		\centering
		\subcaptionbox{Cluster 1 \label{P_CH_1_comp_P_1k_vs_P_tot_diff_SNR_o}}{\vspace{-.2 cm}\includegraphics[width=1.95in,height=1.1in]{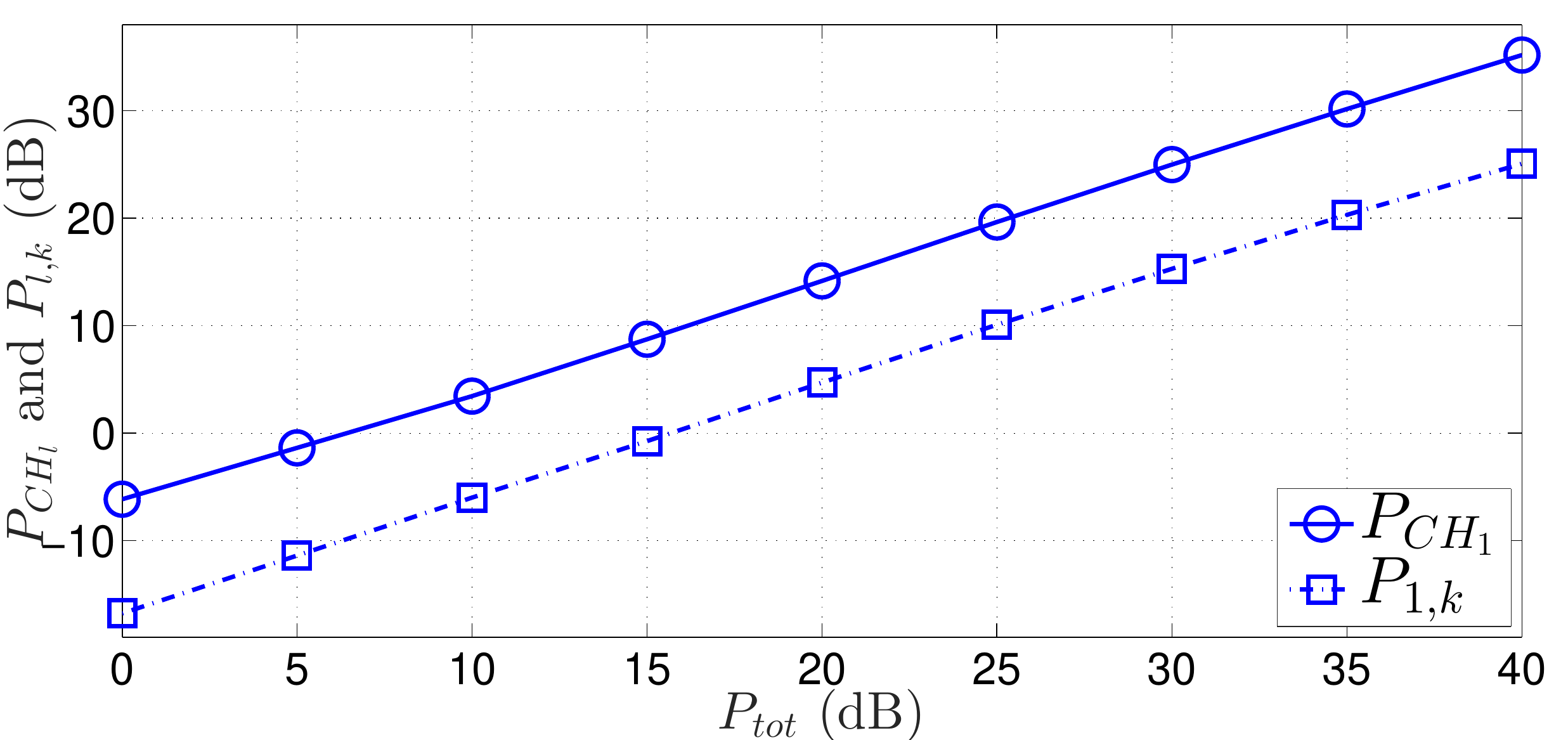}}
	\end{subfigure}%	
	%	
	%\hspace{.4cm}
	\begin{subfigure}[b]{0.25\textwidth}
		\hspace{-1cm}
		\centering
		\subcaptionbox{Cluster 2 \label{P_CH_2_comp_P_2k_vs_P_tot_diff_SNR_o}}{\vspace{-.2 cm}\includegraphics[width=1.95in,height=1.1in]{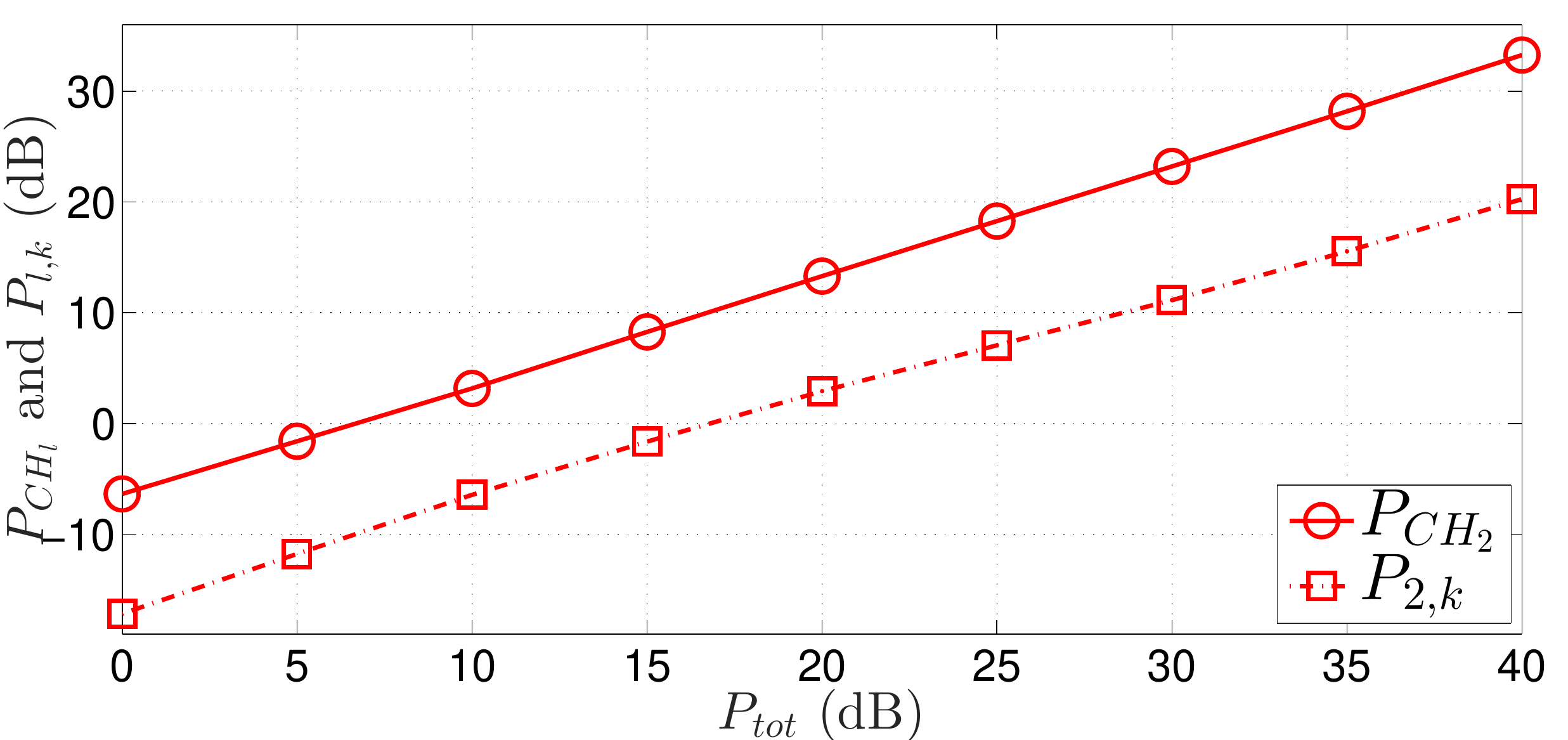}}
	\end{subfigure} 
	%	
	%\hspace{.4cm}
	\begin{subfigure}[b]{0.25\textwidth}
		%\hspace{.6cm}
		\centering
		\subcaptionbox{Cluster 3 \label{P_CH_3_comp_P_3k_vs_P_tot_diff_SNR_o}}{\vspace{-.2 cm}\includegraphics[width=1.95in,height=1.1in]{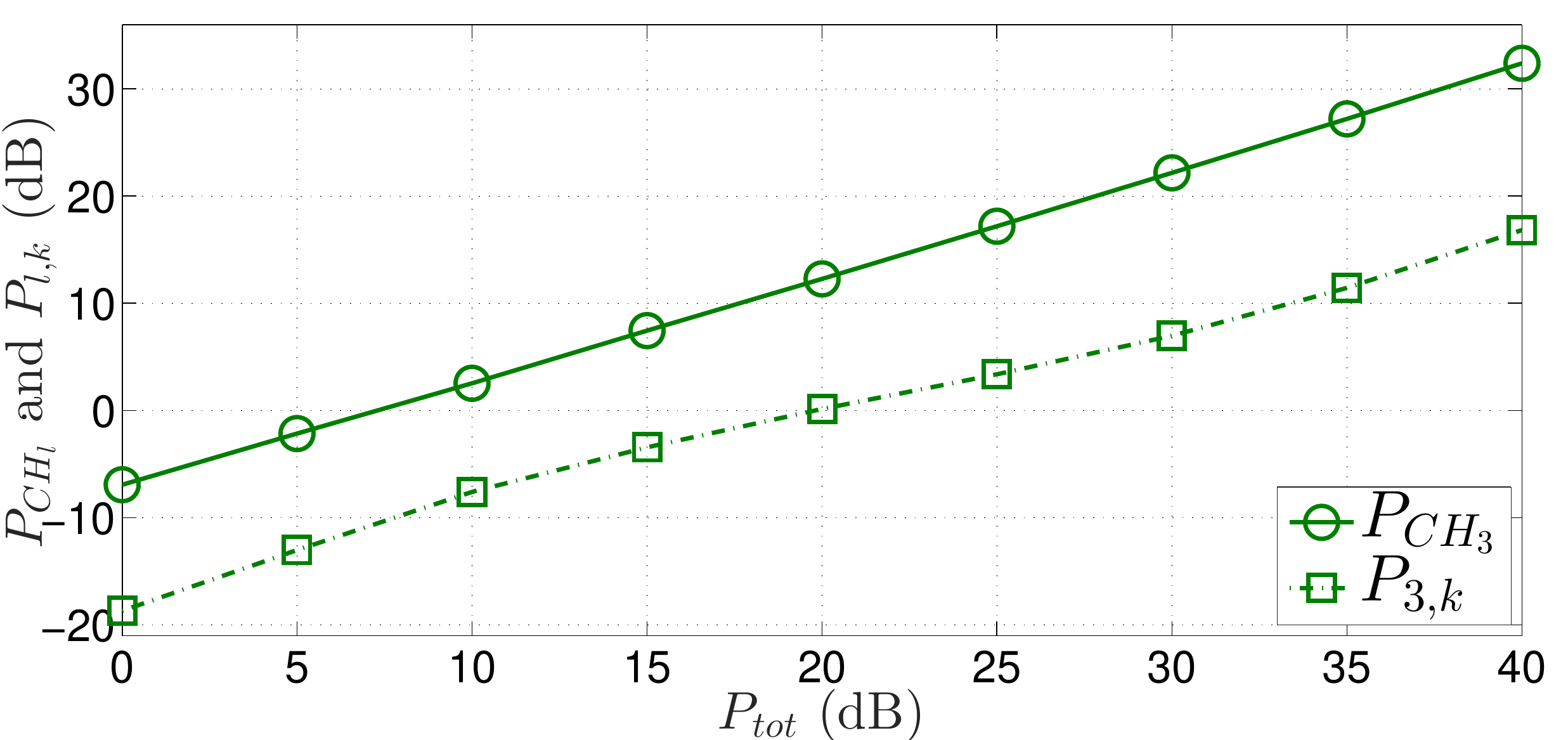}}
	\end{subfigure}
	%
	%\vspace{-.3cm}	
	\caption{{\blue $P_{{\mbox{\tiny CH}}_l}={\cal P}_l+ \psi_l$ and $P_{l,k}=\frac{P_l}{K_l}$ versus $P_{tot}$(dB) when $\{{\gamma}_l^c\!=\!5\ \text{dB}, {\gamma}_l^d\!=\!5\ \text{dB}\}_{l=1}^3$ and ${\gamma}_1^o\!>\!{\gamma}_2^o\!>\!{\gamma}_3^o$.}}   
	\label{P_CH_l_comp_P_lk_vs_P_tot_diff_SNR_o}
\end{figure*}

{\blue One may wonder given our proposed power allocation scheme, how the powers allocated to a CH and a sensor would be different. To answer this question, we let $P_{{\mbox{\tiny CH}}_l}={\cal P}_l+ \psi_l$ denote the sum of power that CH$_l$ consumes for transmitting its fused signal $y_l$ as well as its training symbol to the FC. 
%
%does the power allocation between CH$_l$ and sensor $k$ in cluster $l$ look like'', i.e., how do $P_{{CH}_l}$ and $P_{l,k}$ behave with respect to each other as $P_{tot}$ increases? 
%
Fig.~\ref{P_CH_l_comp_P_lk_vs_P_tot_diff_SNR_o} plots  $P_{{\mbox{\tiny CH}}_l}$ and $P_{l,k}$  versus $P_{tot}$, using the same setup parameters of Fig.~\ref{heterog wrt gamma_o}. We observe that for all clusters $P_{{\mbox{\tiny CH}}_l}\!>>\!P_{l,k}, k=1, ..., K_l$, i.e., the power allocated to each sensor is much smaller than the power allocated to each CH.
}
%=======================================
%=======================================
%=======================================
\vspace{-.4cm}
\section{Conclusions} \label{conclusions}
\vspace{-.1cm}
We studied distributed estimation of a random source in a hierarchical power constrained WSN, where CHs linearly fuse the received signals from sensors within their clusters, and transmit over orthogonal fading channels to the FC. Prior to data transmission, CHs send pilot symbols to the FC to enable channel estimation at the FC. 
We derived the MSE $D$ corresponding to the LMMSE estimator of the source at the FC, and established lower bounds on $D$, including the Bayesian CRB. We addressed constrained minimization of $D$ under the constraint on $P_{tot}$, where the optimization variables are: i) training power $P_{trn}$ and $\{\!\psi_l\!\}_{l=1}^L$, ii) sensor-CH data transmission powers $\{\!P_l\!\}_{l=1}^L$, iii) CH-FC data transmission powers $\{\!{\cal P}_l\!\}_{l=1}^L$. %Through simulations we observed that $D$ and all its established bounds decrease as $P_{tot}$ increases. 
We demonstrated the superior performance of our proposed power allocation scheme, 
%with respect to each optimization variable via comparing it with the scheme that allocates power equally with respect to that optimization variable and optimally with respect to the other two optimization variables. 
comparing with schemes obtained from solving special case problems where subsets of these variables are optimized. 
Our simulations revealed that 
%
%1) $P_{trn}$ and ${\cal P}_l$ optimization are always beneficial for low-region of $P_{tot}$, and $P_l$ optimization is beneficial for low-region to moderate-region of $P_{tot}$, 
%
1) when CNR corresponding to CH$_l$-FC link varies across clusters, $\psi_l, P_l$ allocation follow water filling fashion in all regions of $P_{tot}$, ${\cal P}_l$ follows (inverse of) water filling fashion in (high-region) low-region of $P_{tot}$, 
2) when CNR corresponding to sensors-CH$_l$ links varies across clusters, $P_l$ allocation follows (inverse of) water filling fashion in (high-region) low-region of $P_{tot}$, ${\cal P}_l$ allocation follows water filling fashion in all regions of $P_{tot}$, 
3) when observation SNR varies across clusters, both $P_l, {\cal P}_l$ allocation follow water filling fashion in all regions of $P_{tot}$, and they diverge from uniform power allocation scheme as $P_{tot}$ increases.
{\blue Leveraging on this work, we discuss three future research directions as follows. First direction is considering a coherent multiple access channel model (instead of orthogonal channels) for intra-cluster communication, where sensors within a cluster transmit their amplified measurements to their CH simultaneously. %
%Second is  investigations could consider collaboration among sensors within each cluster before they send their observations to the CH, and the energy consumption caused by sensor collaboration. It would be interesting to examine the optimal power allocation between collaboration, channel estimation, and message transmission (from CHs to the FC). 
%
Second direction is exploring distributed estimation of a random vector source with correlated components. Similar to our work, all sensors can make noisy measurements of a common vector source, or sensors of different clusters can make partial observations of the vector source. Third direction is studying a system where the FC is equipped with multiple antennas (MIMO system model).   

%the extension of our proposed power allocation scheme to the vector parameter case, i.e. the parameter to be estimated is a vector instead of a scalar, and the elements of the random vector parameter could be mutually independent or correlated, or each cluster can only observe one element of the random vector that is correlated with the elements being observed by the other clusters. 
%In this case, we can treat the scalar parameters individually and place individual power constraint on each scalar parameter.
}
%=======================================
%=======================================
%=======================================
\vspace{-.2cm}
\appendix
\vspace{-.1cm}
\subsection{Derivation of {\red Bayesian} CRB}\label{deriving crb}
\vspace{-.1cm}
 %%%%%%%%%%%%%%%%%%%%%%%%%%%%%%%%%%%%
Using the Bayes' rule $f(\boldsymbol{z},\hat{\boldsymbol{h}},\theta)\!=\!f(\boldsymbol{z},\hat{\boldsymbol{h}}|\theta)f(\boldsymbol{\theta})$, we can decompose $G$ into two terms \cite{Van_Trees_Bayesian_bnd_book}:
\vspace{-.2cm}
\begin{align} \label{FI in decomposed form}
G\!=\!\mathbb{E}\{\underbrace{-\frac{\partial^2 \ln f(\theta)}{\partial \theta^2}}_{=G_1(\theta)}\}\!+\!\mathbb{E}\{\underbrace{-\mathbb{E}\{\frac{\partial^2 \ln f(\boldsymbol{z},\hat{\boldsymbol{h}}|\theta)}{\partial \theta^2}\}}_{=G_2(\theta)}\},
\vspace{-.1cm}
\end{align}
%
%Therefore, $\boldsymbol{J}$ is decomposed into two terms:
%
%\begin{equation} \label{FIM in decomposed form}
%\boldsymbol{J}=\mathbb{E}\{\boldsymbol{\Omega}(\boldsymbol{\theta})\}
%+\mathbb{E}\{\boldsymbol{\Lambda}(\boldsymbol{\theta})\},
%\end{equation}
%
in which the outer expectations are taken over the pdf of $\theta$, denoted as $f(\theta)$. Note that $\mathbb{E}\{G_1(\theta)\}$ depends on $f(\theta)$ \cite{Vosoughi2006sp2}. For instance, if $\theta$ is Gaussian with variance $\sigma^2_{\theta}$, we obtain $\mathbb{E}\{G_1(\theta)\}\!=\!\sigma^{-2}_{\theta}$. 
Since $\hat{\boldsymbol{h}}$ and $\theta$ are independent, the Bayes' rule says $f(\boldsymbol{z},\hat{\boldsymbol{h}}|\theta)\!=\!f(\boldsymbol{z}|\hat{\boldsymbol{h}},\theta)f(\hat{\boldsymbol{h}})$, and we can rewrite $G_2(\theta)\!=\!-\mathbb{E}\{\mathbb{E}\{\frac{\partial^2 \ln f(\boldsymbol{z}|\hat{\boldsymbol{h}},\theta)}{\partial \theta^2}\big{|}\hat{\boldsymbol{h}}\}\}$, where the outer and inner expectations are taken over the pdfs $f(\hat{\boldsymbol{h}})$ and $f(\boldsymbol{z}|\hat{\boldsymbol{h}},\theta)$, respectively. 
We note that $G_2(\theta)$ depends on the parameters of the observation model at the sensors as well as the physical layer parameters corresponding to sensors-CHs and CHs-FC links. One can show that $z_l$'s conditioned on $\hat{\boldsymbol{h}},\theta$ are independent, i.e., $f(\boldsymbol{z}|\hat{\boldsymbol{h}},\theta)=\prod_{l=1}^{L}f(z_l|\hat{h}_l,\theta)$. Moreover, since channel estimation is performed independently for each cluster, we have $f(\hat{\boldsymbol{h}})\!=\!\prod_{l=1}^{L}f(\hat{h}_l)$. Hence $G_2(\theta)$ becomes:
\vspace{-0.2cm}
\begin{align*} 
G_2(\theta)\!&=\!-\!\int_{\hat{\boldsymbol{h}}}\int_{\boldsymbol{z}}\!\{\sum\limits_{l=1}^{L}[\frac{\partial^2 f(z_l|\hat{h}_l,\theta)}{\partial \theta^2}\!-\!\frac{1}{f(\!z_l|\hat{h}_l,\theta)}{(\!\frac{\partial f(\!z_l|\hat{h}_l,\theta)}{\partial \theta}\!)}^2\!]\notag\\
\!&\times f(\hat{h}_l)\}\prod_{\underset{i\neq l}{i=1}}^{L}f(z_i|\hat{h}_i,\theta)f(\hat{h}_i)d\boldsymbol{z}d\hat{\boldsymbol{h}}.
\vspace{-.1cm}
\end{align*}
Using the following two facts:
\vspace{-0.3cm}
\begin{align*} %\label{two fact for deriving FIM}
&\int\limits_{\hat{h}_1}\!\dots\!\int\limits_{\hat{h}_{l-1}}\int\limits_{\hat{h}_{l+1}}\!\dots\!\int\limits_{\hat{h}_L}\int\limits_{z_1}\!\dots\!\int\limits_{z_{l-1}}\int\limits_{z_{l+1}}\!\dots\!\int\limits_{z_L}\prod_{\underset{i\neq l}{i=1}}^{L}\!\!f(z_i|\hat{h}_i,\theta)f(\hat{h}_i)\times\\
&dz_1\dots dz_{l-1}dz_{l+1}\dots dz_{L}d\hat{h}_1\dots d\hat{h}_{l-1}d\hat{h}_{l+1}\dots d\hat{h}_{L}=1,
\end{align*}
\vspace{-.5cm}
\begin{align*}
\!\!\!\!\sum_{l=1}^{L}\int_{z_l}\frac{\partial^2 f(z_l|\hat{h}_l,\theta)}{\partial \theta^2}dz_l=\sum_{l=1}^{L}\frac{\partial^2}{\partial \theta^2}(\underbrace{\int_{z_l}f(z_l|\hat{h}_l,\theta)}_{=1})=0,
\vspace{-.1cm}
\end{align*}
%
%\vspace{-0.1cm}
we find that $G_2(\theta)$ reduces to \eqref{G_2(theta) distributed}.
%
%\vspace{-0.25cm}
%\begin{equation} \label{G_2(theta) distributed}
%G_2(\theta)=\sum_{l=1}^{L}\int_{\hat{h}_l}\int_{z_l}\frac{f(\hat{h}_l)}{f(z_l|\hat{h}_l,\theta)}{(\frac{\partial f(z_l|\hat{h}_l,\theta)}{\partial \theta})}^2dz_ld\hat{h}_l.
%\end{equation}
%
Examining \eqref{G_2(theta) distributed} we realize that we need to find two terms in order to fully characterize $G_2(\theta)$: the conditional pdf $f(z_l|\hat{h}_l,\theta)$, and its first derivative with respect to $\theta$, $\partial f(z_l|\hat{h}_l,\theta) / \partial \theta$. In the following, we derive these two terms. 
Using \eqref{z_vec given h_hat} we can write the received signal at the FC from CH$_l$ as:
\vspace{-.2cm}
\begin{equation} \label{z_l to obtain FI}
z_l=(\underbrace{\hat{h}_l+\tilde{h}_l}_{=u_{1_l}})\underbrace{{\boldsymbol{w}_l}^T(\sqrt{\boldsymbol{A}_l}(\theta\boldsymbol{1}_l+\boldsymbol{n}_l)+\boldsymbol{q}_l)}_{=u_{2_l}}+v_l.
\end{equation}
in which $u_{1_l}, u_{2_l}, v_l$ are mutually independent conditioned on $\hat{h}_l, \theta$.  Let $\bar{z}_l\!=\!u_{1_l}u_{2_l}$. Hence, $z_l\!=\!\bar{z}_l+v_l$. Next, we find the conditional pdf of $\bar{z}_l$, conditioned on $\hat{h}_l, \theta$. Considering \eqref{pilot symbol at the FC}, we note that $h_l$, $\nu_l$ are zero-mean independent complex Gaussian, and hence from \eqref{channel estimate and channel estimation error variance} we find that $\hat{h}_l$ is also a zero-mean complex Gaussian. Since $h_l\!=\!\hat{h}_l\!+\!\tilde{h}_l$, we have $\tilde{h}_l\!\sim\!\mathcal{CN}\left(0,\zeta_l^2\right)$. Also, $u_{1_l}\!\sim\!\mathcal{CN}\left(\hat{h}_l,\zeta_l^2\right)$ and $u_{2_l}\!\sim\!\mathcal{N}\left(\bar{\mu}_l,\bar{\sigma}_l^2\right)$ in \eqref{z_l to obtain FI}, where $\bar{\mu}_l\!=\!\theta{\boldsymbol{w}_l}^T\sqrt{\boldsymbol{A}_l}\boldsymbol{1}_l$, $\bar{\sigma}_l^2\!=\!{\boldsymbol{w}_l}^T(\sqrt{\boldsymbol{A}_l}\boldsymbol{\Sigma}_{n_l}\sqrt{\boldsymbol{A}_l}+\boldsymbol{\Sigma}_{q_l})\boldsymbol{w}_l$. 
%The pdf of a complex random variable $X$ is defined as the joint pdf of its real $\mathfrak{R}_X$ and its imaginary $\mathfrak{I}_X$ parts \cite{proakis_book}. 
%Recall $v_l\!\sim\!\mathcal{CN}\left(0,2\sigma_{v_l}^2\right)$. Hence $f(v_l)\!=\!f(v_{l_r},v_{l_i})\!=\!\frac{1}{(2\pi\sigma_{v_l}^2)}\exp{(-\frac{v_{l_r}^2+v_{l_i}^2}{2\sigma_{v_l}^2})}$. 
%Consider $Z\!=\!XY$ when $X\!\sim\!\mathcal{CN}\left(\mu_xe^{j\phi_x},\sigma_x^2\right)$ and $Y\!\sim\!\mathcal{CN}\left(\mu_ye^{j\phi_y},\sigma_y^2\right)$ are independent complex Gaussian random variables. \cite{pdf_of_prod_of_two_complex_Gauss} derived the joint pdf of $\mathfrak{R}_Z$ and $\mathfrak{I}_Z$ as (cf. equation (77) in \cite{pdf_of_prod_of_two_complex_Gauss}):
{\blue 
To find the conditional pdf of $\bar{z}_l$ we use the following lemma from \cite{pdf_of_prod_of_two_complex_Gauss}.
\vspace{-.1cm}
\begin{lemma}\label{lemma for pdf of Z=XY}
\textnormal{
If $X\!\sim\!\mathcal{CN}\left(\mu_xe^{j\phi_x},\sigma_x^2\right)$ and $Y\!\sim\!\mathcal{CN}\left(\mu_ye^{j\phi_y},\sigma_y^2\right)$ are independent complex Gaussian random variables, the pdf of $Z\!=\!XY$ (which is equal to the joint pdf of its real and imaginary parts) is:
\vspace{-.2cm}
\begin{align}\label{pdf of prod of two Gauss}
f(Z)&\!=\!f(z_r,z_i)\!=\!\frac{2}{\pi\sigma_x^2\sigma_y^2}e^{-(k_x^2+k_y^2)}\\
&\!\times\!\sum_{m=0}^{\infty}\sum_{n=0}^{m}\sum_{p=0}^{m-n}\frac{{(2\cos(\angle Z-\phi_x-\phi_y))}^{m-n-p}}{m!n!p!(m-n-p)!}\nonumber\\
&\!\times\!{(\frac{|Z|k_xk_y}{\sigma_x\sigma_y})}^m{(\frac{k_x}{k_y})}^{n-p}K_{n-p}(\frac{2|Z|}{\sigma_x\sigma_y}),\nonumber
\vspace{-.1cm}
\end{align}
where $k_x\!=\!\mu_x/\sigma_x, k_y\!=\!\mu_y/\sigma_y, |Z|\!=\!\sqrt{z_r^2+z_i^2}, \angle Z\!=\!\arctan(z_i/z_r)$, and $K_r(x)$ is the modified Bessel function of the second kind with order $r$ and argument $x$.
}
\end{lemma}
}
Therefore, we can write the conditional joint pdf $f(\bar{z}_{l_r}, \bar{z}_{l_i}|\hat{h}_l,\theta)$ using \eqref{pdf of prod of two Gauss}. Recall $v_l\!\sim\!\mathcal{CN}\left(0,2\sigma_{v_l}^2\right)$. Hence $f(v_l)\!=\!f(v_{l_r},v_{l_i})\!=\!\frac{1}{(2\pi\sigma_{v_l}^2)}\exp{(-\frac{v_{l_r}^2+v_{l_i}^2}{2\sigma_{v_l}^2})}$. Since $\bar{z}_l$ and $v_l$ are independent, the conditional joint pdf $f(z_{l_r},z_{l_i}|\hat{h}_l,\theta)$ is computed as $f(z_{l_r},z_{l_i}|\hat{h}_l,\theta)\!=\!f(\bar{z}_{l_r}, \bar{z}_{l_i}|\hat{h}_l,\theta)*f(v_{l_r},v_{l_i})$, in which $*$ is the operator for two-dimensional convolution. Substituting for $f(\bar{z}_{l_r}, \bar{z}_{l_i}|\hat{h}_l,\theta), f(v_{l_r},v_{l_i})$ from above and defining $b=|b|e^{j\angle b}$, after some mathematical manipulations, we reach 
{\blue 
$f(z_l|\hat{h}_l,\theta)$ and $\frac{\partial f(\!z_l|\hat{h}_l,\theta)}{\partial \theta}$ in \eqref{f(z_l|h_hat_l,theta)} and \eqref{der wrt theta f(z_l|h_hat_l,theta)}, respectively, whose parameters are defined in \eqref{parameters for f and der f}. 
}
Substituting \eqref{f(z_l|h_hat_l,theta)} and \eqref{der wrt theta f(z_l|h_hat_l,theta)} in \eqref{G_2(theta) distributed}, we compute $G_2(\theta)$.
%%%%%%%%%%%%%%%%%%%%%%%%%%%%%%%%
%\vspace{-.1cm}
\subsection{Proof of Proposition \ref{simplified w_l and F_l in term of P_l}: Finding ${\boldsymbol{w}}^{opt}_l, {\cal F}^{opt}_l$ in terms of $P_l$} \label{finding w_opt and J_opt}
%\vspace{-.1cm}
According to \eqref{simplified eig problem 1}, the only non-zero eigenvalue of $\boldsymbol{\cal B}_1$ and its corresponding eigenvector are: 
%
%\vspace{-.2cm}
%\begin{subequations}
\begin{equation} \label{J_opt and s_opt in terms of P_l and v_l}
{\cal F}^{opt}_l={|\boldsymbol{\mu}_l|}^T{\boldsymbol{\cal B}_1}^{-1}|\boldsymbol{\mu}_l|,%\label{J_opt after rank-1 prop}\\
\ \ {\boldsymbol{s}}^{opt}_l={\boldsymbol{\cal B}_1}^{-1}|\boldsymbol{\mu}_l|.%\label{s_opt after rank-1 prop}
\end{equation}
%\end{subequations}
%
Define $\delta_l\!=\!{\cal V}_l-P_l,\ \xi_l\!=\!\frac{\sigma^2_{\theta}}{{|\hat{h}_l|}^2}(\frac{\sigma_{v_{l}}^2}{\delta_l}+\zeta_l^2),\ \boldsymbol{\Sigma}_{\mu_l}\!=\!|\boldsymbol{\mu}_l|{|\boldsymbol{\mu}_l|}^T,\ \phi_l\!=\!{|\hat{h}_l|}^2+\zeta_l^2+\frac{\sigma_{v_{l}}^2}{\delta_l},\ \boldsymbol{\Sigma}_{P_l}\!=\!\boldsymbol{\Sigma}_{q_l}\!+\!P_l\boldsymbol{\Delta}_l,\ \boldsymbol{\Sigma}_{\phi_l}\!=\!\phi_l\boldsymbol{\Sigma}_{P_l}$. 
By substituting $\boldsymbol{\Delta}_l, \boldsymbol{\Pi}_l$ into $\boldsymbol{\Omega}_l$ and $\boldsymbol{B}_l$, and $\boldsymbol{\Omega}_l$ into $\boldsymbol{R}_{t_l}$, 
%and finally $\boldsymbol{R}_{t_l}$ and $\boldsymbol{B}_l$ in \eqref{J_opt and s_opt in terms of P_l and v_l}, 
$\boldsymbol{\cal B}_1$ in \eqref{eig problem} becomes $\boldsymbol{\cal B}_1\!=\!\boldsymbol{\Sigma}_{\phi_l}+\xi_l\boldsymbol{\Sigma}_{\mu_l}$.
%
%\begin{equation*}
%\frac{\sigma_{v_{l}}^2\boldsymbol{R}_{t_l}}{\delta_l}+({|\hat{h}_l|}^2+\zeta_l^2)\boldsymbol{\Sigma}_{q_l}+P_l\boldsymbol{B}_l=\boldsymbol{\Sigma}_{\phi_l}+\xi_l\boldsymbol{\Sigma}_{\mu_l}.
%\end{equation*}
%
Using the Binomial inversion Lemma \cite{Matrix_Analysis} we compute ${\boldsymbol{s}}^{opt}_l$ in \eqref{J_opt and s_opt in terms of P_l and v_l}:
%
%\vspace{-.1cm}
\begin{equation} \label{s_l^opt after binomial inversion}
{\boldsymbol{s}}^{opt}_l=\frac{\boldsymbol{\Sigma}_{\phi_l}^{-1}|\boldsymbol{\mu}_l|}{1+\xi_l{|\boldsymbol{\mu}_l|}^T\boldsymbol{\Sigma}_{\phi_l}^{-1}|\boldsymbol{\mu}_l|}.
\end{equation}
%
%Recall ${\boldsymbol{w}}^{opt}_l\!=\!r_l{\boldsymbol{s}}^{opt}_l$, where the scalar $r_l$ is such that \eqref{active constraint P6 1} is satisfied. Therefore using 
From \eqref{s_l^opt after binomial inversion}, we obtain ${\boldsymbol{w}}^{opt}_l$:
%
%\vspace{-.15cm}
\begin{align} \label{w_l_opt}
&{\boldsymbol{w}}^{opt}_l\!=\!\sqrt{\frac{\delta_l}{{|\boldsymbol{\mu}_l|}^T\boldsymbol{\Sigma}_{\phi_l}^{-1}\boldsymbol{R}_{t_l}\boldsymbol{\Sigma}_{\phi_l}^{-1}|\boldsymbol{\mu}_l|}}\boldsymbol{\Sigma}_{\phi_l}^{-1}|\boldsymbol{\mu}_l|\\
&\!\overset{(a)}{=}\!\sqrt{\frac{\delta_l}{\boldsymbol{\rho}_l^T\boldsymbol{\Sigma}_{P_l}^{-1}\boldsymbol{\rho}_l(1+\sigma^2_{\theta}P_l\boldsymbol{\rho}_l^T\boldsymbol{\Sigma}_{P_l}^{-1}\boldsymbol{\rho}_l)}}\boldsymbol{\Sigma}_{P_l}^{-1}\boldsymbol{\rho}_l\!\overset{(b)}{=}\!\sqrt{\frac{\delta_l}{\tau_l}}{\boldsymbol{R}_{t_l}}^{-1}\boldsymbol{\rho}_l,\nonumber
\end{align}
%
%It is straightforward to rewrite $\boldsymbol{R}_{t_l}$ as $\boldsymbol{R}_{t_l}\!=\!\boldsymbol{\Sigma}_{P_l}\!+\!\frac{\sigma^2_{\theta}}{{|\hat{h}_l|}^2}\boldsymbol{\Sigma}_{\mu_l}$,
where $\tau_l$ is defined in Proposition \ref{simplified w_l and F_l in term of P_l}. To obtain $(a)$ in \eqref{w_l_opt}, we use the fact that ${|\boldsymbol{\mu}_l|}^T\boldsymbol{\Sigma}_{\phi_l}^{-1}\boldsymbol{R}_{t_l}\boldsymbol{\Sigma}_{\phi_l}^{-1}|\boldsymbol{\mu}_l|\!=\!\frac{\epsilon_l}{\phi_l^2}(1+\frac{\sigma^2_{\theta}}{{|\hat{h}_l|}^2}\epsilon_l)$,
%
%\begin{equation*}
%\boldsymbol{R}_{t_l}=\boldsymbol{\Sigma}_{P_l}+\frac{\sigma^2_{\theta}}{{|\hat{h}_l|}^2}\boldsymbol{\Sigma}_{\mu_l}.
%\end{equation*}
%
%and therefore, some mathematical manipulations yield:
%
%\begin{equation} \label{denominator to find w_l_opt}
%{|\boldsymbol{\mu}_l|}^T\boldsymbol{\Sigma}_{\phi_l}^{-1}\boldsymbol{R}_{t_l}\boldsymbol{\Sigma}_{\phi_l}^{-1}|\boldsymbol{\mu}_l|=\frac{\epsilon_l}{\phi_l^2}(1+\frac{\sigma^2_{\theta}}{{|\hat{h}_l|}^2}\epsilon_l),
%\end{equation}
%
where $\epsilon_l\!=\!{|\boldsymbol{\mu}_l|}^T\boldsymbol{\Sigma}_{P_l}^{-1}|\boldsymbol{\mu}_l|$. 
%Substituting \eqref{denominator to find w_l_opt} in \eqref{w_l_opt} and using $\boldsymbol{\Sigma}_{\phi_l}\!=\!\phi_l\boldsymbol{\Sigma}_{P_l}, \boldsymbol{\mu}_l\!=\!\sqrt{P_l}\hat{h}_l\boldsymbol{\rho}_l$ result in:
%
%\begin{equation*}
%{\boldsymbol{w}}^{opt}_l=\sqrt{\frac{\delta_l}{\epsilon_l(1+\frac{\sigma^2_{\theta}}{{|\hat{h}_l|}^2}\epsilon_l)}}\boldsymbol{\Sigma}_{P_l}^{-1}|\boldsymbol{\mu}_l|.
%\end{equation*}
%
%Since $\boldsymbol{\mu}_l\!=\!\sqrt{P_l}\hat{h}_l\boldsymbol{\rho}_l$, we get:
%
%\begin{equation} \label{simplified w_opt}
%{\boldsymbol{w}}^{opt}_l=\sqrt{\frac{\delta_l}{\boldsymbol{\rho}_l^T\boldsymbol{\Sigma}_{P_l}^{-1}\boldsymbol{\rho}_l(1+\sigma^2_{\theta}P_l\boldsymbol{\rho}_l^T\boldsymbol{\Sigma}_{P_l}^{-1}\boldsymbol{\rho}_l)}}\boldsymbol{\Sigma}_{P_l}^{-1}\boldsymbol{\rho}_l.
%\end{equation}
%
%Using the Binomial Inversion Lemma, the following equality holds:
To obtain $(b)$ in \eqref{w_l_opt}, we use ${\boldsymbol{R}_{t_l}}^{-1}\boldsymbol{\rho}_l\!=\!\frac{\boldsymbol{\Sigma}_{P_l}^{-1}\boldsymbol{\rho}_l}{1+\sigma^2_{\theta}P_l\boldsymbol{\rho}_l^T\boldsymbol{\Sigma}_{P_l}^{-1}\boldsymbol{\rho}_l}${\blue, which is established using the Binomial inversion lemma}.
%
%\begin{equation} \label{relation between Q_inv and sigma_P_inv}
%{\boldsymbol{R}_{t_l}}^{-1}\boldsymbol{\rho}_l=\frac{\boldsymbol{\Sigma}_{P_l}^{-1}\boldsymbol{\rho}_l}{1+\sigma^2_{\theta}P_l\boldsymbol{\rho}_l^T\boldsymbol{\Sigma}_{P_l}^{-1}\boldsymbol{\rho}_l}.
%\end{equation}
%
%Defining $\tau_l\!=\!{\boldsymbol{\rho}_l}^T{\boldsymbol{R}_{t_l}}^{-1}\boldsymbol{\rho}_l$ and using \eqref{relation between Q_inv and sigma_P_inv} in \eqref{simplified w_opt}, the optimal fusion vector ${\boldsymbol{w}}^{opt}_l$ is computed as:
%
%\begin{equation*}
%{\boldsymbol{w}}^{opt}_l=\sqrt{\frac{\delta_l}{\tau_l}}{\boldsymbol{R}_{t_l}}^{-1}\boldsymbol{\rho}_l.
%\end{equation*}
%
%To compute ${\cal F}^{opt}_l$ in \eqref{J_opt and s_opt in terms of P_l and v_l}, since 
We have ${\cal F}^{opt}_l\!=\!{|\boldsymbol{\mu}_l|}^T{\boldsymbol{s}}^{opt}_l$. Substituting ${\boldsymbol{s}}^{opt}_l$ from \eqref{s_l^opt after binomial inversion} in \eqref{J_opt and s_opt in terms of P_l and v_l} and using the fact that $1\!-\!\sigma^2_{\theta}P_l\tau_l\!=\!\frac{1}{1+\sigma^2_{\theta}P_l\boldsymbol{\rho}_l^T\boldsymbol{\Sigma}_{P_l}^{-1}\boldsymbol{\rho}_l}$ we reach:
%
%\begin{equation} \label{F_l^opt using simplified s_l^opt}
%{\cal F}^{opt}_l=\frac{{|\boldsymbol{\mu}_l|}^T\boldsymbol{\Sigma}_{\phi_l}^{-1}|\boldsymbol{\mu}_l|}{1+\xi_l{|\boldsymbol{\mu}_l|}^T\boldsymbol{\Sigma}_{\phi_l}^{-1}|\boldsymbol{\mu}_l|}.
%\end{equation}
%
%Defining $\beta_l\!=\!\frac{\sigma_{v_{l}}^2}{{|\hat{h}_l|}^2(1-\sigma^2_{\theta}P_l\tau_l)+\zeta_l^2}$ and substituting $\boldsymbol{\mu}_l, \xi_l, \boldsymbol{\Sigma}_{\phi_l}$ in \eqref{F_l^opt using simplified s_l^opt},
%
%\begin{equation} \label{simplified J_L_opt}
%{\cal F}^{opt}_l=\frac{{|\hat{h}_l|}^2P_l\boldsymbol{\rho}_l^T\boldsymbol{\Sigma}_{P_l}^{-1}\boldsymbol{\rho}_l}{{|\hat{h}_l|}^2+(\zeta_l^2+\frac{\sigma_{v_{l}}^2}{\delta_l})(1+\sigma^2_{\theta}P_l\boldsymbol{\rho}_l^T\boldsymbol{\Sigma}_{P_l}^{-1}\boldsymbol{\rho}_l)}.
%\end{equation}
%
%The following fact can be easily obtained from \eqref{relation between Q_inv and sigma_P_inv}:
%
%\begin{equation} \label{simplifying fact to obtain J_opt}
%1-\sigma^2_{\theta}P_l\tau_l=\frac{1}{1+\sigma^2_{\theta}P_l\boldsymbol{\rho}_l^T\boldsymbol{\Sigma}_{P_l}^{-1}\boldsymbol{\rho}_l}.
%\end{equation}
%
%By using \eqref{simplifying fact to obtain J_opt} in \eqref{simplified J_L_opt}, 
%the optimal objective value ${\cal F}^{opt}_l$ is obtained as:
%
{\blue
\begin{align} \label{final formula for F_l^opt}
{\cal F}^{opt}_l&=\frac{{|\boldsymbol{\mu}_l|}^T\boldsymbol{\Sigma}_{\phi_l}^{-1}|\boldsymbol{\mu}_l|}{1+\xi_l{|\boldsymbol{\mu}_l|}^T\boldsymbol{\Sigma}_{\phi_l}^{-1}|\boldsymbol{\mu}_l|}\\
&=\frac{{|\hat{h}_l|}^2P_l\tau_l}{{|\hat{h}_l|}^2(1-\sigma^2_{\theta}P_l\tau_l)+\zeta_l^2+\frac{\sigma_{v_{l}}^2}{\delta_l}}\!=\!\frac{{|\hat{h}_l|}^2\beta_lP_l\tau_l}{\sigma_{v_{l}}^2(1+\frac{\beta_l}{\delta_l})},\nonumber
\end{align}
}
%
%in which ($a$) in \eqref{final formula for F_l^opt} follows from the fact $1-\sigma^2_{\theta}P_l\tau_l=\frac{1}{1+\sigma^2_{\theta}P_l\boldsymbol{\rho}_l^T\boldsymbol{\Sigma}_{P_l}^{-1}\boldsymbol{\rho}_l}$.
%where we defined $\beta_l\!=\!\frac{\sigma_{v_{l}}^2}{{|\hat{h}_l|}^2(1-\sigma^2_{\theta}P_l\tau_l)+\zeta_l^2}$.
%%%%%%%%%%%%%%%%%%%%%%%%%%%%%%%%%%%%%
%\vspace{-.2cm}
\subsection{Solution of the Problem in \eqref{simplified max problem P5-2}} \label{finding delta_opt}
%\vspace{-.1cm}
%Let $\beta_l\!=\!\frac{\sigma_{v_{l}}^2}{{|\hat{h}_l|}^2(1-\sigma^2_{\theta}P_l\tau_l)+\zeta_l^2}$. 
%The first-order derivative of the function $T$ in \eqref{simplified max problem P5-2} with respect to $\delta_l$ is 
Define $\delta_l\!=\!{\cal V}_l-P_l$ and let $T$ denote the objective function in \eqref{simplified max problem P5-2}. We have $\frac{\partial T}{\partial \delta_l}\!=\!\frac{{|\hat{h}_l|}^2\beta_l^2P_l\tau_l}{\sigma_{v_{l}}^2{(\beta_l+\delta_l)}^2}\!>\!0$, 
%which is always positive for any $\delta_l$. This tells us that $T$ is an increasing function of $\delta_l$, and thus, the solution of \eqref{simplified max problem P5-2} should be determined such that it satisfies the power constraint with equality, that is, 
implying that the solution to \eqref{simplified max problem P5-2} must satisfy the equality constraint $\sum_{l=1}^{L}\delta_l+P_l\!=\!\sigma P_{tot}$. Also,
%The second-order derivatives of $T$ with respect to $\delta_l$'s are given as follows: 
$\frac{\partial^2 T}{\partial \delta_i\partial \delta_j}\!=\!0, \forall i\!\ne\!j$, and $\frac{\partial^2 T}{\partial \delta_l^2}\!=\!\frac{-2{|\hat{h}_l|}^2\beta_l^2P_l\tau_l}{\sigma_{v_{l}}^2(\beta_l+\delta_l)^3}\!<\!0, \forall l$. Thus the Hessian of $T$ with respect to $\delta_l$'s is diagonal and negative definite, proving that $T$ is jointly concave over $\delta_l$'s. Since the constraint is linear in $\delta_l$, the problem in \eqref{simplified max problem P5-2} is concave. The Lagrangian function $\cal L$ associated with \eqref{simplified max problem P5-2} is:
%
%\vspace{-.2cm}
\begin{equation*} %\label{Lagrangian of P7}
\!\!\mathcal L(\lambda,\{\eta_l,\delta_l\}_{l=1}^L)\!=\!\!\sum_{l=1}^{L}\!\frac{{|\hat{h}_l|}^2\beta_lP_l\tau_l}{\sigma_{v_{l}}^2(1\!+\!\frac{\beta_l}{\delta_l})}-\delta_l\left(\lambda\!-\!\eta_l\right)\!+\!\lambda(\sigma P_{tot}-\!\sum_{l=1}^{L}\!P_l),
\end{equation*}
%
%in \eqref{Lagrangian of P7},
%========single column equation===============================
%\begin{figure*}[b]
%\begin{equation} \label{Lagrangian of P7}
%\mathcal L(\lambda,\{\eta_l,\delta_l\}_{l=1}^L)\!=\!\sum_{l=1}^{L}\!\frac{{|\hat{h}_l|}^2\beta_lP_l\tau_l}{\sigma_{v_{l}}^2\!+\!\frac{\sigma_{v_{l}}^2\beta_l}{\delta_l}}\!-\!\delta_l\left(\lambda\!-\!\eta_l\right)\!+\!\lambda(\sigma P_{tot}\!-\!\sum_{l=1}^{L}\!P_l),
%\end{equation}
%\end{figure*}
%=============================================================
where $\lambda, \eta_l$'s are the Lagrange multipliers. The KKT optimality conditions are:
%
%\vspace{-.15cm}
{\blue
	\vspace{+.05cm}
\begin{subequations} \label{KKT cond. for problem P7}
	\begin{align} 
	&\frac{{|\hat{h}_l|}^2\beta_l^2P_l\tau_l}{\sigma_{v_{l}}^2{(\beta_l+\delta_l)}^2}-\lambda+\eta_l=0,\ \forall l,\label{KKT cond. for problem P7-1}\\
	&\lambda \left(\sum_{l=1}^{L}\delta_l+P_l-\sigma P_{tot}\right)=0,\ \lambda\geq 0,\label{KKT cond. for problem P7-2}\\
	&\eta_l\delta_l=0,\ \eta_l\geq 0,\ \delta_l\geq 0,\ \forall l.\label{KKT cond. for problem P7-3}
	\end{align}
	%\vspace{-.15cm}
\end{subequations}
The condition \eqref{KKT cond. for problem P7-3} implies $\eta_l\!=\!0$ for active clusters with $\delta_l\!>\!0$. From \eqref{KKT cond. for problem P7-1} we infer:
}
\vspace{-.25cm}
\begin{equation} \label{delta_l opt}
\delta_l^{opt}=\left[\beta_l(\frac{|\hat{h}_l|}{\sigma_{v_{l}}}\sqrt{\frac{P_l\tau_l}{\lambda}}-1)\right]^{+},
\vspace{-.15cm}
\end{equation}
in which $\left[x\right]^{+}\!=\!max\{x,0\}$. Having $\delta_l^{opt}$, we find ${\cal V}_l^{opt}\!=\!\delta_l^{opt}\!+\!P_l$ given in \eqref{opt_delta_l}. Substituting \eqref{delta_l opt} in the active constraint condition $\sum_{l=1}^{L}\!\delta_l\!+\!P_l\!=\!\sigma P_{tot}$, the Lagrange multiplier $\lambda$ becomes equal to the expression given in \eqref{opt lambda}, 
%can be computed as:
%
%\begin{equation} \label{lambda for P7}
%\lambda=(\frac{\sum_{l\in{\cal A}}\frac{|\hat{h}_l|\beta_l\sqrt{P_l\tau_l}}{\sigma_{v_{l}}}}{\sigma P_{tot}\!-\!\sum_{l\in{\cal A}}P_l\!+\!\sum_{l\in{\cal A}}\beta_l})^2,
%\end{equation}
% 
in which $\cal A$ is the set of active clusters. To uniquely determine $\cal A$, we carry out the following procedure. Let $L_{\cal A}\!=\!|{\cal A}|$ where $L_{\cal A}\!\leq\!L$. 
%According to \eqref{delta_l opt}, these $\hat{S}$ clusters must be $\{1, 2, ..., \hat{S}\}$ where $\{i\}$ is a set of indices such that 
Suppose the clusters are indexed in the descending order of $\frac{{|\hat{h}_1|}^2P_1\tau_1}{\sigma^2_{v_{1}}}\!\geq\!\frac{{|\hat{h}_2|}^2P_2\tau_2}{\sigma^2_{v_{2}}}\!\geq\!...\!\geq\!\frac{{|\hat{h}_L|}^2P_L\tau_L}{\sigma^2_{v_{L}}}$. 
%Therefore, $\lambda$ can be obtained from \eqref{lambda for P7} where we assumed only first $\hat{S}$ clusters are active. Now we substitute $\lambda$ back into \eqref{delta_l opt} 
Choosing an $L_{\cal A}$ value we find $\lambda$ and 
%use \eqref{delta_l opt} to obtain $\{\delta_1^{opt},\delta_2^{opt},...,\delta_L^{opt}\}$. 
compute $\delta_l^{opt}\!=\!\beta_l(\frac{|\hat{h}_l|}{\sigma_{v_{l}}}\sqrt{\frac{P_l\tau_l}{\lambda}}\!-\!1), \forall l$.
If $\delta_l^{opt}\!>\!0,\ l\!=\!1,...,L_{\cal A}$ and $\delta_l^{opt}\!\leq\!0,\ l\!=\!L_{\cal A}\!+\!1,...,L$, then 
%it is the true solution of optimization problem in \eqref{simplified max problem P5-2} and we conclude ${\cal A}\!=\!\{1,...,\hat{S}\}$; otherwise we have to choose another $\hat{S}$ to repeat the above procedure. Furthermore, it 
we have identified the set of active clusters $\cal A$ with their corresponding $P_l, l\!\in\!{\cal A}$. Otherwise, we repeat this process for another $L_{\cal A}$ value. It is proved that the solution always exists and is unique{\blue\cite{Goldsmith_2006}}. 
%Having $\delta_l^{opt}$, the optimal value ${\cal V}_l^{opt}$ is computed as $\!{\cal V}_l^{opt}\!=\delta_l^{opt}+P_l$.
%%%%%%%%%%%%%%%%%%%%%%%%%%%%%%%%
\vspace{-.25cm}
\subsection{$\!$Proof of Concavity of sub-problem $\!(\!a\!)$ of \eqref{max problem to minimize D_d} over $P_l\!$'s} \label{prove P13_b is concave} 
\vspace{-.15cm}
We rewrite the cost function of sub-problem $(a)$, denoted as ${\cal F}$, as:
\vspace{-.3cm}
\begin{equation} \label{F in P13-c}
{\cal F}=\frac{1}{\sigma^2_{\theta}}\sum_{l=1}^{L}\overbrace{\frac{1}{b_l}(1-\frac{s_l}{s_l+P_lm_l})}^{{\cal F}_l},
\vspace{-.15cm}
\end{equation}
where $b_l\!=\!\frac{1}{{|\hat{h}_l|}^2}(\frac{\sigma_{v_{l}}^2}{{\mathcal P}}\!+\!\zeta_l^2), s_l\!=\!\frac{1+b_l}{\sigma^2_{\theta}b_l}, m_l\!=\!{\boldsymbol{\rho}_l}^T\boldsymbol{\Sigma}_{P_l}^{-1}\boldsymbol{\rho}_l, \boldsymbol{\Sigma}_{P_l}\!=\!\boldsymbol{\Sigma}_{q_l}\!+\!P_l\boldsymbol{\Delta}_l$. We have $b_l, s_l, m_l\!>\!0$ and $\boldsymbol{\Sigma}_{P_l}\!\succ\!\boldsymbol{0}$. 
Also, $\frac{\partial m_l}{\partial P_l}\!=\!-{\boldsymbol{\rho}_l}^T\boldsymbol{\Sigma}_{P_l}^{-1}\boldsymbol{\Delta}_l\boldsymbol{\Sigma}_{P_l}^{-1}\boldsymbol{\rho}_l\!<\!0, \frac{\partial^2 m_l}{\partial P_l^2}\!=\!2{\boldsymbol{\rho}_l}^T\boldsymbol{\Sigma}_{P_l}^{-1}\boldsymbol{\Delta}_l\boldsymbol{\Sigma}_{P_l}^{-1}\boldsymbol{\Delta}_l\boldsymbol{\Sigma}_{P_l}^{-1}\boldsymbol{\rho}_l\!>\!0$. One can obtain $\frac{\partial {\cal F}_l}{\partial P_l}\!=\!\frac{s_l(m_l+P_l\frac{\partial m_l}{\partial P_l})}{b_l(s_l+P_lm_l)^2}$ and prove that $m_l\!+\!P_l\frac{\partial m_l}{\partial P_l}\!>\!0$ 
{\blue which infers $\frac{\partial {\cal F}_l}{\partial P_l}\!>\!0$, i.e.,
\vspace{-.25cm}
\begin{align*} %\label{proof of der F_l wrt P_l is pos}
&\boldsymbol{\Sigma}_{q_l}\!\succ\!\boldsymbol{0}\Rightarrow\boldsymbol{\Sigma}_{P_l}\!\succ\!P_l\boldsymbol{\Delta}_l\Rightarrow\boldsymbol{\Sigma}_{P_l}^{-1}\!\succ\!P_l\boldsymbol{\Sigma}_{P_l}^{-1}\boldsymbol{\Delta}_l\boldsymbol{\Sigma}_{P_l}^{-1}\Rightarrow\\
&{\boldsymbol{\rho}_l}^T\boldsymbol{\Sigma}_{P_l}^{-1}\boldsymbol{\rho}_l-P_l{\boldsymbol{\rho}_l}^T\boldsymbol{\Sigma}_{P_l}^{-1}\boldsymbol{\Delta}_l\boldsymbol{\Sigma}_{P_l}^{-1}\boldsymbol{\rho}_l\!>\!0\Rightarrow m_l\!+\!P_l\frac{\partial m_l}{\partial P_l}\!>\!0.
\vspace{-.15cm}
\end{align*}
${\cal F}$ in \eqref{F in P13-c}
} 
is an increasing function of $P_l$, and thus, the solution of sub-problem $(a)$ of \eqref{max problem to minimize D_d} must satisfy the equality constraint $P_{trn}\!+\!\sum_{l=1}^{L}\{P_l\!+\!{\mathcal P}\}\!=\!P_{tot}$. Furthermore
\vspace{-.2cm}
\begin{equation*}
\frac{\partial^2 {\cal F}_l}{\partial P_l^2}\!=\!\frac{s_l[(s_l+P_lm_l)(2\frac{\partial m_l}{\partial P_l}+P_l\frac{\partial^2 m_l}{\partial P_l^2})-2(m_l+P_l\frac{\partial m_l}{\partial P_l})^2]}{b_l(s_l+P_lm_l)^3}. 
\vspace{-.1cm}
\end{equation*}
The denominator of the right-hand side is positive. The numerator of the right-hand side can be simplified as $\text{num}\!=\!I_1\!+\!I_2\!+\!I_3$, where
\vspace{-.2cm}
\begin{subequations} %\label{numerator of sec der of F_l wrt P_l}
	\begin{align} 
	I_1&=s_l{\boldsymbol{\rho}_l}^T(P_l\boldsymbol{\Sigma}_{P_l}^{-1}\boldsymbol{\Delta}_l\boldsymbol{\Sigma}_{P_l}^{-1}\boldsymbol{\Delta}_l\boldsymbol{\Sigma}_{P_l}^{-1}-\boldsymbol{\Sigma}_{P_l}^{-1}\boldsymbol{\Delta}_l\boldsymbol{\Sigma}_{P_l}^{-1})\boldsymbol{\rho}_l,\nonumber\\ %\label{I_1}\\
	I_2&={\boldsymbol{\rho}_l}^T(P_l\boldsymbol{\Sigma}_{P_l}^{-1}\boldsymbol{\Pi}_l\boldsymbol{\Sigma}_{P_l}^{-1}\boldsymbol{\Delta}_l\boldsymbol{\Sigma}_{P_l}^{-1}-\boldsymbol{\Sigma}_{P_l}^{-1}\boldsymbol{\Pi}_l\boldsymbol{\Sigma}_{P_l}^{-1})\boldsymbol{\rho}_l,\nonumber\\ %\label{I_2}\\
	I_3&=P_l^2{\boldsymbol{\rho}_l}^T(\boldsymbol{\Sigma}_{P_l}^{-1}\boldsymbol{\Pi}_l\boldsymbol{\Sigma}_{P_l}^{-1}\boldsymbol{\Delta}_l\boldsymbol{\Sigma}_{P_l}^{-1}\boldsymbol{\Delta}_l\boldsymbol{\Sigma}_{P_l}^{-1}\nonumber\\
	&-\boldsymbol{\Sigma}_{P_l}^{-1}\boldsymbol{\Delta}_l\boldsymbol{\Sigma}_{P_l}^{-1}\boldsymbol{\Pi}_l\boldsymbol{\Sigma}_{P_l}^{-1}\boldsymbol{\Delta}_l\boldsymbol{\Sigma}_{P_l}^{-1})\boldsymbol{\rho}_l.\nonumber %\label{I_3}
	\end{align}
	%\vspace{-.1cm}
\end{subequations}
One can prove that $I_1\!<\!0, I_2\!<\!0, I_3\!=\!0$. Hence, $\text{num}\!<\!0$ and $\frac{\partial^2 {\cal F}_l}{\partial P_l^2}\!<\!0$. 
{\blue The following sequences of inequalities are easy to verify:
\vspace{-.2cm}
\begin{subequations} %\label{proof of neg for I1, I2 and zero for I3}
	\begin{align} 
	&\boldsymbol{\Sigma}_{q_l}\!\succ\!\boldsymbol{0}\Rightarrow\boldsymbol{\Sigma}_{P_l}\!\succ\!P_l\boldsymbol{\Delta}_l\Rightarrow\boldsymbol{I}\!\succ\!P_l\boldsymbol{\Sigma}_{P_l}^{-1}\boldsymbol{\Delta}_l\Rightarrow\nonumber\\
	&\boldsymbol{\Sigma}_{P_l}^{-1}\boldsymbol{\Delta}_l\boldsymbol{\Sigma}_{P_l}^{-1}\!\succ\!P_l\boldsymbol{\Sigma}_{P_l}^{-1}\boldsymbol{\Delta}_lP_l\boldsymbol{\Sigma}_{P_l}^{-1}\boldsymbol{\Delta}_l\boldsymbol{\Sigma}_{P_l}^{-1}\Rightarrow\boxed{I_1\!<\!0},\nonumber\\
	&\boldsymbol{I}\!\succ\!P_l\boldsymbol{\Sigma}_{P_l}^{-1}\boldsymbol{\Delta}_l\Rightarrow({\boldsymbol{\rho}_l}^T\boldsymbol{\rho}_l)^2\!>\!P_l({\boldsymbol{\rho}_l}^T\boldsymbol{\rho}_l)({\boldsymbol{\rho}_l}^T\boldsymbol{\Sigma}_{P_l}^{-1}\boldsymbol{\Delta}_l\boldsymbol{\rho}_l)\Rightarrow\nonumber\\
	&\boldsymbol{\Pi}_l\!\succ\!P_l\boldsymbol{\Pi}_l\boldsymbol{\Sigma}_{P_l}^{-1}\boldsymbol{\Delta}_l\Rightarrow\boldsymbol{\Sigma}_{P_l}^{-1}\boldsymbol{\Pi}_l\boldsymbol{\Sigma}_{P_l}^{-1}\!\succ\!P_l\boldsymbol{\Sigma}_{P_l}^{-1}\boldsymbol{\Pi}_l\boldsymbol{\Sigma}_{P_l}^{-1}\boldsymbol{\Delta}_l\boldsymbol{\Sigma}_{P_l}^{-1}\nonumber\\
	&\Rightarrow\boxed{I_2\!<\!0},\nonumber\\
	&{\boldsymbol{\rho}_l}^T\boldsymbol{\Sigma}_{P_l}^{-1}\boldsymbol{\Delta}_l\boldsymbol{\rho}_l\!\overset{(a)}{=}\!{\boldsymbol{\rho}_l}^T\boldsymbol{\Delta}_l\boldsymbol{\Sigma}_{P_l}^{-1}\boldsymbol{\rho}_l\Rightarrow({\boldsymbol{\rho}_l}^T\boldsymbol{\rho}_l){\boldsymbol{\rho}_l}^T\boldsymbol{\Sigma}_{P_l}^{-1}\boldsymbol{\Delta}_l\boldsymbol{\rho}_l\!=\!\nonumber\\
	&{\boldsymbol{\rho}_l}^T\boldsymbol{\Delta}_l\boldsymbol{\Sigma}_{P_l}^{-1}\boldsymbol{\rho}_l({\boldsymbol{\rho}_l}^T\boldsymbol{\rho}_l)\Rightarrow\boldsymbol{\Pi}_l\boldsymbol{\Sigma}_{P_l}^{-1}\boldsymbol{\Delta}_l\!=\!\boldsymbol{\Delta}_l\boldsymbol{\Sigma}_{P_l}^{-1}\boldsymbol{\Pi}_l\Rightarrow\nonumber\\
	&\boldsymbol{\Sigma}_{P_l}^{-1}\boldsymbol{\Pi}_l\boldsymbol{\Sigma}_{P_l}^{-1}\boldsymbol{\Delta}_l\boldsymbol{\Sigma}_{P_l}^{-1}\boldsymbol{\Delta}_l\boldsymbol{\Sigma}_{P_l}^{-1}\!=\!\boldsymbol{\Sigma}_{P_l}^{-1}\boldsymbol{\Delta}_l\boldsymbol{\Sigma}_{P_l}^{-1}\boldsymbol{\Pi}_l\boldsymbol{\Sigma}_{P_l}^{-1}\boldsymbol{\Delta}_l\boldsymbol{\Sigma}_{P_l}^{-1}\nonumber\\
	&\Rightarrow\boxed{I_3\!=\!0},\nonumber
	\end{align}
	%\vspace{-.1cm}
\end{subequations}
where ($a$) comes by the fact that ${\boldsymbol{\rho}_l}^T\boldsymbol{\Sigma}_{P_l}^{-1}\boldsymbol{\Delta}_l\boldsymbol{\rho}_l$ is scalar.} 
The Hessian of ${\cal F}$ with respect to $P_l$'s is diagonal and negative definite, which proves that ${\cal F}$ is jointly concave over $P_l$'s. Moreover, the constraint is linear in $P_l$, and therefore finding $P_l$'s in sub-problem $(a)$ of \eqref{max problem to minimize D_d} is jointly concave over $P_l$'s and has a unique solution.
%\section*{Acknowledgment} \label{acknowledgment}
%The authors would like to sincerely acknowledge the contributions by Alireza Sani of University of Central Florida to attain the aims of this paper. 
%This research is supported by NSF under grants CCF-1341966 and CCF-1319770.
%=======================================
%=======================================
%=======================================
%\begin{thebibliography}
%\vspace{-.05cm}
\bibliographystyle{IEEEtran}
\bibliography{myref}

%=======================================
%=======================================
%=======================================
%\end{thebibliography}

%\end{IEEEbiography}


%\newpage



%\vfill


\end{document}